\documentclass[11pt]{article}
\usepackage{epsfig}
\usepackage{verbatim}
\usepackage{amsmath}
\usepackage[pdfmenubar=false, letterpaper=false, pdfpagelabels=true]{hyperref}
\usepackage{makeidx,showidx}
\usepackage{pst-node}  
\usepackage[utf8]{inputenc} 

\newcommand{\code}{\texttt}

\begin{document}

\def\mvec#1{{\bm{#1}}}   

\newcommand{\bm}[1]{\mbox{\boldmath$#1$}}
\newcommand{\unbfm}[1]{\mbox{\boldmath$#1$}}

\def\etal{{\it et al }}      
\def\kmh#1#2{\underline{#1} \{#2\}}  
\def\given{\,|\,}   
\def\intd{\,\mbox{d}}    
\def\mvec#1{{\bm{#1}}}   
\def\mod{\mbox{mod}}     
\def\emod{\mbox{{\footnotesize mod}}}     

\newcommand{\hl}[1]{{\phantomsection\label{#1}}}  
\newcommand{\Rin}[1]{{\red \code{>\ {#1}}}}
\newcommand{\Rinc}[1]{{\red \code{+\ {#1}}}}
\newcommand{\Rout}[1]{{\blue \,\code{[1]\ {#1}}}}
\newcommand{\Routn}[1]{{\blue \code{{#1}}}}
\newcommand{\RoutD}[1]{{\blue \,\code{\ {#1}}}} 

\title{Learning about probabilistic inference and forecasting\\
by playing with multivariate normal distributions\footnote{Note 
based on lectures 
to PhD students in Rome.}\\
(with examples in R)
}
\author{G.~D'Agostini \\
Universit\`a ``La Sapienza'' and INFN, Roma, Italia \\
{\small (giulio.dagostini@roma1.infn.it,
 \url{http://www.roma1.infn.it/~dagos})}
}

\date{}

\maketitle

\begin{abstract}
The properties of the normal distribution under linear transformation,
as well the easy way to compute the covariance matrix of
marginals and conditionals,
offer a unique opportunity to get an insight about several
aspects of uncertainties in measurements.  
The way to build the overall covariance matrix in a few, 
but conceptually relevant 
cases is illustrated: several observations made with (possibly)
different instruments measuring the same quantity; 
effect of systematics (although limited to {\em offset}, in order
to stick to linear models) on the determination of the `true value',
as well in the prediction of future observations; 
correlations which arise when 
different quantities are measured 
with the same instrument affected by an offset uncertainty; 
inferences and predictions based on averages; inference about 
constrained values; fits under some assumptions 
(linear models with known standard deviations).
Many numerical examples are provided, exploiting the ability
of the R language to handle large matrices and to produce
high quality plots.
Some of the results are framed in the general problem of 
`propagation of evidence', crucial in analyzing graphical models
of knowledge.
\end{abstract}

\mbox{} \vspace{0.1cm}
{\small
\begin{flushright}
{\sl ``So far as the theories of mathematics are about reality,
they are not certain; }\\
{\sl  so far as they are certain,  they are not about reality.}\\
{(A. Einstein)} \\
\mbox{}\\
{\sl ``If we were not ignorant there would be no probability,} \\
{\sl there could only be certainty. }\\
{\sl But our ignorance cannot be absolute,}\\
{\sl for then there would be no longer any probability at all.''}\\
{(H. Poincaré)} \\
\mbox{}\\
{\sl ``Probability is good sense reduced to a calculus''} \\
{(S. Laplace)} \\
\mbox{}\\
{\sl ``All models are wrong but some are useful''} \\
{(G. Box)} \\
\end{flushright}
}

\thispagestyle{empty}

\section{Introduction}
The opening quotes set up the frame in which this
paper has been written: in the sciences we always
deal with uncertainties; being in condition on uncertainty
we can only state `somehow' how much we believe something;
in order to do that we need to build up probabilistic models
based on good sense. 
For example, if we are uncertain about 
the value we are going to {\em read on} an instrument, 
we can make probabilistic assessments about it. 
But in general our interest is the {\em numerical value 
of a physics quantity}. We are usually in great condition 
of uncertainty before the measurement, but we still remain
with some degree of uncertainty after the measurement
has been performed. Models enter in the construction
of the the causal network
which connects physics quantities to what we can observe
on the instruments. They are also important because it is convenient
to use, whenever it is possible, probability distributions,
instead than to assign individual probabilities to each individual
`value' (after suitable discretization) that a physics quantity
might assume.  

As we know, there are good reasons why in many
cases the Gaussian distribution (or {\em normal} distribution) 
offers a {\em reasonable} and {\em convenient} description of the 
probability that the quantity of interest lies within
some bounds. But it is important to remember that, 
\underline{as it was clear to Gauss}~\cite{Gauss} 
when he derived the famous
distribution for the measurement errors, one should not
take literally the fact that the variable appearing in the formula
can range from minus infinite to plus infinite: an apple cannot
have infinite mass, or a negative one! 

Sticking hereafter to Gaussian distributions, it is 
clear that if we are only interested to the probability density
function (pdf) of a variable at the time, we can only describe 
our uncertainty about that quantity, and nothing more.
The game becomes interesting when we study the joint 
distribution of several variables, because this is the
way we can learn about some of them assuming the values
of the others. For example, if we assume 
the joint pdf $f(x_1,x_2\,|\,I)$ of variables $X_1$ and $X_2$
under the state of information $I$ (on which we ground our 
assumptions), we can evaluate $f(x_1\,|\,x_2\,,I)$, that is
the pdf adding the extra condition  $X_2=x_2$, which is 
usually not the same as  $f(x_1\,|\,I)$, that is the pdf of $X_1$
for any value $X_2$ might assume.\footnote{The pdf 
$f(x_1\,|\,I)$ is called {\em marginal}, although 
there is never special about this name, since all 
distributions of a single variable can be thought as being `marginal'
to all other possible quantities which we are not interested about.
 $f(x_1\,|\,x_2\,,I)$ is instead `called' {\em conditional},
although it is a matter of fact that \underline{all} distributions
are conditional to a given state of information, here indicated by
$I$. Note that throughout this paper will shall use
the same symbol $f()$ for all pdf's, as it is customary among 
physicists -- I have met mathematics oriented guys getting mad
by the equation $f(x,y) = f(x|y)\cdot f(y)$ because, they say, 
``the three functions cannot be the same''...
} 

\begin{figure}
\begin{center}
\epsfig{file=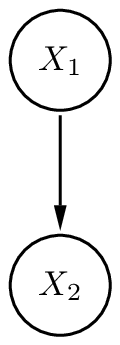,clip=}\hspace{1.5cm}
\epsfig{file=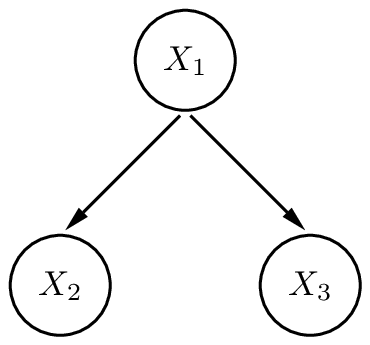,clip=}\hspace{1.0cm}
\epsfig{file=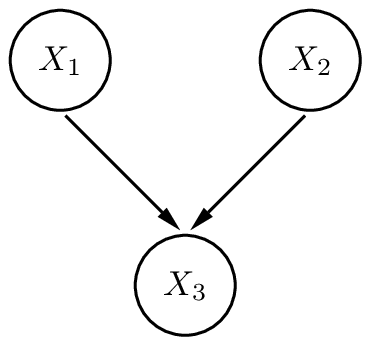,clip=}
\end{center}
\caption{Basic models of joint probabilities}
\label{fig:modelli_base}
\end{figure}

Let us take for example the three diagrams of 
Fig.~\ref{fig:modelli_base}
to which we give a physical interpretation:
\begin{enumerate}
\item In the diagram on the left the variable $X_1$ might represent 
      the numerical value of a physics quantity, on which we
      are in condition on uncertainty, modelled by 
      \begin{eqnarray}
       X_1 \sim {\cal N}(X_0, \sigma_{1})\,,
      \end{eqnarray} 
      where $X_0$ and $\sigma_1$ are suitable parameters
      to state our `ignorance' about $X_1$ (`complete ignorance',
      if it does ever exist, is recovered in the limit
      $\sigma_1\rightarrow \infty$). 
      Instead, $X_2$ is then what we read on an instrument
      when we apply it to $X_1$. 
      That is, even if we knew $X_1$, 
      we are still uncertain about what we can read
      on the instrument, as it is well understood. 
      Modelling this uncertainty by a normal distribution
      we have, for any value of $X_1$
      \begin{eqnarray}
      \left.X_2\right|_{X_1} \sim {\cal N}(X_1, \sigma_{2|1})\,, 
      \end{eqnarray} 
      where $\sigma_{2|1}$ is a compact symbol for 
      $\sigma(\left.X_2\right|_{X_1})$ and which is in general
      different from $\sigma_2 \equiv \sigma(X_2)$. In fact
      our uncertainty about $X_2$ (for any possible
      value of $X_1$ ) must be larger than that about
      $X_1$ itself, for obvious reasons -- we shall see later the details.
\item In the diagram on the center $X_3$ might represent
      a second observation done {\em independently} applying in general
      a second (possibly different) 
      instrument to the identical value $X_1$.
      This means that $\left.X_2\right|_{X_1}$ and $\left.X_3\right|_{X_1}$
      are independent, although  $X_2$ and $X_3$ are \underline{not}, 
      as we shall see. 
\item In the diagram on the right $X_3$ is the observation 
      read on the instrument applies to $X_1$, but possibly
      influenced by $X_2$, that might then represent 
      a kind of {\em systematics}. 
\end{enumerate}
Note, how it has been precisely stated, that $X_2$ of the first and 
of the second diagrams, as well as $X_3$ of the other two,
are the {\em readings} on the instruments and \underline{not}
the result of the measurement! This is because  by 
``result of the measurement'' we mean statements about the quantity
of interest and not about the quantities read on the instruments
(think for example at the an experiment measuring the Higgs 
boson mass, making use of the information recorded by the detector!).
In this case the ``result of the measurement'' would
be $f(x_1\,|\,\mbox{\bf data}\,,I)$ where {\bf data}
stands for the set of observed variables.

The diagrams of the figure can be complicated, using sets of 
data, with systematics effects common to observations 
in each subset. The aim of this
paper is to help in developing some intuition of what is going on in
problems of this kind, with the only simplification that 
all pdf's of interest are normal. 

\newpage
\section{Technical premises (with some exercises)}
We assume that the reader is familiar with some basic
concepts related to {\em uncertain numbers} and {\em uncertain vectors},
usually met under the name of ``random variables''. 

\subsection{Normal (Gaussian) distribution}\label{ss:normal_distribution}
$X\sim {\cal N}(\mu,\sigma)$: \\
\vspace{-0.3cm}
\begin{equation}
f(x\,|\,{\cal N}(\mu,\sigma))
=\frac{1}{\sqrt{2\,\pi}\,\sigma}\,
\exp\left[-\frac{(x-\mu)^2}{2\,\sigma^2}\right]
\end{equation}

with 
\begin{eqnarray}
\mbox{E}[X] &=& \mu \\
\mbox{Var}[X] &=& \sigma^2 \\
\sigma[X] = \sqrt{\mbox{Var}[X]} &=& \sigma\,.
\end{eqnarray}
(We remind that in most physics applications 
$x\rightarrow\pm\infty$ simply means $|x-\mu|/\sigma \gg 1$.)

\begin{figure}[t]
\centering\epsfig{file=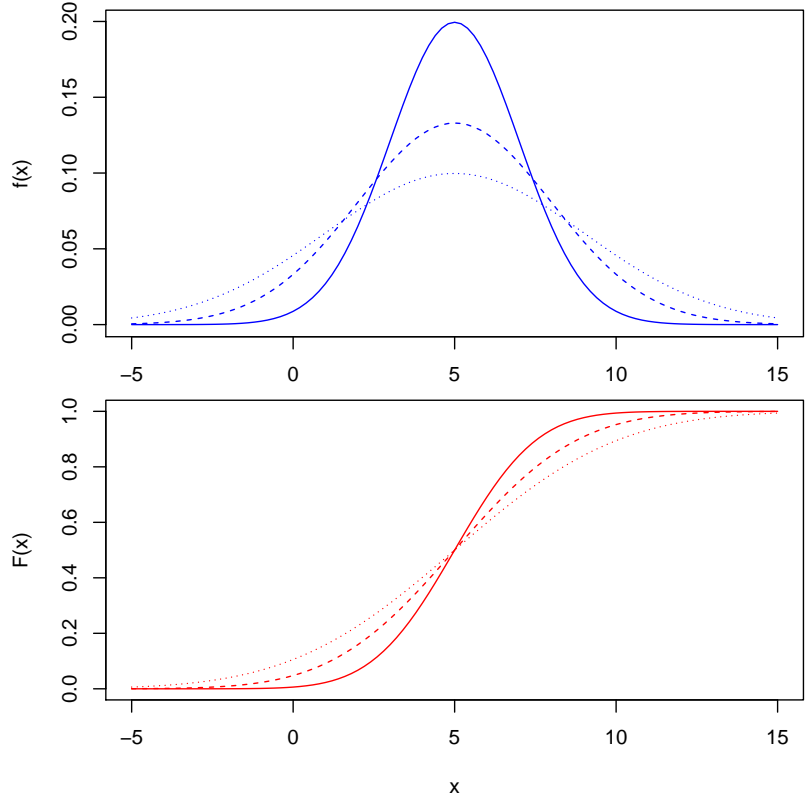,clip=,width=0.7\linewidth}
\caption{Gaussian probability density function (above) and cumulative
function (below) for $\mu=5$ and $\sigma=2$, 3 and 4
(solid, dashed and pointed).}
\label{fig:gaussian_f-F}
\end{figure}

In the R language~\cite{R} there are functions 
(\code{dnorm()}, \code{pnorm()} and \code{qnorm()}, respectively)
to calculate the pdf, 
the cumulative function, usually indicated with
``$F(x)$'', as well as its inverse, as shown 
in the following, self explaining examples\footnote{For information about
the language see one of the many tutorial available on the web.
Most functions we shall use here have self explaining names.
For an help, for example about \code{dnorm()}, just enter\\
\Rin{?dnorm}
 } (`$>$' is the R console prompt):\\
\Rin{dnorm(0, 0, 1)}\\
\Rout{0.3989423}\\
\Rin{1/sqrt(2*pi)  \ \ \# (just a check) } \\
\Rout{0.3989423}\\
\Rin{pnorm(0, 0, 1)}\\
\Rout{0.5}\\
\Rin{pnorm(7, 5, 2) - pnorm(3, 5, 2)}\\
\Rout{0.6826895}\\
\Rin{qnorm(0.5, 5, 2)}\\
\Rout{5}\\
\Rin{qnorm(1, 5, 2)}\\
\Rout{Inf}\\
\Rin{qnorm(0, 5, 2)}\\
\Rout{-Inf}\\
Note the capability of the language to handle infinities,
as it can be cross checked by\\
\Rin{pnorm(Inf, 5, 2)}\\
\Rout{1}\\
And here are the instructions to produce the plots of 
figure \ref{fig:gaussian_f-F}.
\mbox{}\vspace{-0.2cm}
{\small
\begin{verbatim}
mu <- 5; sigma <- 2; x <- seq(mu-5*sigma, mu+5*sigma, len=101)
plot(x, dnorm(x, mu, sigma), ty='l', ylab='f(x)', col='blue')
points(x, dnorm(x, mu, sigma*1.5), ty='l', lty=2, col='blue')
points(x, dnorm(x, mu, sigma*2), ty='l', lty=3, col='blue')
plot(x, pnorm(x, mu, sigma), ty='l', ylab='F(x)', col='red')
points(x, pnorm(x, mu, sigma*1.5), ty='l', lty=2, col='red')
points(x, pnorm(x, mu, sigma*2), ty='l', lty=3, col='red')
\end{verbatim}
} \noindent

\subsection{Bivariate and multivariate normal distribution}
The joint distribution of a bivariate normal distribution is 
given by 
\begin{eqnarray}
f(\bm{x}\,|\,{\cal N}(\bm{\mu},\bm{V})) &=&
\frac{1}{2\,\pi\,\sigma_1\,\sigma_2\,\sqrt{1-\rho_{12}^2}}\,  
  \exp \left\{
                -\frac{1}{2\,(1-\rho_{12}^2)} 
                 \left[  \frac{(x_1-\mu_1)^2}{\sigma_1^2} \right.\right.
 \nonumber \\
&& \left.\left.   - 2\,\rho_{12}\,\frac{(x_1-\mu_1)(x_2-\mu_2)}{\sigma_1\,\sigma_2}
        + \frac{(x_2-\mu_2)^2}{\sigma_2^2}
                 \right]
         \right\} \,,
\label{eq:bivar}
\end{eqnarray}
where
\begin{eqnarray}
\bm{x}   &=& (x_1, x_2) \\
\bm{\mu} &=& (\mu_1, \mu_2) \\
\mbox{E}[X_i] &=& \mu_i \\
\mbox{Var}[X_i] &=& \sigma_i^2 \\
\sigma[X_i] \equiv \sqrt{\mbox{Var}[X_i]} &=& \sigma_i\\
\rho_{12}&=& \frac{\mbox{Cov}[X_1,X_2]}{\sigma_1\,\sigma_2}\,, 
\end{eqnarray}
with variances and covariances forming the {\em covariance matrix} 
\begin{eqnarray}
\bm{V} &=& \left(\!\begin{array}{cc}
            \mbox{Var}[X_1] & \mbox{Cov}[X_1,X_2] \\
             & \\
             \mbox{Cov}[X_1,X_2] & \mbox{Var}[X_2] 
            \end{array}
      \!\right) =
      \left(\!\begin{array}{cc}
            \sigma_1^2 & \rho_{12}\,\sigma_1\,\sigma_2 \\
             & \\
            \rho_{12}\,\sigma_1\,\sigma_2 &  \sigma_2^2
            \end{array}
      \!\right) 
\end{eqnarray}
The bivariate pdf (\ref{eq:bivar}) can be rewritten 
in a compact form as 
\begin{eqnarray}
f(\mvec x\,|\,{\cal N}(\bm{\mu},\bm{V}) ) &=& (2 \pi)^{-n/2}|\bm{V}|^{-1/2}\,
            \exp \left[-\frac{1}{2}\,(\mvec{x}-\mvec{\mu})^T\,\bm{V}^{-1}
            \,(\mvec{x}-\mvec{\mu})  
            \right]
            \,,  \label{eq:normale_multivariata_gen}
\end{eqnarray}
where $|\bm{V}|$ stands for det($\bm{V}$). This 
expression is valid for any number $n$ of  variables
and it turns, \underline{in the case $\bm{V}$ is  diagonal}, 
into
\begin{equation}
\frac{1}{(2 \pi)^{n/2}\,\prod_i\sigma_i}
            \exp \left[-\frac{1}{2}\,\frac{\sum_i(x_i-\mu_i)^2}{\sigma_i^2}
            \right]\,. \label{eq:normale_multivariata_ind}
\end{equation}
(For an extensive, although mathematically oriented
treatise on multivariate distribution see Ref.~\cite{Eaton},
freely available online.) 

\vspace{0.5cm}
\subsubsection{Multivariate normals in R}
Functions to calculate multivariate normal pdf's, 
as well as cumulative functions and random generators
are provided in R via the package 
\code{mnormt}\footnote{\url{http://cran.r-project.org/web/packages/mnormt/}} 
that needs first to be installed\footnote{For all technical details
about R (open source and multi-platform!) see 
the R web site~\cite{R}.}
issuing\\ 
\Rin{install.packages("mnormt")}\\ 
and then loaded
by the command\\ 
\Rin{library(mnormt)}\\ 
Then we have to define the values of the parameters
and built up the vector of the central values and
the covariance matrix. Here is an example:\\
\Rin{m1=0.4; m2=2; s1=1; s2=0.5; rho=0.6}\\ 
\Rin{mu <- c(m1, m2)}\\ 
\Rin{( V  <- rbind( c( s1\^{}2, rho*s1*s2), c(rho*s1*s2, s2\^{}2) ) )}\\ 
\Routn{\ \ \ \ \ [,1] [,2]}\\
\Routn{[1,]\  1.0 0.30}\\
\Routn{[2,]\  0.3 0.25}\\
Then we can evaluate the joint pdf in a point $(x_1,x_2)$, e.g.\\
\Rin{dmnorm(c(0.5, 1.5), mu, V)}\\ 
\Rout{0.1645734}\\
Or we can evaluate $P(X_1 \le 0.5\  \&\   X_2 \le 1.5)$, or 
 $P(X_1 \le \mu_1\  \&\   X_2 \le \mu_2)$, respectively, with\\
\Rin{pmnorm(c(0.5, 1.5), mu, V)}\\
\Rout{0.140636}\\
and\\
\Rin{pmnorm(mu, mu, V)}\\
\Rout{0.3524164}\\

\subsection{Graphical representation of normal bivariates}
\begin{figure}
\begin{center}
\epsfig{file=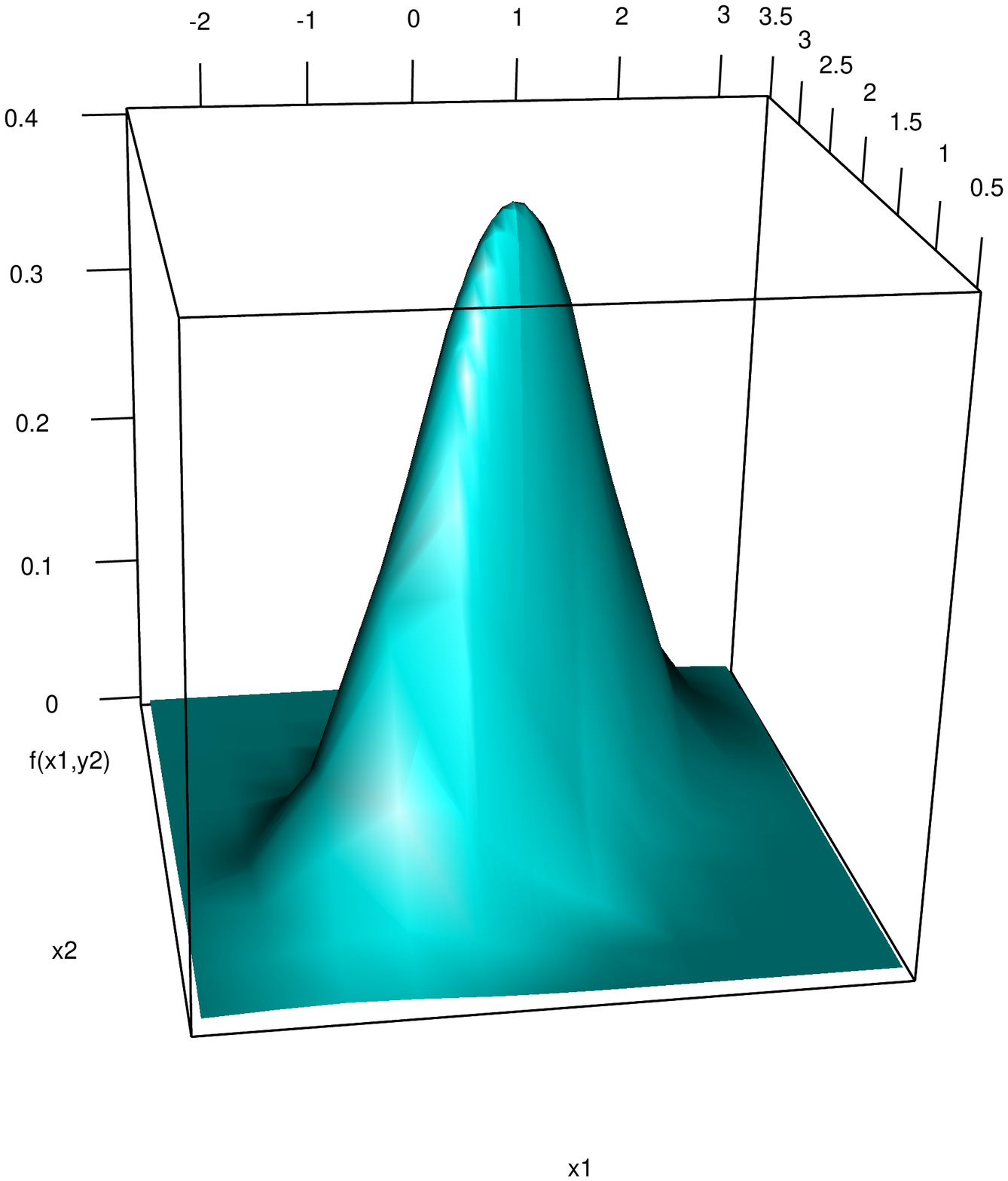,width=0.45\linewidth,clip=}
\epsfig{file=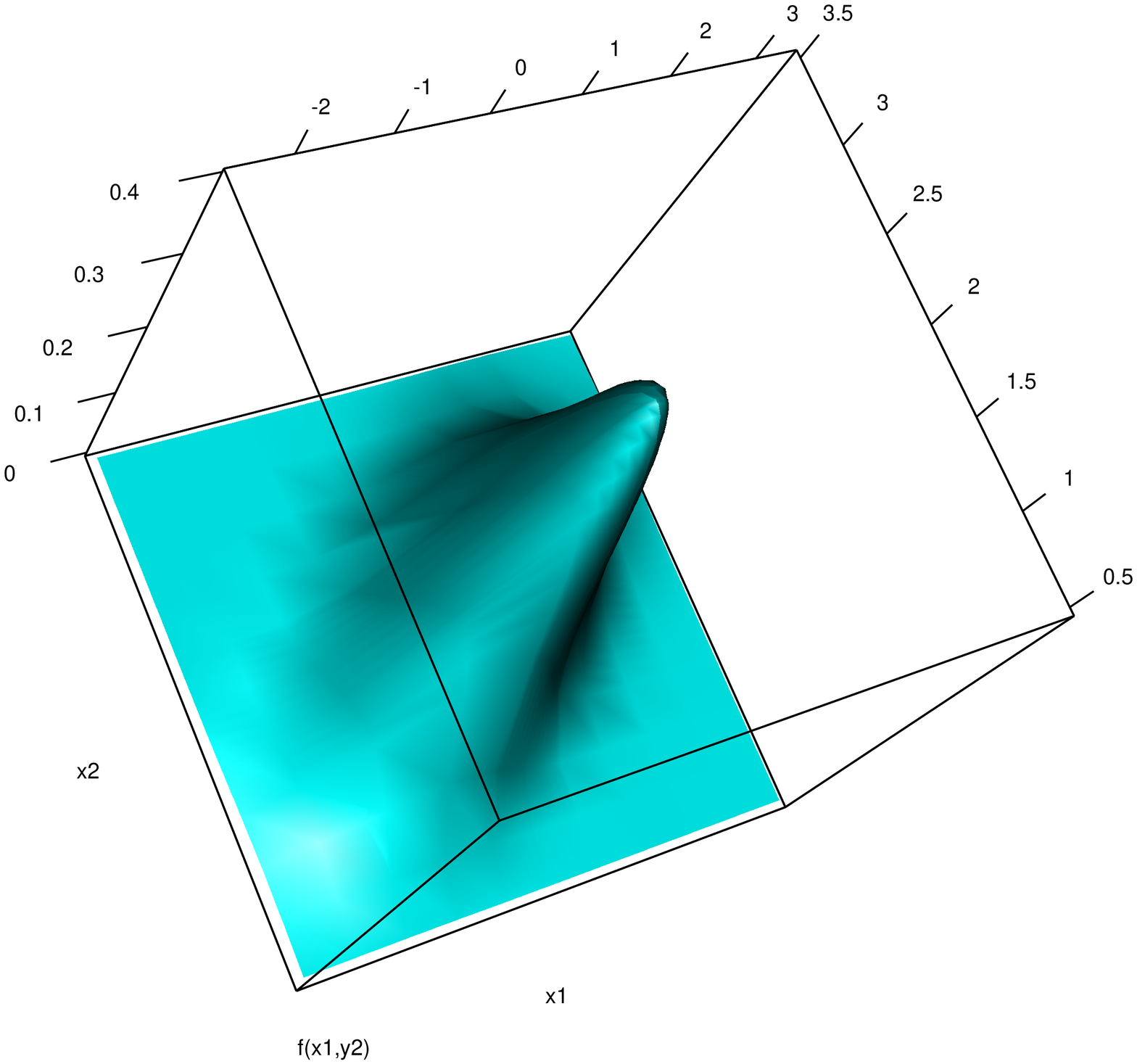,width=0.48\linewidth,clip=}
\mbox{}\vspace{1.0cm}\mbox{}\\
\epsfig{file=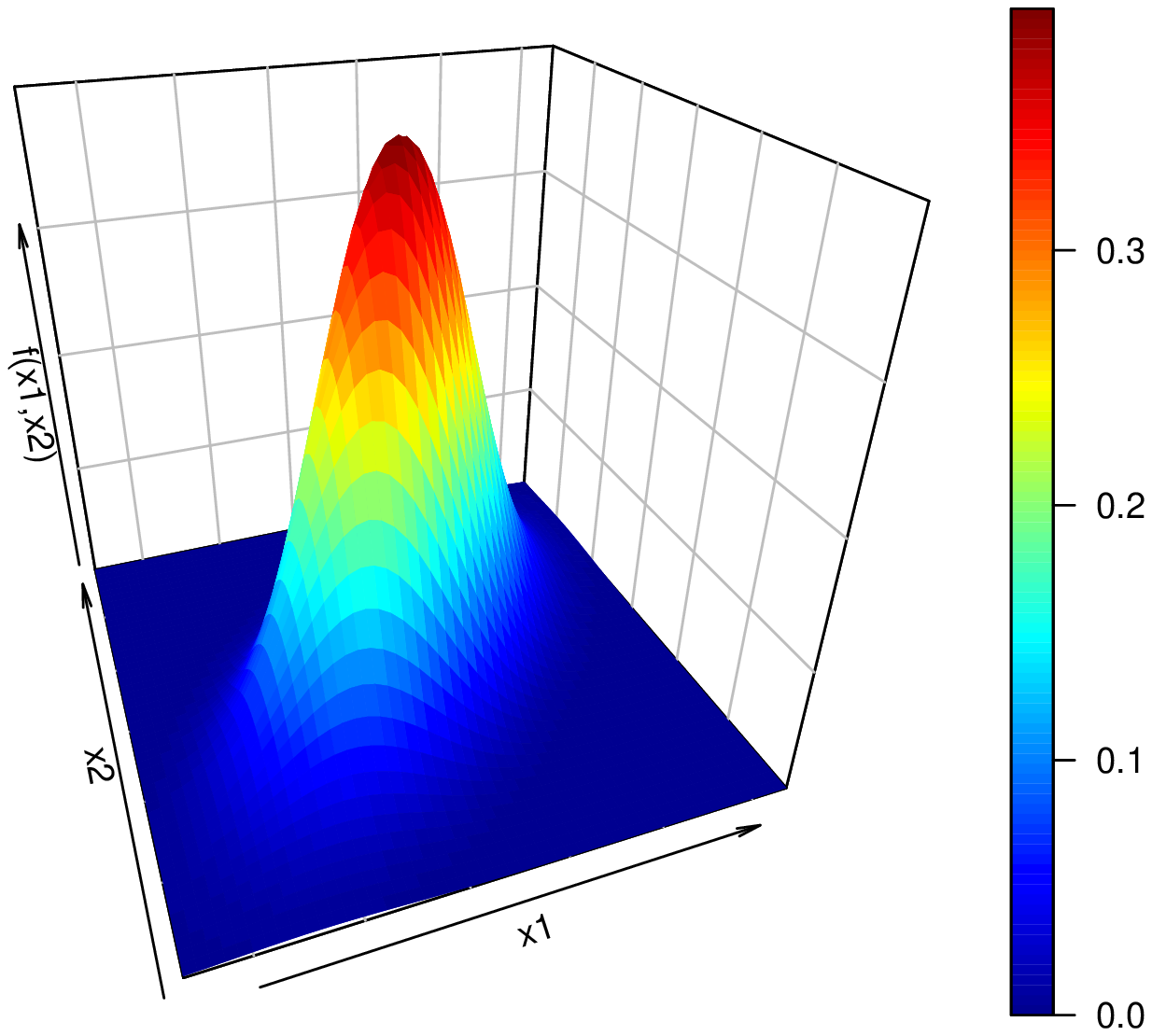,width=0.65\linewidth,clip=}
\end{center}
\caption{Three views of the bivariate normal distribution 
obtained with the R code provided in the text
($\mu_1=0.4$, $\mu_2=2$, $\sigma_1=0.1$,  $\sigma_2=0.5$, $\rho_{12}=0.6$). 
The above two are 
obtained by \code{perp3d()} of the package {\em rgl}, 
producing interactive 3D plots. The one below is produced by 
\code{surf3D()} of the package \code{plot3D}.}
\label{fig:normale_bivariata}
\end{figure}

If we like to visualize the
joint distribution we need a 3D graphical package, for example 
\code{rgl}\footnote{\url{https://r-forge.r-project.org/projects/rgl/}}
or \code{plot3D}.\footnote{http://www.r-bloggers.com/3d-plots-in-r/}
We need to evaluate the joint pdf on a grid
of values `$x$' and `$y$' and provide them to the suited function. 
Here are the instructions that use the \code{persp3d()} of the
\code{rgl} package:\\
\Rin{library(rgl)}\\
\Rin{fun <- function(x1,x2) dmnorm(cbind(x1, x2), mu, V)}\\
\Rin{x1 <- seq(m1-3*s1, m1+3*s1, len=51)  }\\
\Rin{x2 <- seq(m2-3*s2, m2+3*s2, len=51)}\\
\Rin{f <- outer(x1, x2, fun)}\\
\Rin{persp3d(x1, x2, f, col='cyan', xlab="x1", ylab="x2", zlab="f(x1,y2)")}\\
After the plot is shown in the graphics window, the window can be enlarged
and the plot rotated at wish. Figure \ref{fig:normale_bivariata}
shows in the upper two plots two views of the
same distribution.

Here are also the instructions to use \code{plot3D()}:\\
\Rin{library(plot3D)}\\
\Rin{M <- mesh(x1, x2)}\\
\Rin{surf3D(M\$x, M\$y, f, bty='b2', phi = 30, theta = -20,}\\
\Rinc{xlab='x1', ylab='x2', zlab='f(x1,x2)') }\\
The result is shown in the lower 
plot of Fig.~\ref{fig:normale_bivariata}.

Another convenient and often used representation 
of normal bivariates is to draw iso-pdf contours, i.e.
lines in correspondence of the points in the plane 
$(x_1,x_2)$ such as $f(x_1,x_2\,|\,I)=\mbox{const}$. 
This requires that the {\em quadratic form} 
at the exponent of Eq.~(\ref{eq:bivar})  [that is what is written in general as 
$(\mvec{x}-\mvec{\mu})^T\,\bm{V}^{-1}
            \,(\mvec{x}-\mvec{\mu})$]
has a fixed value. In the two dimensional case of Eq.~(\ref{eq:bivar}) 
we recognize the expression of an ellipse. We have in R the 
convenient package 
\code{ellipse}\footnote{\url{http://cran.r-project.org/web/packages/ellipse/}} to
evaluate the points of such an ellipse, given the vector of
expected values, the covariance matrix and the probability
that a point falls inside it. 
Here is the script that applies the function 
to the same bivariate normal of Fig.~\ref{fig:normale_bivariata}, thus
producing the contour plots of Fig.~\ref{fig:normale_bivariata_ellipse}:
\begin{figure}[t]
\begin{center}
\epsfig{file=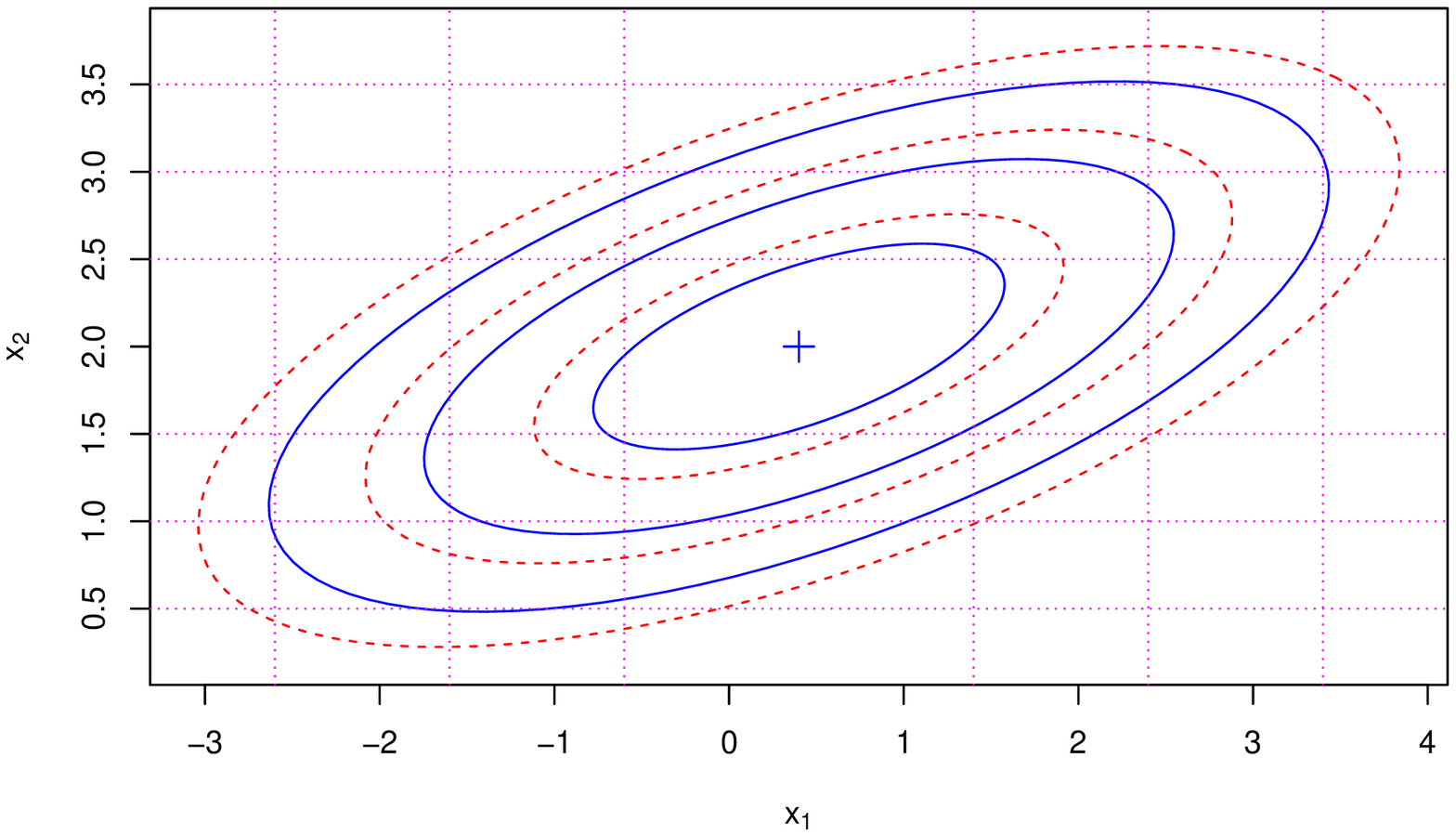,width=\linewidth,clip=}
\end{center}
\caption{Contour plots of the same bivariate normal of 
Fig.~\ref{fig:normale_bivariata}. The solid lines show the ellipses
inside which there is, from the smaller to the larger, 
50\%, 90\% and 99\% probability that a point $(x_1,x_2)$ falls
inside them. 
The dashed ellipses define instead the 68.3\%, 95.5\% and 
99.73\% probability contours [these are the (in-)famous 1-$\sigma$, 
2-$\sigma$ and 3-$\sigma$ contours, not simply related to the
standard deviations of the individual variable, whose 
1-$\sigma$, 2-$\sigma$ and 3-$\sigma$ bounds are indicated by the
dotted vertical and horizontal lines].}
\label{fig:normale_bivariata_ellipse}
\end{figure}
\mbox{}\vspace{-0.1cm}
{\small
\begin{verbatim}
plot( ellipse(V, centre=mu, level=0.9973), ty='l', lty=2, col='red',
     asp=1, xlab=expression(x[1]), ylab=expression(x[2]) )
points( ellipse(V, centre=mu, level=0.99), ty='l', col='blue')
points( ellipse(V, centre=mu, level=0.954), ty='l', lty=2, col='red')
points( ellipse(V, centre=mu, level=0.5), ty='l', col='blue')
points( ellipse(V, centre=mu, level=0.683), ty='l',  lty=2, col='red')
points( ellipse(V, centre=mu, level=0.90), ty='l', col='blue')
points(mu[1], mu[2], pch=3, cex=1.5, col='blue')
for(k in 1:3) {
  abline(v=mu[1]-k*sqrt(V[1,1]), lty=3, col='magenta')
  abline(v=mu[1]+k*sqrt(V[1,1]), lty=3, col='magenta')
  abline(h=mu[2]-k*sqrt(V[2,2]), lty=3, col='magenta')
  abline(h=mu[2]+k*sqrt(V[2,2]), lty=3, col='magenta')
}
\end{verbatim}
} \noindent
The probability to find a point inside the ellipse contour
is defined by the argument \code{level}. 
The ellipses drawn 
with solid lines define, in order of size, 
50\%, 90\% and 99\% contours. For comparison there are
also the contours at 68.3\%, 95.5\% and 
99.73\%, which define the {\em highly confusing}
1-$\sigma$ , 2-$\sigma$ and 3-$\sigma$ contours. 
Indeed, the probability that each of the variable
falls in the interval of $\mbox{E}[X_i]\pm k\,\sigma[X_i]$ 
has little to do with these ellipses. If we are interested
to the probability that a point falls in a rectangles
defined by  
$(\mbox{E}[X_1]\pm k\,\sigma[X_1]\, \& \,\mbox{E}[X_2]\pm k\,\sigma[X_2])$ 
the probability needs to be calculated making the integral 
of the joint distribution inside the rectangle
(some of these rectangles are shown in 
Fig.~\ref{fig:normale_bivariata_ellipse} by the dotted lines,
that indicate 1-$\sigma$ , 2-$\sigma$ and 3-$\sigma$ bound in 
the individual variable).  

Let us see how to evaluate in R the probability that a point
falls in a rectangle, making use of the cumulative probability
function \code{pmnorm()}. In fact the probability in a rectangle
is related to the cumulative distribution by the following relation
\\ \mbox{}\vspace{-0.7cm}
\begin{eqnarray}
P[\,(x_{1_m} \le X_1 \le x_{1_M})\, \&\, (x_{2_m} \le X_2 \le x_{2_M})\,] 
&=&   P[\,( X_1 \le x_{1_M})\, \&\, ( X_2 \le x_{2_M}) ]  \nonumber \\
&&  - P[\,( X_1 \le x_{1_M})\, \&\, ( X_2 \le x_{2_m}) ]  \nonumber \\
&&     - P[\,( X_1 \le x_{1_m})\, \&\, ( X_2 \le x_{2_M}) ]  \nonumber \\
&&   +  P[\,( X_1 \le x_{1_m})\, \&\, ( X_2 \le x_{2_1}) ]\,, \hspace{0.9cm}
\end{eqnarray}
\\ \mbox{}\vspace{-1.0cm} \\
that can be implemented in an R function:
\mbox{}\vspace{-0.2cm}
{\small
\begin{verbatim}
p.rect.norm <- function(xlim, ylim, mu, V, sigmas=FALSE, ...) {
  # The argument '...' might be useful to pass extra arguments to pmnorm.    
  if ( (length(mu) != 2) | sum( dim(V) != c(2,2) )   # some check
      | (length(xlim) != 2) | (length(ylim) != 2) ) {
    print("wrong dimensions in one of parameters")
    return(NULL)
  } else if ( sum( eigen(V)$values <= 0 ) >  0) {
    cat( sprintf("V is not positively defined\n") )
    return(NULL)
  } 
  # If argument 'sigmas' is TRUE: 
  if( sigmas ) { # rectangular defined in units of individual sigma around mu
    xlim <- mu[1] + xlim * sqrt(V[1,1]) 
    ylim <- mu[2] + ylim * sqrt(V[2,2])
  }
  library(mnormt)
  p.rect <- pmnorm( c(xlim[2], ylim[2]), mu, V, ...) -
            pmnorm( c(xlim[2], ylim[1]), mu, V, ...) -
            pmnorm( c(xlim[1], ylim[2]), mu, V, ...) +
            pmnorm( c(xlim[1], ylim[1]), mu, V, ...)  
  return(p.rect)
}
\end{verbatim}
} \noindent
For example\footnote{For Monte Carlo oriented guys, here is how to
cross check the results (don't expect to reproduce 51313!):\\
\Rin{xy <- rmnorm(100000, mu, V)}\\
\Rin{length( xy[,1][ xy[,1]  > m1 - s1 \& xy[,1]  < m1+s1 \& xy[,2]  > m2 - s2 \& xy[,2]  < m2+s2 ] )}\\
\Rout{51313}
}\\
\Rin{p.rect.norm(c(m1-s1, m1+s1), c(m2-s2, m2+s2), mu, V)}\\
\Rout{0.5138685}\\
\Rin{p.rect.norm(c(-1, 1), c(-1, 1), mu, V, sigmas=TRUE)}\\
\Rout{0.5138685}\\
As a cross check, let us calculate the probabilities in strips 
of plus/minus one standard deviations around the 
averages (the `strips' provide a good intuition of what
a `marginal' is):\\
\Rin{p.rect.norm(c(-1, 1), c(-10, 10), mu, V, sigmas=TRUE)}\\
\Rout{0.6826895}\\
\Rin{p.rect.norm(c(-10, 10), c(-1, 1), mu, V, sigmas=TRUE)}\\
\Rout{0.6826895}

\subsection{Marginals (and `multivariate marginals') of multivariate normals}
A nice feature of the multivariate normal distribution is that
if we are just interested to a subset of variables alone, 
neglecting which value the other ones can take (`marginalizing'),
we just drop from $\mvec{\mu}$ and from $V$ 
the uninteresting values, or the relative rows and columns, respectively. 
For example, if we have -- see subsection \ref{sss:syst_X3} --
\begin{equation}
\bm{\mu} = \left(\! \begin{array}{c}
                   1.96 \\ 0.02 \\ 1.98
                   \end{array} 
          \!\right)
\hspace{1.0cm}
\bm{V} = \left(\! \begin{array}{ccc}
                  1.96 & -0.98 & 0.98 \\
                  -0.98 & 0.99 & 0.01 \\
                  0.98  & 0.01 & 1.99 
                   \end{array} 
          \!\right)
\end{equation}
marginalizing over the second variable (i.e. being
only interested in the first and the third) we obtain
\begin{equation}
\bm{\mu}' = \left(\! \begin{array}{c}
                   1.96  \\ 1.98
                   \end{array} 
          \!\right)
\hspace{1.0cm}
\bm{V}' = \left(\! \begin{array}{cc}
                  1.96 &   0.98 \\
                  0.98  &  1.99 
                   \end{array} 
          \!\right)
\end{equation}
Here is a function that returns expected values and variance
of the multivariate `marginal'
{\small
\begin{verbatim}
marginal.norm <- function(mu, V, x.m) {
  # x.m is a vector with logical values (or non zero) indicating
  # the elements on which to marginalise (the others are 0, NA or FALSE)
  x.m[is.na(xm)] <- FALSE
  v <- which( as.logical(x.m) )
  list(mu=mu[v], V=V[v, v])
}
\end{verbatim}
} \noindent
(Note how the function has been written in a very compact form,
exploiting some peculiarities of the R language. In particular,
the elements of \code{x.m} to which we are interested can be
 \code{TRUE}, or can be a numeric value different from zero;
the others can be \code{FALSE}, \code{0} or \code{NA}.)

\subsection{Conditional distribution of a variable, given its
bivariate distribution with another variable}
A different problem is the pdf of one of variables, say $X_1$,
for a given value of the other. This is not as straightforward
as the marginal (and for this reason in this subsection we only consider
the bivariate case). Fortunately the distribution is still a
Gaussian, with {\em shifted central value} and {\em squeezed width}:
\begin{equation}
\left.X_1\right|_{x_2} \sim {\cal N}\left( \mu_1+\rho_{12}\,\frac{\sigma_1}{\sigma_2}
                  \,\left(x_2 - \mu_2\right),\,
                 \sigma_1\sqrt{1-\rho_{12}^2}\right)\,,
\label{eq:x1_cond1}
\end{equation}
i.e. 
\begin{eqnarray}
\mbox{E}[X_1] &=& \mu_1 +\rho_{12}\,\frac{\sigma_1}{\sigma_2}\, (x_2-\mu_2) 
\label{eq:x1_cond1_E} \\
\mbox{Var}[X_1] &=& \sigma_1^2\cdot(1-\rho_{12}^2) \label{eq:x1_cond1_Var} \\
\sigma[X_1] &=& \sigma_1 \cdot\sqrt{1-\rho_{12}^2}\,. 
\end{eqnarray}
And, by symmetry, 
\begin{equation}
\left.X_2\right|_{x_1} \sim {\cal N}\left( \mu_2+\rho_{12}\,\frac{\sigma_2}{\sigma_1}
                  \,\left(x_1 - \mu_1\right),\,
                 \sigma_2\sqrt{1-\rho_{12}^2}\right)\,.
\end{equation}
Mnemonic rules to remember Eqs.~(\ref{eq:x1_cond1_E}) and 
 (\ref{eq:x1_cond1_Var}) are
\begin{itemize}
\item the shift of the expected value depends linearly on the 
correlation coefficient as well on the difference between the 
value of the conditionand ($x_2$) and its expected value
($\mu_2$); the ratio $\sigma_1/\sigma_2$ can be seen as a
minimal dimensional factor in order to get a quantity
that has the same dimensions of $\mu_1$ 
(remember that $X_1$ and $X_2$ have in general 
different physical dimensions);
\item the variance is reduced by a factor which depends on 
      the absolute value of the correlation coefficient,
      but not on its sign. In particular it goes to zero
      if $|\rho_{12}|\rightarrow 1$, limit in which the two quantities
      become linear dependent, while it does not change
      if $\rho_{12}\rightarrow 0$, since the two variables become
      independent and they cannot effect each other. (In general
      independence implies $\rho=0$. For the normal bivariate
      it is also true the other way around.)
\end{itemize}
\begin{figure}
\centering\epsfig{file=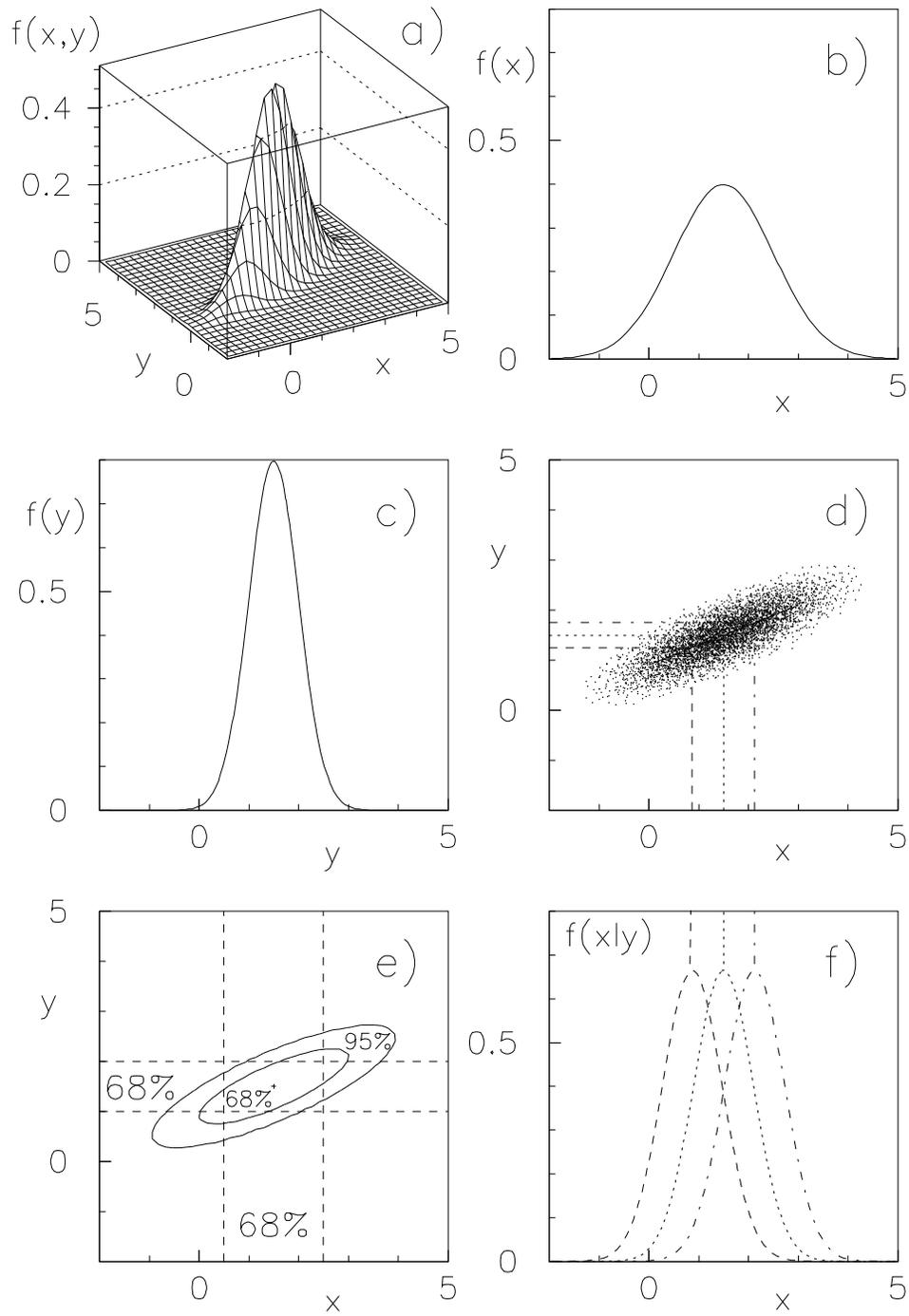,width=0.86\linewidth,clip=}
\caption{Example of bivariate normal distribution.}
\label{fig:bivar}
\end{figure}
An example of a bivariate distribution (from \cite{BR}, with
$x_1$ and $x_2$ indicated as customary with $x$ and $y$) is 
given in Fig.~\ref{fig:bivar}, which shows also the marginals
and some conditionals.

\subsubsection{Evaluation of a conditional from a given bivariate normal}
As an exercise, lets prove (\ref{eq:x1_cond1}), with the purpose
of show some useful tricks to simplify the calculations.
If we take literally the rule to evaluate $f(x_1\,x_2\,|\,I)$ knowing
that $f(x_1,x_2\,|\,I)$ is given by (\ref{eq:bivar}) we need to calculate
\begin{eqnarray}
f(x_1\,|\,x_2,\,I) &=& \frac{f(x_1,x_2\,|\,I)}{f(x_2\,|\,I)}\,.
\end{eqnarray}
The trick is to make the calculations neglecting all 
irrelevant multiplicative
factors, starting from the whole denominator $f(x_2\,|\,I)$,
which \underline{is a number}
given $X_2=x_2$ (whatever its value might be!). 

Here are the details (note that additive terms in the exponential 
are factors in the function of interest!):\footnote{Essentially
the trick consists in observing that if we have a pdf
proportional to $\exp{[-h^2\,(x^2+\alpha\,x)]}$, then it is
also proportional to 
  $$\exp{\left[-h^2\,\left(x^2+2\,\frac{\alpha}{2}\,x + 
\left(\frac{\alpha}{2}\right)^2\right)\right]} = 
\exp{\left[-h^2\,\left(x- \left(-\frac{\alpha}{2}\right) \right)^2\right]}\,,$$
that is a Gaussian with $\mu=-\alpha/2$ and $\sigma^2=1/(2\,h^2)$.
}
\begin{eqnarray}
f(x_1\,|\,x_2,\,I) \!\! &\propto & \!\! f(x_1,x_2\,|\,I) \nonumber \\
\!\!&\propto&\!\! 
\exp \left\{  -\frac{1}{2\,(1-\rho_{12}^2)} 
                 \left[  \frac{(x_1-\mu_1)^2}{\sigma_1^2}
  - 2\,\rho_{12}\,\frac{(x_1-\mu_1)(x_2-\mu_2)}{\sigma_1\,\sigma_2}
        + \frac{(x_2-\mu_2)^2}{\sigma_2^2}
                 \right]
         \right\}  \nonumber \\
\!\! &\propto&\!\!  \exp \left\{ 
 -\frac{1}{2\,(1-\rho_{12}^2)\,\sigma_1^2} 
  \left[ (x_1-\mu_1)^2
   - 2\,\rho_{12}\,\frac{\sigma_1}{\sigma_2}(x_1-\mu_1)(x_2-\mu_2)
                 \right]
 \right\}  \nonumber  \\
\!\! &\propto&\!\!  \exp \left\{ 
 -\frac{1}{2\,(1-\rho_{12}^2)\,\sigma_1^2} 
  \left[ x_1^2-2\,\mu_1x_1+ \mu_1^2
   - 2\,\rho_{12}\,\frac{\sigma_1}{\sigma_2}\,(x_2-\mu_2)\,x_1
                 \right]
 \right\} \,, \nonumber \\
\!\! &\propto&\!\!  \exp \left\{ 
 -\frac{1}{2\,(1-\rho_{12}^2)\,\sigma_1^2} 
  \left[ x_1^2-2\,x_1\,[ \mu_1
   + \rho_{12}\,\frac{\sigma_1}{\sigma_2}\,(x_2-\mu_2)]
                 \right]
 \right\} \nonumber\\
\!\! &\propto&\!\!  \exp \left\{ 
 -\frac{1}{2\,(1-\rho_{12}^2)\,\sigma_1^2} 
  \left[ x_1^2-2\,x_1\,[ \mu_1
   + \rho_{12}\,\frac{\sigma_1}{\sigma_2}\,(x_2-\mu_2)]
             + [ \mu_1  
   + \rho_{12}\,\frac{\sigma_1}{\sigma_2}\,(x_2-\mu_2)]^2   \right]
 \right\}  \nonumber  \nonumber\\
\!\! &\propto&\!\!  \exp \left\{ 
 -\frac{1}{2\,(1-\rho_{12}^2)\,\sigma_1^2} 
  \left( x_1  - [ \mu_1  
   + \rho_{12}\,\frac{\sigma_1}{\sigma_2}\,(x_2-\mu_2)]   \right)^2
 \right\}  
\end{eqnarray}
in which we recognize a Gaussian with expected value 
$ \mu_1  
   + \rho_{12}\,\frac{\sigma_1}{\sigma_2}\,(x_2-\mu_2)$
and standard deviation $\sigma_1\sqrt{1-\rho_{12}^2}$
(and therefore the normalization factor can be
obtained without any calculation).

\subsection{Linear combinations}\label{ss:LinearCombinations}
Linear transformations of variables are important because
there are several practical problems to which they apply.
There are also other cases in which the transformation is not 
rigorously linear, but it can be still approximately
 linearized in the region
of interest, where the probability mass is concentrated.
There are well known theorems that relate expected values
and covariance matrix of the {\em input quantities} 
to expected values and covariance matrix of the 
{\em output quantities}. The most famous case is when
a single output quantity $Y$ depends on several variables
$\bm{X}$. So, given
\begin{eqnarray}
Y &=& \sum_i c_i\, X_i\,,
\end{eqnarray}
there is a relation which always holds, no matter
if the $X_i$ are independent or not and whichever are the pdf's
 which  describe them:
\begin{eqnarray}
\mbox{E}[Y] &=& \sum_i c_i\, \mbox{E}[X_i]\,.
\end{eqnarray}
In the special case that the $X_i$ are also {\bf independent}, we have
\begin{eqnarray}
\mbox{Var}[Y] &=& \sum_i c_i^2\, \mbox{Var}[X_i].
\end{eqnarray}
Instead it is not always simple to calculate
the pdf of $Y$ in the most general case. 
There are however two remarkable cases,
which we assume known and just recall them here,
in which is $Y$ is normally distributed:
\begin{enumerate}
\item {\bf linear combinations of normally distributed variables} 
      are still normal;
\item the {\bf Central Limit Theorem} states that if we have 
      `many'\footnote{The theorem says ``for $n$ that goes to infinity''! Some
       practice is then needed to judge when it is large enough --
       often $n$ around 10 is can be considered `large', in other cases even
       $10^6$ is not enough! (Think of one million
       of variables described by a Poisson distribution
       with $\lambda=10^{-6}$.)} {\bf independent variables} 
      their linear combination is normally distributed with
      variance equal to $\sum_i c_i^2\, \mbox{Var}[X_i]$ if none of 
      the non-normal components dominates the overall 
      variance, i.e. if $c_j^2\, \mbox{Var}[X_j] \ll 
      \sum_i c_i^2\, \mbox{Var}[X_i]$, where $j$ denotes any 
      of those non-normal components. 
\end{enumerate}
Since in this paper we only stick to normal pdf's,
the only task will be to evaluate the covariance matrix
of the set of variables of interest, depending on the problem.

\begin{figure}[t]
\centering\epsfig{file=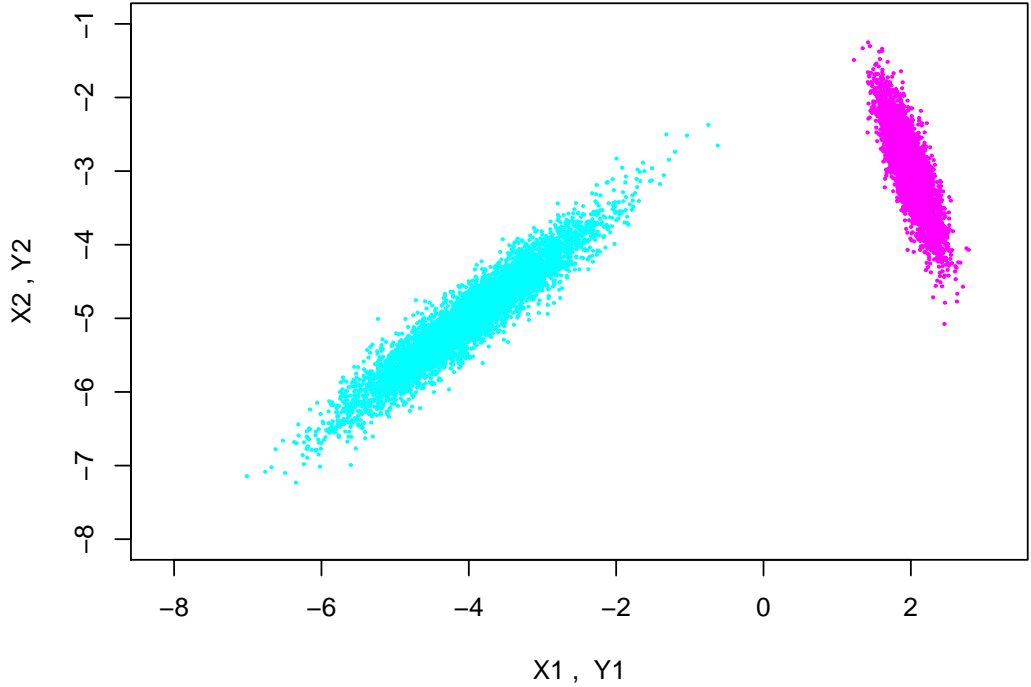,width=0.9\linewidth,clip=}
\caption{Monte Carlo sampling of two multivariate
normal distributions (see text).}
\label{fig:esempio_tras.eps}
\end{figure}

The general transformation from $n$ input variables to $m$ output
variable is given by\footnote{We neglect a possible 
extra constant term in the linear combination
because this plays no role 
in the uncertainty.}
\begin{eqnarray}
Y_{i} &=& c_{ij} X_j\,, 
\end{eqnarray}
or, in a compact form that use the {\em transformation matrix} $\bm{C}$,
whose elements are the $c_{ij}$,
\begin{eqnarray}
\bm{Y} &=& \bm{C}\, \bm{X}\,.
\end{eqnarray}
Expected value and covariance matrix of the output quantities 
are given by
\begin{eqnarray}
\mbox{E}[\bm{Y}] &=&  \bm{C}\, \mbox{E}[\bm{X}] \\
\bm{V}_Y &=& \bm{C}\,V_X\,\bm{C}^T
\end{eqnarray}
For example, if $\bm{\mu}_X = (2, -3)$, with 
$\sigma_{X_1}= 0.2$,  $\sigma_{X_2}= 0.5$ and $\rho_{X_{12}} = -0.8$,
and the transformation rule is given by 
\begin{eqnarray}
Y_1 &=& X_1 + 2\,X_2 \\
Y_2 &=& -X_1 + X_2\,,
\end{eqnarray}
i.e. \vspace{-0.3cm}
\begin{eqnarray}
\bm{C} & = & \left(\! \begin{array}{cc}
                  1 &   2 \\
                  -1  &  1 
                   \end{array} 
          \!\right)
\end{eqnarray}
we get in R: \footnote{The function \code{outer()} produces
by default a matrix which is by default is the {\em outer}
product of two vectors, i.e. $\bm{v}_1\,\bm{v}_2^T$. But 
it has a third parameter \code{FUN} which which it is possible
to evaluate different function on the `grid' defined by
the Cartesian product of the two vector. Try for example\\
\Rin{outer(1:3, 1:3, '+')}\\
\Rin{outer(1:3, 1:3, function(x,y) x + y\^{}2))}\\
\Rin{round( outer(0:10, 0:10, function(x,y) sin(x)*cos(y)), 2 )}\\
}
\mbox{}\vspace{-0.2cm}
{\small
\begin{verbatim}
> mu.X <- c(2, -3)
> s.X <- c(0.2, 0.5)
> rho.X   <- -0.8
> V.X  <- outer(s.X, s.X)
> V.X[1,2] <- V.X[2,1] <- V.X[1,2]*rho.X
> V.X
      [,1]  [,2]
[1,]  0.04 -0.08
[2,] -0.08  0.25
> ( cor.X <- V.X / outer(s.X,s.X) )
     [,1] [,2]
[1,]  1.0 -0.8
[2,] -0.8  1.0
> ( C <- rbind( c(1,2), c(-1,1) ) )
     [,1] [,2]
[1,]    1    2
[2,]   -1    1
> ( mu.Y <- as.vector( C %*% mu.X ) )
[1] -4 -5
> ( V.Y <- C %*% V.X %*% t(C) )
     [,1] [,2]
[1,] 0.72 0.54
[2,] 0.54 0.45
> ( s.Y   <- sqrt(diag(V.Y)) )
[1] 0.8485281 0.6708204
> ( cor.Y <- V.Y / outer(s.Y,s.Y) )
          [,1]      [,2]
[1,] 1.0000000 0.9486833
[2,] 0.9486833 1.0000000
\end{verbatim}
} \noindent
Let us get a visual representation of the probability
distribution of $\bm{X}$ and  $\bm{Y}$ using this time, instead
of iso-pdf ellipses, points in the $X-Y$ plane produced by
the random generator provided by the package mnormt 
(see result in Fig.~\ref{fig:esempio_tras.eps}):
\mbox{}\vspace{-0.2cm}
{\small
\begin{verbatim}
> n=5000; r.X <- rmnorm(n, mu.X, V.X); r.Y <- rmnorm(n, mu.Y, V.Y)
> plot(r.X, col='magenta', xlim=c(-7,2), ylim=c(-8,-1), cex=0.2,
+  asp=1, xlab='X1 ,  Y1',  ylab='X2 , Y2')
> points(r.Y, col='cyan', cex=0.2)
\end{verbatim}
} \noindent

\subsection{Conditional distributions in many dimensions}
Instead, a less known rule is  that which gives the covariance matrix
of a conditional distribution with a number of variables above two.
For example we might have 5 variables $X_1,X_2,\ldots X_5$ 
and could be interested in the expected values and the 
covariance matrix of 
$(X_1,\, X_4\,\, X_5)$, given $(X_2,\, X_3)$. Problems of this kind
might look a mere mathematical curiosity, but they are indeed important 
to understand how we learn from data and we 
make probabilistic predictions using probability theory.

Compact formulae to solve this problems can be found
in Ref.~\cite{Eaton}. If we partition $\bm{\mu}$ and $\bm{V}$ 
into the the subsets of variable on which we want to condition
and the other ones, i.e.
\begin{eqnarray}
\bm{\mu} &=& \left(\!\!\begin{array}{c}
                   \bm{\mu}_1 \\
                   \bm{\mu}_2  
                   \end{array}
           \!\!\right) \\
&& \nonumber \\
\bm{V} &=& \left(\!\! \begin{array}{cc}
                 \bm{V}_{11} & \bm{V}_{12} \\
                 \bm{V}_{21} & \bm{V}_{22}   
                 \end{array}
      \!\!\right)
\end{eqnarray}
the result is
\begin{eqnarray}
\mbox{E}\left[\left.\bm{X}_1\right|_{\bm{X}_2 = \bm{a}}\right] 
&=& \bm{\mu}_1 +\bm{V}_{12}\,\bm{V}_{22}^{-1}\,(\bm{a}-\bm{\mu}_2)
\label{eq:Eaton_E} \\
&& \nonumber \\
\bm{V}\left[\left.\bm{X}_1\right|_{\bm{X}_2}\right] &=& 
\bm{V}_{11} - \bm{V}_{12}\,\bm{V}_{22}^{-1}\,\bm{V}_{21}\label{eq:Eaton_V}
\end{eqnarray}
(And analogous formulae for 
$\mbox{E}\left[\left.\bm{X}_2\right|_{\bm{X}_1 = \bm{b}}\right] $
and $\mbox{Var}\left[\left.\bm{X}_2\right|_{\bm{X}_1 = \bm{b}}\right]$.)

In the case of a bivariate distributions we recover easily
Eqs.~(\ref{eq:x1_cond1_E})-(\ref{eq:x1_cond1_Var}), as it follows.
\begin{description}
\item[Expected value:] $\bm{V}_{12}$ is the off-diagonal term
      $\rho_{12}\sigma_1\sigma_2$, while   $\bm{V}_{22}$ is equal to 
      $\sigma_2^2$. Eq.~(\ref{eq:Eaton_E}) becomes then
      \begin{eqnarray}
      \mbox{E}[\left.X_1\right|_{X_2}] &=& 
        \mu_1 + \rho_{12}\,\sigma_1\,\sigma_2\, \frac{1}{\sigma_2^2}
                \,(a - \mu_2) \nonumber \\
 &=&    \mu_1 + \rho_{12}\,\frac{\sigma_1}{\sigma_2}
                \,(a - \mu_2)
      \end{eqnarray}
\item[Variance:] The remaining two terms of interest are also very simple:
            $\bm{V}_{11}$ is $\sigma_1^2$, while  $\bm{V}_{21}$, equal
            to  $\bm{V}_{12}$, is $\rho_{12}\sigma_1\sigma_2$. It follows
      \begin{eqnarray}
       \mbox{Var}[\left.X_1\right|_{X_2}] &=& \sigma_1^2 - 
                   \rho_{12}\,\sigma_1\,\sigma_2\, \frac{1}{\sigma_2^2}
                  \,  \rho_{12}\,\sigma_1\,\sigma_2 \nonumber \\
   &=&  \sigma_1^2 - \rho_{12}^2\,\sigma_1^2 \nonumber \\
  & =&  \sigma_1^2\,(1-\rho_{12}^2).
       \end{eqnarray}
\end{description}
{\bf Note} that, while the conditioned
expected value depends on the conditionand vector
$\bm{a}$, the conditioned variance \underline{does not}. 

\section{R implementation of the rule to condition 
multivariate normal distributions}
At this point, having set up all our tools, here is 
the R function which implements the above 
formulae:\footnote{As it will be mentioned
in the footnote \ref{fn:Choleski} 
of Sec.~\ref{sec:fits}, a more numerically 
stable way to invert a matrix in R would be 
using the Choleski decomposition, but for the purpose
of this note the difference is slightly appreciable.}
 \mbox{}\vspace{-0.2cm}
{\small
\begin{verbatim}
norm.mult.cond <- function(mu, V, x.c, full=TRUE) {
  out <- NULL
  n <- length(mu)

  # Checks dimensions of mu and V
  if ( sum(dim(V) != n) ) {
    cat( sprintf("dimensions of V incompatible with length of mu\n") )
    return(out)
  }

  # number of conditionand variables
  nc <- length(x.c[!is.na(x.c)])
  # peculiar/anomalous cases
  if( (length(x.c) > n) | (nc > n) ) {
    cat( sprintf("x.c has more elements than mu\n") )
    return(out)
  } else if (nc == 0) {  # No condition
    out$mu <- mu
    out$V  <- V
    return(out)
  } else if(nc == n) {
    out$mu <- x.c    # exact values
    out$V  <- NULL   # covariance matrix is meaningless
    return(out)
  }

  # Apply Eaton's formulae
  v.c <- which(!is.na(x.c))  # conditioning variables
  v   <- which(is.na(x.c))   # variables of interest 
  V11 <- V[v, v]      
  V22 <- V[v.c, v.c]  
  V12 <- V[v, v.c]    
  V21 <- V[v.c, v]    
  mu.cond <- mu[v] + V12 %*% solve(V22) %*%  (x.c[!is.na(x.c)] - mu[v.c])
  V.cond  <- V11 - V12 %*% solve(V22) %*% V21
  if(!full) { # returns only interesting part
    out$mu <- as.vector(mu.cond)
    out$V  <- V.cond
  } else {    # returns all (better to understand!!)
    mu1 <- mu
    V1 <- V
    mu1[v]   <- mu.cond
    mu1[v.c] <- x.c[!is.na(x.c)]
    V1[v, v] <- V.cond
    V1[v.c, v.c] <- 0
    V1[v, v.c]   <- 0
    V1[v.c, v]   <- 0
    out$mu <- as.vector(mu1)
    out$V  <- V1
  }
  return(out)
}
\end{verbatim}
} \noindent
The conditionand vector $\code{x.c}$ has to contain numbers
in the positions corresponding to the variables on which we
want to condition, and \code{NA}, that is `not available'
or `unknown', in the others, as we shall see in the examples.
The code of parameter \code{full} is to return the vector
of expectation and the covariance having the initial dimensionality.
The expectation of the variable used as condition is 
the condition itself. All elements of the covariance matrix
related to conditionals are instead zero, and the utility of
this convention will be clear going through the examples.

Let us try with a simple case of two normal quantities
$\bm{\mu}_X = (2, -3)$ of section \ref{ss:LinearCombinations}.
The question is how our uncertainty on $\mu_{X_1}$ change if 
{\em we assume} $\mu_{X_2}= -2$:
\mbox{}\vspace{-0.2cm}
{\small
\begin{verbatim}
> ( V.X.cond <- norm.mult.cond(mu.X, V.X, c(NA, -2)) )
$mu
[1]  1.68 -2.00

$V
       [,1] [,2]
[1,] 0.0144    0
[2,] 0.0000    0

> sqrt(diag(V.X.cond$V))
[1] 0.12 0.00
\end{verbatim}
} \noindent
The  effect of the conditions to shift
the expected value of  $\mu_{X_1}$ 
from 2 to 1.68 and to squeeze its {\em standard uncertainty} 
to 0.12. If we provide our result
in the conventional form ``expected value $\pm$ standard uncertainty'',
the assumption (or `knowledge') $X_2=-2$ updates our `knowledge'
about $X_1$ from `$2.00\pm 0.20$' to  `$1.67\pm 0.12$'.

\newpage
\section{The `simplest experiment'}
Let us go back to the first diagram of
Fig.~\ref{fig:modelli_base}, that we repeat
here for convenience:
\begin{center}
\epsfig{file=x1-x2.eps,clip=}\hspace{1.5cm}
\end{center}
This diagram describes the situation in which we have the physical
quantity $X_1$, that is a parameter of our physical model
of reality, and the reading on an instrument, $X_2$,
{\em caused} by $X_1$. 

The instrument has been well calibrate,
such to give $X_2$ around $X_1$, but it is not perfect,
as usual. 
In other words, even if we knew exactly the value $x_1$
we were not sure about the value $x_2$ we would read.
For simplicity, let us model this uncertainty
by a normal distribution, i.e.
\begin{eqnarray}
\left.X_2\right|_{X_1} &\sim& {\cal N}(X_1, \sigma_{2|1})\,.
\end{eqnarray}
But we usually do not know $X_1$, and therefore we are even more 
uncertain about what we shall read on the instrument. In fact
we are dealing with a joint distribution describing the joint
uncertainty about the two quantities, that is
\begin{eqnarray}
f(x_1,\,x_2\,|\,I) &=& f(x_2\,|\,x_1,\,I)\cdot f(x_1\,|\,I)\,.
\end{eqnarray}
Our knowledge about $X_2$ will be given, instead, 
by $f(x_2\,|\,I) = \int_{\{x_1\}}f(x_1,\,x_2\,|\,I)\,dx_1$, 
a distribution characterized by 
$\mbox{Var}[X_2] \ne \mbox{Var}[\left.X_2\right|_{X_1}]$.

It is convenient to model our uncertainty about $X_1$ 
with a normal distribution, with a standard deviation $\sigma_1$ 
much larger than $\sigma_{2|1}$ -- if we make a measurement
we want to gain knowledge about that quantity! -- and centered
around the values we roughly expect.\footnote{For extensive
discussions about modelling prior knowledge of physical quantities
see Ref.~\cite{BR} and references therein. As a practical example,
think at the width of the table at which a sit 
in the very moment you read these lines (or any other
object), and about the reading on a ruler when you try
to measure it.}

In order to simplify the calculations, in the exercise that follows
let us assume that $X_1$ is centered around zero. We shall see later
how to get rid of this limitation. 

The joint distribution $f(x_1,\,x_2\,|\,I)$ is then given by
\begin{eqnarray}
f(x_1,\,x_2\,|\,I) &=& \frac{1}{\sqrt{2\,\pi}\,\sigma_{2|1}}\,
 \exp\left[-\frac{(x_2-x_1)^2}{2\,\sigma_{2|1}^2}\right]\times
\frac{1}{\sqrt{2\,\pi}\,\sigma_1}\,
 \exp\left[-\frac{x_1^2}{2\,\sigma_1^2}\right]
\label{eq:joint_x1x1}
\end{eqnarray}
As an exercise,
let us see how to evaluate $f(x_1,\,x_2\,|\,I)$. 
The trick, already applied before, is 
to manipulate the terms in the exponent in order
to recover a well known pattern. Here are the details,
starting from (\ref{eq:joint_x1x1}) rewritten dropping all
irrelevant factors:
\begin{eqnarray}
f(x_1,\,x_2\,|\,I)  
&\propto& \exp\left[ -\frac{(x_2-x_1)^2}{2\,\sigma_{2|1}^2} -
\frac{x_1^2}{2\,\sigma_1^2}  \right]\\
&\propto& \exp\left[ -\frac{1}{2}\left(
                       \frac{x_2^2-2\,x_1x_2+x_1^2}{\sigma_{2|1}^2}
                      + \frac{x_1^2}{\sigma_1^2}   
                     \right)\right] \\
&\propto& \exp\left[  -\frac{1}{2}\left(
          \frac{x_2^2}{\sigma_{2|1}^2} 
         -\frac{2\,x_1x_2}{\sigma_{2|1}^2}
         +x_1^2\cdot\left(\frac{1}{\sigma_{2|1}^2} +\frac{1}{\sigma_{1}^2}\right) 
         \right)\right] \\
&\propto& \exp\left[  -\frac{1}{2}\left(
          \frac{x_2^2}{\sigma_{2|1}^2} 
         -\frac{2\,x_1x_2}{\sigma_{2|1}^2}
         +x_1^2\cdot\frac{\sigma_{2|1}^2+\sigma_{1}^2}
                        {\sigma_{2|1}^2\cdot\sigma_{1}^2}
         \right)\right] \\
&\propto& \exp\left[  -\frac{1}{2}\,\frac{\sigma_{2|1}^2+\sigma_{1}^2}
                                         {\sigma_{2|1}^2}
          \,\left(
          \frac{x_2^2}{\sigma_{2|1}^2+\sigma_1^2} 
         -\frac{2\,x_1x_2}{\sigma_{2|1}^2+\sigma_1^2}
         + \frac{x_1^2}{\sigma_{1}^2}
         \right)\right] \\
&\propto& \exp\left[  -\frac{1}{2}\,\frac{1}{\frac{\sigma_{2|1}^2}
                                            {\sigma_{2|1}^2+\sigma_{1}^2}
                                         }
          \,\left(
          \frac{x_2^2}{\sigma_{2|1}^2+\sigma_1^2} 
         -\frac{2\,x_1x_2}{\sigma_{2|1}^2+\sigma_1^2}
         + \frac{x_1^2}{\sigma_{1}^2}
         \right)\right]
\end{eqnarray}
In this expression we recognize a bivariate distribution centered
around $(0,0)$, provided we interpret
\begin{eqnarray}
 \sigma_{2|1}^2+\sigma_1^2  &=& \sigma_2^2\\
\frac{\sigma_{2|1}^2}{\sigma_{2|1}^2+\sigma_{1}^2}  &=&   1-\rho_{12}^2 \,,
\end{eqnarray}
and after having checked the consistency of the terms multiplying
$x_1\,x_2$. Indeed we have 
\begin{eqnarray}
\rho_{12}^2 &=& 1 - \frac{\sigma_{2|1}^2}{\sigma_{2|1}^2+\sigma_{1}^2} = 
          \frac{\sigma_1^2}{\sigma_{2|1}^2+\sigma_{1}^2} \\
&& \nonumber \\
\rho_{12} &=&  \frac{\sigma_1}{\sqrt{\sigma_{2|1}^2+\sigma_{1}^2}} =
          \frac{\sigma_1}{\sigma_2} 
\end{eqnarray}
and then the second term within parenthesis can be rewritten
as
\begin{eqnarray}
\frac{2\,x_1x_2}{\sigma_{2|1}^2+\sigma_1^2} &=& 
\frac{2\,x_1x_2}{\sigma_2\cdot\sigma_2} = 
\frac{2\,\rho_{12}\,x_1x_2}{\sigma_1\cdot\sigma_2}.
\end{eqnarray}
Then 
\begin{eqnarray}
f(x_1,\,x_2\,|\,I) 
&\propto& \exp\left[ -\frac{1}{2\,(1-\rho_{12}^2)}
\left(\frac{x_1^2}{\sigma_1^2} - \frac{2\,\rho_{12}\,x_1x_2}{\sigma_1\cdot\sigma_2} + \frac{x_2^2}{\sigma_2^2}
\right)\right]
\end{eqnarray}
is definitively a bivariate normal distribution with
\begin{eqnarray}
\bm{\mu} &=& \left(\!\! \begin{array}{c} 0 \\ 0  
                        \end{array} \!\!\right) \\
&& \nonumber \\
\bm{V} &=& \left(\!\! \begin{array}{cc} \sigma_1^2 & \sigma_1^2 \\ 
                               \sigma_1^2 & \sigma_1^2 + \sigma_{2|1}^2 
                      \end{array} \!\!\right)
\end{eqnarray}
As a cross check, let us evaluate expected value and variance
of $X_2$ if we assume a certain value of $X_1$, for example
$X_1=x_1$:
\begin{eqnarray}
\mbox{E}[\left.X_2\right|_{X_1=x_1}] &=& 0 + 
\frac{\sigma_1^2}{\sigma_1^2}\cdot(x_1-0)
                                  = x_1\\
\mbox{Var}[\left.X_2\right|_{X_1=x_1}] &=&  \sigma_1^2 + \sigma_{2|1}^2  
-   \frac{\sigma_1^2}{\sigma_1^2} \,\sigma_1^2 =  \sigma_{2|1}^2 \,,
\end{eqnarray} 
as it should be: provided we know the value
of $X_1$ our expectation of $X_2$ is around its value, with standard
uncertainty $\sigma_{2|1}$.

More interesting is the other way around, that is indeed the purpose
of the experiment: how our knowledge about $X_1$ is modified
by $X_2=x_2$:
\begin{eqnarray}
\mbox{E}[\left.X_1\right|_{X_2=x_2}] &=& 0 + 
\frac{\sigma_1^2}{\sigma_1^2+\sigma_{2|1}^2}\cdot(x_2-0)
                    = x_2 \cdot \frac{1}{1+\sigma_{2|1}^2/\sigma_1^2}
\label{eq:E_X1|X2} \\
\mbox{Var}[\left.X_1\right|_{X_2=x_2}] &=&  \sigma_1^2   
-   \frac{\sigma_1^2}{\sigma_1^2 + \sigma_{2|1}^2} \,\sigma_1^2 
= \sigma_{1|2}^2  \cdot \frac{1}{1+\sigma_{2|1}^2/\sigma_1^2} \,,
\label{eq:Var_X1|X2}
\end{eqnarray} 
Contrary to the first case, this second result is initially 
not very intuitive: the expected value of $X_1$ is not
exactly equal to the `observed' value $x_2$, unless $\sigma_1$, 
that models our {\em prior standard uncertainty} about $X_1$,
is much larger than the experimental 
resolution $\sigma_{2|1}$. Similarly, the {\em final standard
uncertainty} is in general a   smaller than 
$\sigma_{2|1}$, unless, again, 
$\sigma_{1|2}/\sigma_1\ll 1$.\footnote{You might recognize
in Eq.~(\ref{eq:Var_X1|X2}) 
\begin{eqnarray*}
\frac{1}{\mbox{Var}[\left.X_1\right|_{X_2=x_2}]} &=&
\frac{1}{\sigma_1^2}  + \frac{1}{\sigma_{2|1}^2}\,,
\end{eqnarray*}
which stems naturally from probability theory
and tells how a new observation 
squeezes the uncertainty on the true value of a quantity. 
Indeed, it easy to show that Eq.~\ref{eq:E_X1|X2} can be
written, as rather well known (see e.g. Ref.~\cite{BR}), as
\begin{eqnarray*}
\mbox{E}[\left.X_1\right|_{X_2=x_2}] &=&
\frac{\mbox{E}[X_1]\cdot\sigma_1^{-2} + x_2\,\sigma_{2|1}^{-2}}
      {\sigma_1^{-2} + \sigma_{2|1}^{-2}}\,,
\end{eqnarray*}
weighted average of the prior expected value and
observation, with weights equal to the prior
variance and the instrument variance.

{\em En passent} we can rewrite 
Eqs.~(\ref{eq:E_X1|X2})-(\ref{eq:Var_X1|X2}) as
\begin{eqnarray*}
\mbox{E}[\left.X_1\right|_{X_2=x_2}] &=& \mbox{E}[X_1] +
\frac{\sigma_1^{2}}{\sigma_1^{2} + \sigma_{2|1}^{2}}\,(x_2-\mbox{E}[X_1]) \\
\mbox{Var}[\left.X_1\right|_{X_2=x_2}] &=& \sigma_1^2 -
\frac{\sigma_1^{2}}{\sigma_1^{2} + \sigma_{2|1}^{2}}\,\sigma_1^2\,,
\end{eqnarray*} 
or
\begin{eqnarray*}
\mbox{E}[\left.X_1\right|_{X_2=x_2}] &=& \mbox{E}[X_1] +
k \,(x_2-\mbox{E}[X_1]) \\
\mbox{Var}[\left.X_1\right|_{X_2=x_2}] &=& 
\sigma_1^2 - k\,\sigma_1^2 = \sigma_1^2\,(1-k)\,,
\end{eqnarray*} 
with $$k=\frac{\sigma_1^{2}}{\sigma_1^{2}+\sigma_{2|1}^{2}}\,,$$ 
in order to emphasize the Kalman filter's updating rules.~\cite{Kalman}
\label{fn:media_pesata}
}
Although initially surprising, these result are in qualitative
agreement with the good sense of experienced physicists~\cite{BR}.

\section{Several independent measurements on the same physics quantity}
The next step is to see what happens when we are in the conditions
to make several {\em independent measurements} on the same
quantity $X_1$, possibly with different instruments, each one
characterized by a conditional standard uncertainty
$\sigma_{i|1}$ and perfectly calibrated, that is 
$\mbox{E}[\left.X_i\right|_{X_1=x_1}] = x_1$. The situation
can be illustrated with the diagram at the center of
Fig.~\ref{fig:modelli_base}, reported here for convenience, extended
to other observations:
\begin{center} 
\epsfig{file=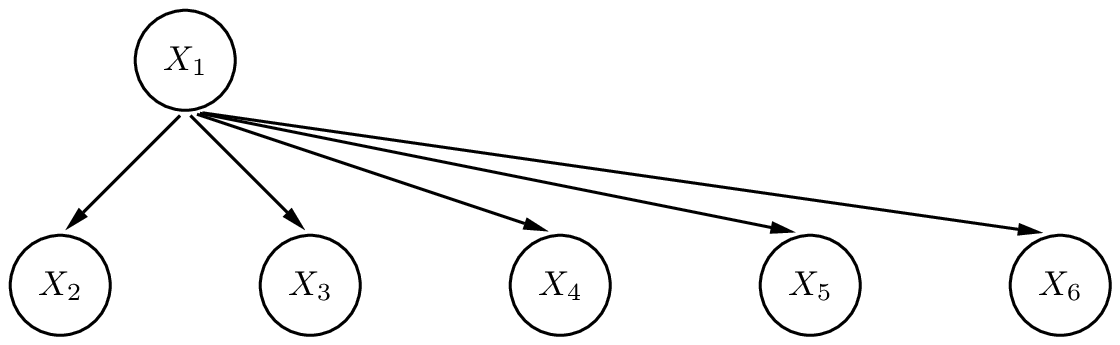,clip=}\hspace{1.5cm}
\end{center}
We have learned that if we are able to build up the covariance
matrix of the joint distribution $f(x_1,x_2,x_3,\ldots\,|\,I)$ the problem
is readily solved, at least in the normal approximations
we are using throughout the paper.

In principle we should repeat the previous exercise to 
evaluate, sticking to the first two observations $x_2$ and $x_3$, 
\begin{eqnarray}
f(x_1,x_2,x_3\,|\,I) &=& f(x_1\,|\,I)\cdot f(x_2\,|\,x_1,\,I) 
\cdot f(x_3\,|\,x_1,x_2\,I) \\
&=& f(x_1\,|\,I)\cdot f(x_2\,|\,x_1,\,I) 
\cdot f(x_3\,|\,x_1,\,I)\,,   
\end{eqnarray}
where in the last step we have made explicit that 
 $f(x_3\,|\,x_1\,I)$ does not depend on $X_2$, once $X_1$
is known. But this does not implies that $X_2$ and $X_3$
are independent, as we shall see later! They are simply 
{\em conditionally independent}, i.e. independent under 
the \underline{condition} (to be meant in general
as an {\em hypothesis}) that $X_1$ has a precisely known value.

In reality we do not need to go through 
a similar derivation,
that indeed {\em was just an exercise}. 
The easy solution arises, going back
to the previous case, noting that the {\em observation}
$o_i$ is the sum of the  value of the physics quantity 
 $v$ and the instrumental {\em error} $e_i$ (a `noise',
as you might like to see it), i.e. 
\begin{eqnarray}
o_i &=& v + e_i\,, 
\end{eqnarray}
with $e_i$ modelled, as usual,  
by a normal distribution, that is, in general
\begin{eqnarray}
e_i &\sim& {\cal N}(0, \sigma_{e_i})\,. 
\end{eqnarray}
The general uncertain 
vector $\bm{X}$ will be then $\bm{X} = (v, o_1, o_2)$,
as clarified by the following diagram:
\begin{center} 
\epsfig{file=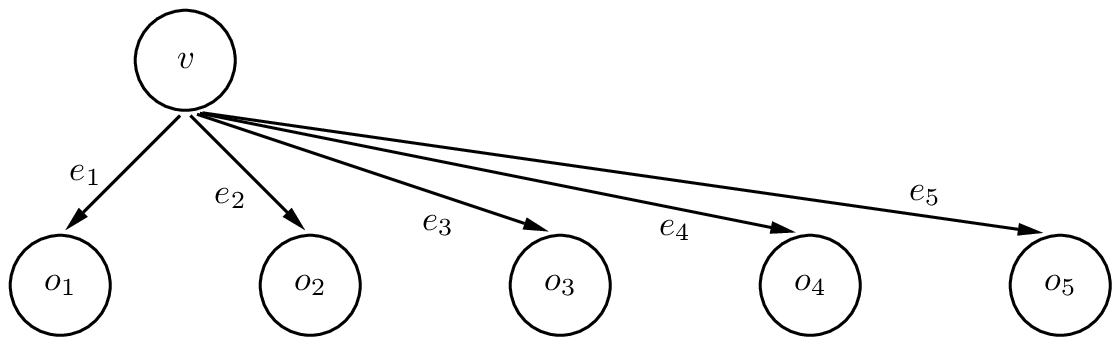,clip=}\hspace{1.5cm}
\end{center}
These are then the first terms the transformation rules:
\begin{eqnarray}
X_1 &=& v \\
X_2 &=& o_1 = v + e_1 \label{eq:X2-O1E1} \\ 
X_3 &=& o_2 = v + e_2  \label{eq:X3-O2E2} 
\end{eqnarray}
from which the calculation of the covariance matrix is 
straightforward: 
\begin{itemize}
\item the $i$-th element diagonal is given by the variance of $X_i$,
      that is $\sigma_1^2$, $(\sigma_1^2+\sigma_{e_1}^2)$, and so on;
\item the off-diagonal elements are all equal to $\sigma_1^2$,
      because the only element in common in 
      all linear combinations is $v$.
\end{itemize}
Hence here is the covariance matrix of interest:
\begin{eqnarray}
\bm{V} &=&  \left(\! \begin{array}{ccc} 
 \sigma_1^2 & \sigma_1^2 & \sigma_1^2 \\ 
 \sigma_1^2 & \sigma_1^2 + \sigma_{2|1}^2 &   \sigma_1^2 \\
 \sigma_1^2 & \sigma_1^2 &  \sigma_1^2 + \sigma_{3|1}^2 
                      \end{array} \!\right) \label{eq:CovMatX1-X2_9}
\end{eqnarray}

\subsection{Getting some insights with numerical examples}
\label{ss:divergent_R}
At this point, instead of trying to get analytic formulae
for all conditional probabilities of interest, we prefer to
use the properties of the multivariate normal distribution
implemented in the function \code{norm.mult.cond()} seen before.
And, since the game is now automatic, we enlarge our space
to 6 variables, $X_1$ for the `true value' and 
$X_2$\,-$X_6$ for four possible `readings'. Although it is
not any longer needed,
 we still set out prior central value about $X_1$
around $0$, which is equivalent to set to 0 all 
expected values. (For didactic purposes we have set $\sigma_1$
{\em only} 10 larger than the experimental resolutions $\sigma_{i|1}$,
as we shall discuss commenting the results.)
Here is the R code, with some comments:
\mbox{}\vspace{-0.2cm}
{\small
\begin{verbatim}
> n=6; muX1=0; sigmaX1=10                 # set size and initial uncertainty on X1
> mu <- rep(muX1, n)                      # set expected values (all equal!)
> ( sigma <- c(sigmaX1, rep(1,n-1)) )     # standard deviations 
[1] 10  1  1  1  1  1
> V <- matrix(rep(sigma[1]^2, n*n), c(n,n))
> diag(V)[2:n] <- diag(V)[2:n] + sigma[2:n]^2
> V                                               # covariance matrix
     [,1] [,2] [,3] [,4] [,5] [,6]
[1,]  100  100  100  100  100  100
[2,]  100  101  100  100  100  100
[3,]  100  100  101  100  100  100
[4,]  100  100  100  101  100  100
[5,]  100  100  100  100  101  100
[6,]  100  100  100  100  100  101
> (su <- sqrt(diag(V)))                                 # standard deviations
[1] 10.00000 10.04988 10.04988 10.04988 10.04988 10.04988
> V/outer(su,su)                                        # correlation matrix
          [,1]      [,2]      [,3]      [,4]      [,5]      [,6]
[1,] 1.0000000 0.9950372 0.9950372 0.9950372 0.9950372 0.9950372
[2,] 0.9950372 1.0000000 0.9900990 0.9900990 0.9900990 0.9900990
[3,] 0.9950372 0.9900990 1.0000000 0.9900990 0.9900990 0.9900990
[4,] 0.9950372 0.9900990 0.9900990 1.0000000 0.9900990 0.9900990
[5,] 0.9950372 0.9900990 0.9900990 0.9900990 1.0000000 0.9900990
[6,] 0.9950372 0.9900990 0.9900990 0.9900990 0.9900990 1.0000000
\end{verbatim}
} \noindent
As we can see, all variables are correlated! The
reason is very simple: any precise information
we get about one of them changes the pdf of all others.
In physics terms, a reading on a instrument changes
our opinion about the value of the quantity of interest
as well as of all other readings we have not yet done
(or we not yet aware of their values -- in probability theory what matters
is not time ordering, but ignorance).

Let us now see what happens if we condition on a 
{\bf precise value of the true value }
 $\mathbf{X_1}$, for example \underline{$X_1=2$}:
\mbox{}\vspace{-0.2cm}
{\small
\begin{verbatim}
> ( mu.c <- c(2, rep(NA, n-1)) )                    # conditionand
[1]  2 NA NA NA NA NA
> ( out<- norm.mult.cond(mu, V, mu.c) )             # resulting multivariate
$mu
[1] 2 2 2 2 2 2

$V
     [,1] [,2] [,3] [,4] [,5] [,6]
[1,]    0    0    0    0    0    0
[2,]    0    1    0    0    0    0
[3,]    0    0    1    0    0    0
[4,]    0    0    0    1    0    0
[5,]    0    0    0    0    1    0
[6,]    0    0    0    0    0    1
\end{verbatim}
} \noindent
As we see, the expected values are all equal,
$X_1$ is not longer uncertain, and all other
variables become {\em independent}, more precisely
``conditional independent'' 

\begin{figure}[t]
\centering\epsfig{file=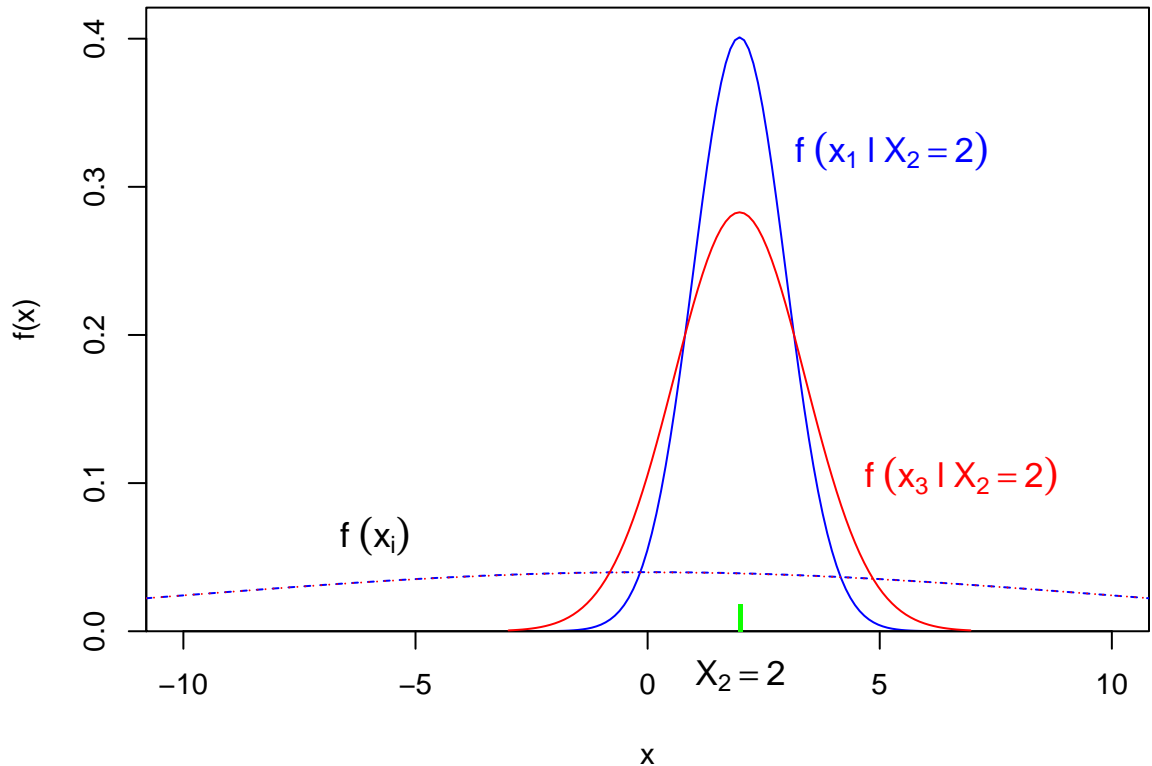,clip=}
\caption{Normal distributions describing our uncertainty
about $X_1$ and $X_3$ before (dashed line) and after (solid line)
the observation $X_2=2$ (see text).}
\label{fig:norm_mult_cond_exp_X2}
\end{figure}

Let's now see what happens if we condition instead 
on the {\bf observation } \underline{$X_2=2$}:
\mbox{}\vspace{-0.2cm}
{\small
\begin{verbatim}
> ( mu.c <- c(NA, 2, rep(NA, n-2)) )
[1] NA  2 NA NA NA NA
> ( out<- norm.mult.cond(mu, V, mu.c) )
$mu
[1] 1.980198 2.000000 1.980198 1.980198 1.980198 1.980198

$V
         [,1] [,2]     [,3]     [,4]     [,5]     [,6]
[1,] 0.990099    0 0.990099 0.990099 0.990099 0.990099
[2,] 0.000000    0 0.000000 0.000000 0.000000 0.000000
[3,] 0.990099    0 1.990099 0.990099 0.990099 0.990099
[4,] 0.990099    0 0.990099 1.990099 0.990099 0.990099
[5,] 0.990099    0 0.990099 0.990099 1.990099 0.990099
[6,] 0.990099    0 0.990099 0.990099 0.990099 1.990099

> ( out.s <- sqrt(diag(out$V)) )                 # standard deviations 
[1] 0.9950372 0.0000000 1.4107087 1.4107087 1.4107087 1.4107087
> out$V / outer(out.s, out.s)                    # correlation matrix (besides NaN)
          [,1] [,2]      [,3]      [,4]      [,5]      [,6]
[1,] 1.0000000  NaN 0.7053456 0.7053456 0.7053456 0.7053456
[2,]       NaN  NaN       NaN       NaN       NaN       NaN
[3,] 0.7053456  NaN 1.0000000 0.4975124 0.4975124 0.4975124
[4,] 0.7053456  NaN 0.4975124 1.0000000 0.4975124 0.4975124
[5,] 0.7053456  NaN 0.4975124 0.4975124 1.0000000 0.4975124
[6,] 0.7053456  NaN 0.4975124 0.4975124 0.4975124 1.0000000
\end{verbatim}
} \noindent
The `measurement' has had the effect of changing
all our expectations, becoming all `practically equal'
to the observed value of $2$. But the uncertainties
about the possible 'future observations' 
are different than that of the true value $X_1$. They are in fact
larger by a factor $\sqrt{2}$ (see also Fig.~\ref{fig:norm_mult_cond_exp_X2}). 
The reason is that $X_2$
and $X_3$ (i.e. $o_1$ and $o_2$) and all other possible readings
 $o_3$, $o_4$ and $o_5$ `communicate' each other via $X_1$: 
their uncertainty is than the combination (quadratic combination!)
of that assigned to $X_1$ and that of the readings $X_i$ 
if we new exactly $X_1$ (that is $\sigma_{e_i}$).

Let us see if we add {\bf another observation}, e.g.
\underline{$X_3=1$}, that is we recondition now simultaneously
on $X_2=2$ and $X_2=1$
\mbox{}\vspace{-0.2cm}
{\small
\begin{verbatim}
> mu.c <- c(NA, 2, 1, NA, NA, NA)
> ( out<- norm.mult.cond(mu, V, mu.c) )
$mu
[1] 1.492537 2.000000 1.000000 1.492537 1.492537 1.492537

$V
          [,1] [,2] [,3]      [,4]      [,5]      [,6]
[1,] 0.4975124    0    0 0.4975124 0.4975124 0.4975124
[2,] 0.0000000    0    0 0.0000000 0.0000000 0.0000000
[3,] 0.0000000    0    0 0.0000000 0.0000000 0.0000000
[4,] 0.4975124    0    0 1.4975124 0.4975124 0.4975124
[5,] 0.4975124    0    0 0.4975124 1.4975124 0.4975124
[6,] 0.4975124    0    0 0.4975124 0.4975124 1.4975124

> ( out.s <- sqrt(diag(out$V)) )
[1] 0.7053456 0.0000000 0.0000000 1.2237289 1.2237289 1.2237289

> out$V / outer(out.s, out.s)
          [,1] [,2] [,3]      [,4]      [,5]      [,6]
[1,] 1.0000000  NaN  NaN 0.5763904 0.5763904 0.5763904
[2,]       NaN  NaN  NaN       NaN       NaN       NaN
[3,]       NaN  NaN  NaN       NaN       NaN       NaN
[4,] 0.5763904  NaN  NaN 1.0000000 0.3322259 0.3322259
[5,] 0.5763904  NaN  NaN 0.3322259 1.0000000 0.3322259
[6,] 0.5763904  NaN  NaN 0.3322259 0.3322259 1.0000000
\end{verbatim}
} \noindent
As we can see, after the second observation
the expected values are {\bf practically} equal to 1.5,
average between the two readings. The uncertainty
about the true value has decreased by a factor 1.41,
that is $\sqrt{2}$, while the uncertainties about the forecasting
decrease only by a factor 1.15, going from 1.41 to 1.22.
This latter number can be understood as
$\sqrt{0.705^2+1^2}=1.22$,
as it will be justified in a while.

Let us see what happens if we suppose that also 
$X_1$ is precisely known, namely $X_1=3$
(different from  $X_1=2$ previously used, not only ``just to change''
but also to use a value different from that of $X_2$ and $X_3$):
\mbox{}\vspace{-0.2cm}
{\small
\begin{verbatim}
> mu.c <- c(3, 2, 1, NA, NA, NA) 
> ( out<- norm.mult.cond(mu, V, mu.c) )
$mu
[1] 3 2 1 3 3 3

$V
     [,1] [,2] [,3]          [,4]          [,5]          [,6]
[1,]    0    0    0  0.000000e+00  0.000000e+00  0.000000e+00
[2,]    0    0    0  0.000000e+00  0.000000e+00  0.000000e+00
[3,]    0    0    0  0.000000e+00  0.000000e+00  0.000000e+00
[4,]    0    0    0  1.000000e+00 -2.302158e-12 -2.302158e-12
[5,]    0    0    0 -2.302158e-12  1.000000e+00 -2.302158e-12
[6,]    0    0    0 -2.302158e-12 -2.302158e-12  1.000000e+00
\end{verbatim}
} \noindent
If $X_1$ is perfectly known the observations
$X_2$ and $X_3$ are irrelevant, 
as it has to be.\footnote{And if $X_2$ and $X_3$ are
`very far' from  $X_1$? In this simple model we are using, there
is little to do, because any observation from 
minus infinite to plus infinite is never incompatible
with a any Gaussian. But we know by experience that something
strange might be happened. It this case we need
to put in mathematical form the model we have in mind.}

Finally, going back to the physical case
of interest, in which $X_1$ is unknown,
let us add a {\bf third observation}, e.g.
 \underline{$X_4=0$}
\mbox{}\vspace{-0.2cm}
{\small
\begin{verbatim}
> mu.c <- c(NA, 2, 1, 0, NA, NA) 
> ( out<- norm.mult.cond(mu, V, mu.c) )
$mu
[1] 0.9966777 2.0000000 1.0000000 0.0000000 0.9966777 0.9966777

$V
          [,1] [,2] [,3] [,4]      [,5]      [,6]
[1,] 0.3322259    0    0    0 0.3322259 0.3322259
[2,] 0.0000000    0    0    0 0.0000000 0.0000000
[3,] 0.0000000    0    0    0 0.0000000 0.0000000
[4,] 0.0000000    0    0    0 0.0000000 0.0000000
[5,] 0.3322259    0    0    0 1.3322259 0.3322259
[6,] 0.3322259    0    0    0 0.3322259 1.3322259

> ( out.s <- sqrt(diag(out$V)) ) 
[1] 0.5763904 0.0000000 0.0000000 0.0000000 1.1542209 1.1542209
> out$V / outer(out.s, out.s)
          [,1] [,2] [,3] [,4]      [,5]      [,6]
[1,] 1.0000000  NaN  NaN  NaN 0.4993762 0.4993762
[2,]       NaN  NaN  NaN  NaN       NaN       NaN
[3,]       NaN  NaN  NaN  NaN       NaN       NaN
[4,]       NaN  NaN  NaN  NaN       NaN       NaN
[5,] 0.4993762  NaN  NaN  NaN 1.0000000 0.2493766
[6,] 0.4993762  NaN  NaN  NaN 0.2493766 1.0000000
\end{verbatim}
} \noindent
As we can see, the value of $X_1$ is with very good approximation
the average of the three observations, that is 1, with a  
the standard uncertainty  decreasing
with $1/\sqrt{n}$, passing from 1.00 to 0.71 to 0.58. 
This is because the three pieces of information enter with the same 
weight, since $\sigma_{i|1}$, related to the `precision
of the instrument', is the same in all cases and equal to 1. 

As far as the prediction of future observations, obviously
they must be centered around the value we believe $X_1$ is, at the
best of our knowledge, a value which changes with the observations.
As far as uncertainty and correlation coefficient are concerned, 
they decrease as follows (starting from the very beginning, before
any observation):
\begin{description}
\item[Standard uncertainty:] 10.05, 1.41, 1.22, 1.15. \\
     We can see that they are a quadratic combination 
     of the uncertainty with which we know $X_1$ and 
     that with which we expect the observation given a precise
     value of $X_1$. If we indicate the state of information
     at time $t$ as $I(t)$, the rule is 
     \begin{eqnarray}
     \mbox{Var}[X_i\,|\,I(t)] &=& \mbox{Var}[X_1\,|\,I(t)] + \sigma_{i|1}^2\,. 
     \end{eqnarray} 
     Asymptotically, when after many measurements the determination
     of $X_1$ is very accurate, it only remains $\sigma_{e_i}^2$,
     as it has to be.
\item[Correlation coefficient:] 0.990, 0.50, 0.33, 0.25. \\
     It is initially very high because any new observation 
     changes dramatically our expectation about the
     others. But then, when we have already made several observations,
     a new one has only very little effect on our forecasting.
     Asymptotically, when we have made a very large number of observations
     and $X_1$ is very well `determined', all future observations 
     become essentially ``conditionally independent''.
\end{description}

\subsection{Follows up}
At this point the game can be continued with different
options. One has only to re-build the initial covariance matrix
and play changing the conditions. 

An interesting exercise 
is certainly that of increasing $\sigma_1$, for example to 100, i.e. 
100 times large than the `precision' of our instrument, or even 1000, 
to see how our conclusions change. The result will be that true value
and future measurements are `practically' only determined by the observations.

It could also interesting to see what happens if the different observations
come from instruments having different precisions.

Finally, one could produce a (relatively) large random sample 
of observations measuring the same true value. Being $m$ the number
of observations, the dimensionality of our problem will be
$m=n+3$, because we have to add -- obviously -- $X_1$
and we want to have at least two future observations 
($X_{1+m+1}$ and $X_{1+m+1}$) in order to check 
their degree of correlation. Here is the R session in which we have
been playing with a sample of 100 observations (for obvious reasons
we shall focus only on the uncertain variables, i.e. $X_{1}$, $X_{101}$
and  $X_{102}$):
\mbox{}\vspace{-0.2cm}
{\small
\begin{verbatim}
> m <- 100; n <- m + 3  # dimensionality of the problem
> mu <- rep(0,n) 
> sigma <- c(10, rep(1,n-1))
> V <- matrix(rep(sigma[1]^2, n*n), c(n,n))
> diag(V)[2:n] <- diag(V)[2:n] + sigma[2:n]^2
> ( X1.God <- 2 )  # the exact value of the quantity we are going to measure
[1] 2
> sample <- rnorm(m, X1.God, sigma[2])   # random sample
> mean(sample)                           # sample mean
[1] 2.094649
> mu.c <- c(NA, sample, NA, NA)
> out <- norm.mult.cond(mu, V, mu.c)     # no printouts, for obvious reasons
> ( out <- marginal.norm(out$mu, out$V, c(1, rep(0, m), 1, 1)) ) # interesting part
$mu
[1] 2.094439 2.094439 2.094439

$V
         [,1]     [,2]     [,3]
[1,] 0.009999 0.009999 0.009999
[2,] 0.009999 1.009999 0.009999
[3,] 0.009999 0.009999 1.009999

> ( su <- sqrt(diag(out$V)) )
[1] 0.099995 1.004987 1.004987

> out$V /outer(su,su)
           [,1]       [,2]       [,3]
[1,] 1.00000000 0.09949879 0.09949879
[2,] 0.09949879 1.00000000 0.00990001
[3,] 0.09949879 0.00990001 1.00000000
\end{verbatim}
} \noindent
Expected values of the true value and of the future measurements
are now equal to the average of the sample, with excellent 
approximation. This is due to the fact that the initial uncertainty
of 10 is in this case much larger than the final one of 0.10. 
This value is indeed equal to $\sigma_{i|1}/\sqrt{n} = 1/10$, 
the famous standard deviation of the mean. This means that the standard
deviation of the sample, that is\\
\Rin{sd(sample)}\\
\Rout{0.08263812}\\
is not used. This is not a surprise, since in our model
$\sigma_{i|1}$ are assumed to be perfectly known.\footnote{A model that
would allow to infer the $\sigma_{i|1}$'s is not any longer
linear, thus going beyond the purpose of this note.}

We see that the uncertainty on the future observations is a bit
larger than that on the true value, as it must be. This is because
they depends on the uncertain value of the true value and 
the experimental resolution, combining in quadrature 
($\sqrt{0.1^2+1^2} = 1.00499$). The correlations become
small, in particular those among the future observations,
which practically become `conditionally independent'. Indeed,
the covariance matrix is that shown in Eq.~(\ref{eq:CovMatX1-X2_9}),
with $\sigma_1$ replaced by  $\sigma_{1|\mbox{sample}}$ (what matters
is the uncertainty about $X_1$, not its source!).

\section{Adding a simple systematic effect (`offset' type)}
Let us now move to the second diagram 
of Fig.~\ref{fig:modelli_base}, which we repeat her for convenience, in
\begin{center} 
\epsfig{file=x1x2-x3.eps,clip=}\hspace{1.5cm}
\end{center}
which $X_3$ is caused
by both $X_1$ and $X_2$:
This diagram can model the presence of a systematic effect, because
we expect that the possible values of $X_3$ are caused
by both $X_1$ and $X_2$, and it will be then 
influenced by how uncertain is the quantity $X_2$ that acts
as a systematic. The simplest case of systematic effect
is an additive one, of unknown value, but with expected
value 0 (the instrument has been calibrated `at the best'!)
and a  standard uncertainty $\sigma_2$. Needless to say,
we also model this uncertainty with a normal distribution, 
with much simplification in the calculations
(and also because this is often the case).

\begin{figure}
\begin{center}
\epsfig{file=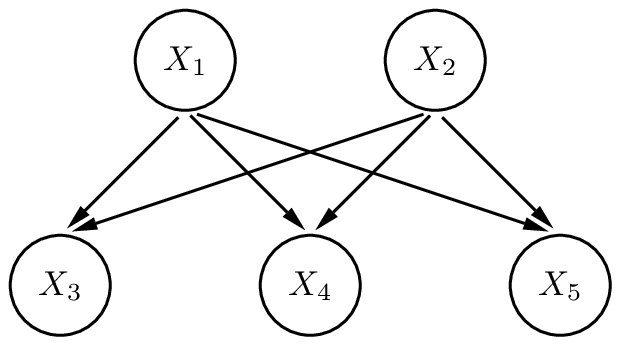,clip=}\hspace{1.5cm}
\epsfig{file=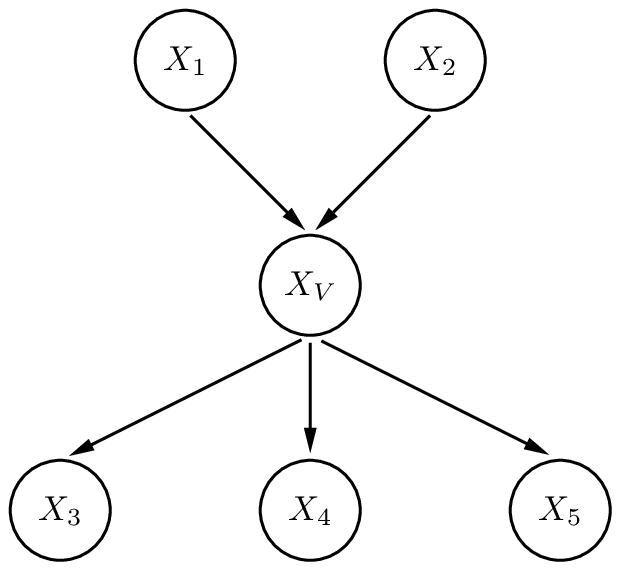,clip=}
\end{center}
\caption{Diagrams to model a systematic effect.}
\label{fig:modelli_syst}
\end{figure}
The model can be extended to several observations, as shown
in the left diagram of Fig.~\ref{fig:modelli_syst}. In the figure
it is also shown a different interpretation of the effect of
the systematic error, which is very close to the physicist intuition.
The observations $X_3$, $X_4$ and $X_5$ are normally distributed
around a kind of `virtual state' $X_V$ determined by the {\em unknown} 
true value $X_1$ and the {\em unknown} offset $X_2$, 
i.e. the true `zero' of the instrument. The transformation rule
to build the initial covariance matrix will be then, starting 
from symbols that have a physical meaning [value $v$, `zero' $z$, 
and the others as in Eqs.~(\ref{eq:X3-O2E2})-(\ref{eq:X3-O2E2}]
\begin{eqnarray}
X_1 &=& v \label{eq:Y1_X1} \\
X_2 &=& z \\
    & & (X_V = v+z= X_1 + X_2) \\
X_3 &=& o_1 = X_V + e_3 =  X_1 + X_2 + e_3 \\
X_4 &=& o_2 = X_V + e_4 =  X_1 + X_2 + e_4 \\
X_5 &=& o_3 = X_V + e_5 =  X_1 + X_2 + e_5 \label{eq:Y5_X5} 
\end{eqnarray}
The calculation of the variances is trivial. As far as the 
covariances we have
\begin{eqnarray}
\mbox{Cov}[X_1, X_2] &=& 0  \\
\mbox{Cov}[X_1, X_i] &=& \sigma_1^2\hspace{1.4cm}(i > 2) \\
\mbox{Cov}[X_2, X_i] &=& \sigma_2^2\hspace{1.4cm}(i > 2) \\
\mbox{Cov}[X_i, X_j] &=& \sigma_1^2 + \sigma_2^2 
    \hspace{0.5cm}(i>2,\ j>2)
\end{eqnarray}
This is then the covariance matrix of interest, limited to the five 
variables shown in the figure (and than it is easy to continue):
\begin{eqnarray}
\bm{V} &=&  \left(\! \begin{array}{ccccc} 
 \sigma_1^2 & 0           & \sigma_1^2 & \sigma_1^2 & \sigma_1^2      \\ 
 0          & \sigma_2^2  &  \sigma_2^2 &  \sigma_2^2   &  \sigma_2^2  \\
 \sigma_1^2 & \sigma_2^2  &  \sigma_1^2 + \sigma_2^2 + \sigma_{e_3}^2 &
                           \sigma_1^2 + \sigma_2^2  &  \sigma_1^2 + \sigma_2^2\\
 \sigma_1^2 & \sigma_2^2  &  \sigma_1^2 + \sigma_2^2 &
                            \sigma_1^2 + \sigma_2^2 + \sigma_{e_4}^2 &
                             \sigma_1^2 + \sigma_2^2\\
 \sigma_1^2 & \sigma_2^2 &  \sigma_1^2 + \sigma_2^2 &  \sigma_1^2 + \sigma_2^2 &
                             \sigma_1^2 + \sigma_2^2 + \sigma_{e_5}^2  
                      \end{array} \!\right)\,,    
\end{eqnarray}
where $\sigma_{e_3}$ stands for $\sigma(e_3)$, i.e. 
$\sigma(\left.X_3\right|_{X_1+X_2})$, and so on, later also 
indicated with the short hand $\sigma_{i|1,2}$. 
 
From such an interesting matrix we can expect interesting
results, useful to {\em train our intuition}. But before
analyzing some cases, as done in the previous section, 
let us make the exercise to build up the covariance matrix
in a different way. The transformation rules 
(\ref{eq:Y1_X1})-(\ref{eq:Y5_X5}) can be rewritten using
the transformation matrix 
\begin{eqnarray}
\bm{C} &=&  \left(\! \begin{array}{ccccc} 
            1 & 0 & 0 & 0 & 0 \\
            0 & 1 & 0 & 0 & 0 \\
            1 & 1 & 1 & 0 & 0 \\
            1 & 1 & 0 & 1 & 0 \\
            1 & 1 & 0 & 0 & 1 
            \end{array} \!\right)    
\end{eqnarray}
to be applied to the diagonal matrix of the independent variables,
\begin{eqnarray}
\bm{V}_0 &=&  \left(\! \begin{array}{ccccc} 
 \sigma_1^2 & 0           & 0 & 0      \\ 
 0          & \sigma_2^2  &  0  &  0   &  0  \\
 0  &  0 &  \sigma_{e_3}^2 & 0 \\
 0  &  0  &  0 &  \sigma_{e_4}^2 & 0\\
 0 & 0 & 0 &  0  & \sigma_{e_5}^2  
                      \end{array} \!\right)    
\end{eqnarray}
Applying then the transformation rule of the covariance matrix
we reobtain the above result --
an implementation in R will be shown in the next subsection.

\subsection{Numerical examples}
Let set up the covariance matrix for 5 possible `observations'
\mbox{}\vspace{-0.2cm}
{\small
\begin{verbatim}
> n=7; muX1=0; sigmaX1=10; muZ=0; sigmaZ=1      # set parameters
> mu <- c(muX1, muZ, rep(muX1+muZ, n-2))        # set expected values
> ( sigma <- c(sigmaX1, sigmaZ, rep(1,n-2)) )   # standard deviations  
[1] 10  1  1  1  1  1  1
> V <- matrix(rep(0, n*n), c(n,n))              # cov matr  # step 0
> V[(1:n)[-2], (1:n)[-2]] <-  sigma[1]^2                    # step 1
> V[(2:n), (2:n)]         <-  V[(2:n), (2:n)] + sigma[2]^2  # step 2
> diag(V)[3:n] <- diag(V)[3:n] + sigma[3:n]^2               # step 3
> V
     [,1] [,2] [,3] [,4] [,5] [,6] [,7]
[1,]  100    0  100  100  100  100  100
[2,]    0    1    1    1    1    1    1
[3,]  100    1  102  101  101  101  101
[4,]  100    1  101  102  101  101  101
[5,]  100    1  101  101  102  101  101
[6,]  100    1  101  101  101  102  101
[7,]  100    1  101  101  101  101  102
> (su <- sqrt(diag(V)))
[1] 10.0000  1.0000 10.0995 10.0995 10.0995 10.0995 10.0995
> round( V/outer(su,su), 4)
       [,1]  [,2]   [,3]   [,4]   [,5]   [,6]   [,7]
[1,] 1.0000 0.000 0.9901 0.9901 0.9901 0.9901 0.9901
[2,] 0.0000 1.000 0.0990 0.0990 0.0990 0.0990 0.0990
[3,] 0.9901 0.099 1.0000 0.9902 0.9902 0.9902 0.9902
[4,] 0.9901 0.099 0.9902 1.0000 0.9902 0.9902 0.9902
[5,] 0.9901 0.099 0.9902 0.9902 1.0000 0.9902 0.9902
[6,] 0.9901 0.099 0.9902 0.9902 0.9902 1.0000 0.9902
[7,] 0.9901 0.099 0.9902 0.9902 0.9902 0.9902 1.0000
\end{verbatim}
} \noindent
Let us also show the {\bf alternative way to build up
the covariance matrix}
{\small
\begin{verbatim}
> C <- matrix(rep(0, n*n), c(n,n))              # transf. matrix
> C[,1] <- c(1, 0, rep(1, n-2))
> C[,2] <- c(0, rep(1, n-1))
> diag(C) <- rep(1, n)
> C
     [,1] [,2] [,3] [,4] [,5] [,6] [,7]
[1,]    1    0    0    0    0    0    0
[2,]    0    1    0    0    0    0    0
[3,]    1    1    1    0    0    0    0
[4,]    1    1    0    1    0    0    0
[5,]    1    1    0    0    1    0    0
[6,]    1    1    0    0    0    1    0
[7,]    1    1    0    0    0    0    1
> V0 <- matrix(rep(0, n*n), c(n,n))     # initial diagonal matrix
> diag(V0) <-  sigma^2
> ( V <- C %*% V0 %*% t(C) )            # joint covariance matrix
     [,1] [,2] [,3] [,4] [,5] [,6] [,7]
[1,]  100    0  100  100  100  100  100
[2,]    0    1    1    1    1    1    1
[3,]  100    1  102  101  101  101  101
[4,]  100    1  101  102  101  101  101
[5,]  100    1  101  101  102  101  101
[6,]  100    1  101  101  101  102  101
[7,]  100    1  101  101  101  101  102
\end{verbatim}
} \noindent
As we see the result is identical to that obtained 
setting the elements `by hand'. 

Then let us now repeat the steps previously followed
without systematic offset. 

\subsubsection{Condition on $X_1=2$ (``known true value'')}
{\small
\begin{verbatim}
> ( mu.c <- c(2, rep(NA, n-1)) )
[1]  2 NA NA NA NA NA NA
> ( out <- norm.mult.cond(mu, V, mu.c) )
$mu
[1] 2 0 2 2 2 2 2

$V
     [,1] [,2] [,3] [,4] [,5] [,6] [,7]
[1,]    0    0    0    0    0    0    0
[2,]    0    1    1    1    1    1    1
[3,]    0    1    2    1    1    1    1
[4,]    0    1    1    2    1    1    1
[5,]    0    1    1    1    2    1    1
[6,]    0    1    1    1    1    2    1
[7,]    0    1    1    1    1    1    2
> ( out.s <- sqrt(diag(out$V)) )
[1] 0.000000 1.000000 1.414214 1.414214 1.414214 1.414214 1.414214
> round( out$V / outer(out.s, out.s), 3) 
     [,1]  [,2]  [,3]  [,4]  [,5]  [,6]  [,7]
[1,]  NaN   NaN   NaN   NaN   NaN   NaN   NaN
[2,]  NaN 1.000 0.707 0.707 0.707 0.707 0.707
[3,]  NaN 0.707 1.000 0.500 0.500 0.500 0.500
[4,]  NaN 0.707 0.500 1.000 0.500 0.500 0.500
[5,]  NaN 0.707 0.500 0.500 1.000 0.500 0.500
[6,]  NaN 0.707 0.500 0.500 0.500 1.000 0.500
[7,]  NaN 0.707 0.500 0.500 0.500 0.500 1.000
\end{verbatim}
} \noindent
The condition on the `true value' changes
the values of the observables to its value,
but it does not affect the offset, which 
has a role in the uncertainty of the future observations
as well in their correlation. In fact, contrary to the case
see in the previous section without uncertain offset,
they are not any longer independent. They would become 
independent if also the offset were known
(try for example with ``\code{mu.c <- c(2, 0, rep(NA, n-2))}''
to see the difference, or even better with 
 ``\code{mu.c <- c(2, 1, rep(NA, n-2))}'').

\subsubsection{Condition on $X_3=2$ (``single observation'')}\label{sss:syst_X3}
{\small
\begin{verbatim}
> ( mu.c <- c(NA, NA, 2, rep(NA, n-3)) )
[1] NA NA  2 NA NA NA NA
> out <- norm.mult.cond(mu, V, mu.c)
> round( out$mu, 4)
[1] 1.9608 0.0196 2.0000 1.9804 1.9804 1.9804 1.9804
> round( out$V, 4)
        [,1]    [,2] [,3]   [,4]   [,5]   [,6]   [,7]
[1,]  1.9608 -0.9804    0 0.9804 0.9804 0.9804 0.9804
[2,] -0.9804  0.9902    0 0.0098 0.0098 0.0098 0.0098
[3,]  0.0000  0.0000    0 0.0000 0.0000 0.0000 0.0000
[4,]  0.9804  0.0098    0 1.9902 0.9902 0.9902 0.9902
[5,]  0.9804  0.0098    0 0.9902 1.9902 0.9902 0.9902
[6,]  0.9804  0.0098    0 0.9902 0.9902 1.9902 0.9902
[7,]  0.9804  0.0098    0 0.9902 0.9902 0.9902 1.9902
> round( out.s <- sqrt(diag(out$V)), 4 ) 
[1] 1.4003 0.9951 0.0000 1.4107 1.4107 1.4107 1.4107
> round( out$V / outer(out.s, out.s), 3)
       [,1]   [,2] [,3]  [,4]  [,5]  [,6]  [,7]
[1,]  1.000 -0.704  NaN 0.496 0.496 0.496 0.496
[2,] -0.704  1.000  NaN 0.007 0.007 0.007 0.007
[3,]    NaN    NaN  NaN   NaN   NaN   NaN   NaN
[4,]  0.496  0.007  NaN 1.000 0.498 0.498 0.498
[5,]  0.496  0.007  NaN 0.498 1.000 0.498 0.498
[6,]  0.496  0.007  NaN 0.498 0.498 1.000 0.498
[7,]  0.496  0.007  NaN 0.498 0.498 0.498 1.000
\end{verbatim}
} \noindent
To understand the result we need to compare it
with the case without uncertainty uncertainty.
In that case we had $X_1=1.98$. Now
we have $X_1=1.96$. The difference, although practically
irrelevant, is conceptually important. 
It is indeed equal to the expected value
of the offset (precisely $0.0196$). This is because the role
of the observation is to give us an information
about $X_1 + X_2$, sum of the true value and the offset.
The fact that we use the observations to update
our knowledge on the true value is simply because
the offset is a priori better known that the true value,
as it is  well understood by experienced physicists:
if the calibration is poor the instrument cannot
be used for `measurements'. Note also the correlation
that now appears between $X_1$ and $X_2$, and in particular
its negative sign: the value of the true value could increase
at the `expenses' of the offset, and the other way around.

\subsubsection{Condition on $X_3=2$ and $X_4=1$ (``two observations'')}
{\small
\begin{verbatim}
> ( mu.c <- c(NA, NA, 2, 1, rep(NA, n-4)) )
[1] NA NA  2  1 NA NA NA
> out <- norm.mult.cond(mu, V, mu.c) 
> round( out$mu, 4)
[1] 1.4778 0.0148 2.0000 1.0000 1.4926 1.4926 1.4926
> round( out$V, 4)     
        [,1]    [,2] [,3] [,4]   [,5]   [,6]   [,7]
[1,]  1.4778 -0.9852    0    0 0.4926 0.4926 0.4926
[2,] -0.9852  0.9901    0    0 0.0049 0.0049 0.0049
[3,]  0.0000  0.0000    0    0 0.0000 0.0000 0.0000
[4,]  0.0000  0.0000    0    0 0.0000 0.0000 0.0000
[5,]  0.4926  0.0049    0    0 1.4975 0.4975 0.4975
[6,]  0.4926  0.0049    0    0 0.4975 1.4975 0.4975
[7,]  0.4926  0.0049    0    0 0.4975 0.4975 1.4975
> round( out.s <- sqrt(diag(out$V)), 4 ) 
[1] 1.2157 0.9951 0.0000 0.0000 1.2237 1.2237 1.2237
> round( out$V / outer(out.s, out.s), 3) 
       [,1]   [,2] [,3] [,4]  [,5]  [,6]  [,7]
[1,]  1.000 -0.814  NaN  NaN 0.331 0.331 0.331
[2,] -0.814  1.000  NaN  NaN 0.004 0.004 0.004
[3,]    NaN    NaN  NaN  NaN   NaN   NaN   NaN
[4,]    NaN    NaN  NaN  NaN   NaN   NaN   NaN
[5,]  0.331  0.004  NaN  NaN 1.000 0.332 0.332
[6,]  0.331  0.004  NaN  NaN 0.332 1.000 0.332
[7,]  0.331  0.004  NaN  NaN 0.332 0.332 1.000
\end{verbatim}
} \noindent
The only new effect we observe is the 
increase  (in module) of the 
correlation coefficient between true value and offset. 
This is due to the fact that the increased number of observation
has increased the constrain between the two quantities.
It will increase more if we use further observations,
for example conditioning on
"\code{mu.c <- c(NA, NA, 2, 1, 1.5, 2.2, 0.5)}'',
or decreasing the standard deviations $\sigma_{i|1,2}$.
For example if we set all $\sigma_{i|1,2}$ to 0.1, 
the same conditioning on $X_3$ and $X_3$ would produce
a correlation coefficient of $-0.9975$. 
Asymptotically there will be a deterministic constrain
between $X_1$ and $X_2$ of the kind $X_1+X_2=k$, and the two variables
become {\em logically dependent}.

\subsubsection{``Ricalibration of the offset'' 
($X_1=2$; $X_3=2$, $X_4=1$)}
What happens if we instead fix the value of the 
true value and some values of the observables?
In this case we update our information on the offset.
Let us see the case in which we fix the value of the true
value at 2, and the average of the two observations
at 1.5.
{\small
\begin{verbatim}
> ( mu.c <- c(2, NA, 2, 1, rep(NA, n-4)) )
[1]  2 NA  2  1 NA NA NA
> out <- norm.mult.cond(mu, V, mu.c) 
> round( out$mu, 4)
[1]  2.0000 -0.3333  2.0000  1.0000  1.6667  1.6667  1.6667
> round( out$V, 4)   
     [,1]   [,2] [,3] [,4]   [,5]   [,6]   [,7]
[1,]    0 0.0000    0    0 0.0000 0.0000 0.0000
[2,]    0 0.3333    0    0 0.3333 0.3333 0.3333
[3,]    0 0.0000    0    0 0.0000 0.0000 0.0000
[4,]    0 0.0000    0    0 0.0000 0.0000 0.0000
[5,]    0 0.3333    0    0 1.3333 0.3333 0.3333
[6,]    0 0.3333    0    0 0.3333 1.3333 0.3333
[7,]    0 0.3333    0    0 0.3333 0.3333 1.3333
> round( out.s <- sqrt(diag(out$V)), 4 ) 
[1] 0.0000 0.5774 0.0000 0.0000 1.1547 1.1547 1.1547
> round( out$V / outer(out.s, out.s), 3) 
     [,1] [,2] [,3] [,4] [,5] [,6] [,7]
[1,]  NaN  NaN  NaN  NaN  NaN  NaN  NaN
[2,]  NaN  1.0  NaN  NaN 0.50 0.50 0.50
[3,]  NaN  NaN  NaN  NaN  NaN  NaN  NaN
[4,]  NaN  NaN  NaN  NaN  NaN  NaN  NaN
[5,]  NaN  0.5  NaN  NaN 1.00 0.25 0.25
[6,]  NaN  0.5  NaN  NaN 0.25 1.00 0.25
[7,]  NaN  0.5  NaN  NaN 0.25 0.25 1.00
\end{verbatim}
} \noindent
As a result, the expected value of the offset becomes
$-0.33$, with a standard deviation of 0.58, against the
(possible) intuitive guess of $-0.5$ (i.e $1.5-2.0$)
with a standard uncertainty of 0.71 (i.e. $1/\sqrt{2}$).
The reason is that our prior knowledge on the offset 
had a standard uncertainty
of 1, that has to be taken into account. Indeed
it can be easily checked that the `intuitive' result
would have been recovered if we had a very large uncertainty
($\sigma_2 \rightarrow\infty$).
In fact  $-0.33$ is the weighted average of the initial value
0 and  $-0.5$, with weights equal to $1$ and $2$. 
The reason is that the result based on reconditioning
provides automatically the rule of the weighted
average with `inverse of the variances', where the `variance'
associated to $-0.5$ would be that obtained if the prior 
knowledge on the offset was irrelevant 
(i.e. $\sigma_2 \rightarrow\infty$).

\begin{figure} 
\begin{center}
\epsfig{file=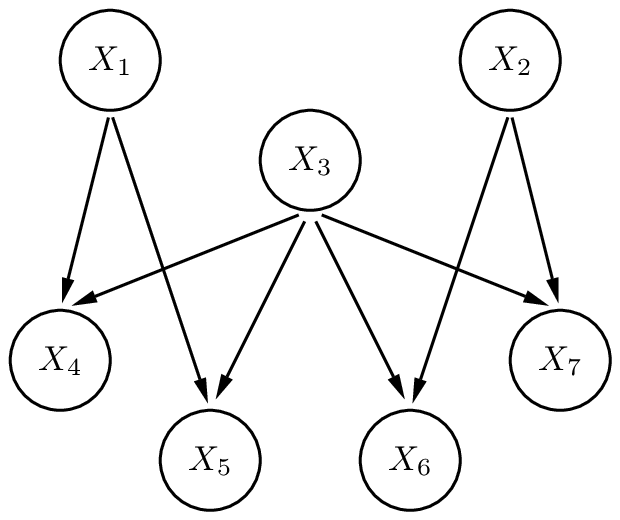,clip=}\hspace{2.0cm}
\epsfig{file=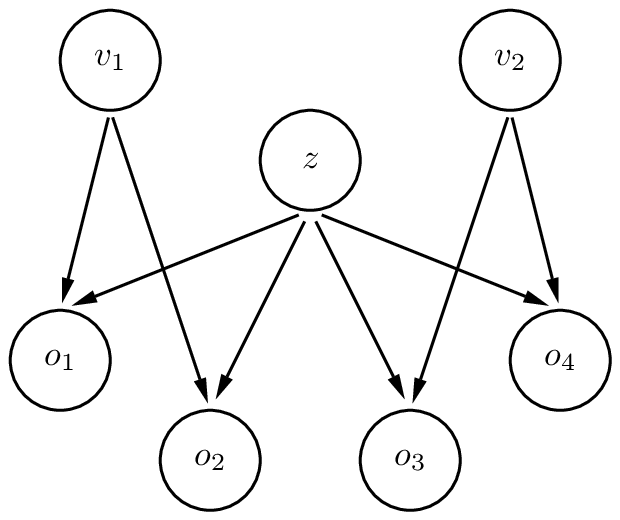,clip=}
\end{center}
\caption{Model to describe the measurements of two quantities
with the same instrument affected by some systematics.}
\label{fig:x1x2x3-x4x5x6x7}
\end{figure}

\section{Measuring two quantities with the same instrument
affected by offset uncertainty}
Another interesting issue, very common in experimental physics,
is when we make several measurements on homogeneous quantities
using the same instrument that, as all instruments, has
unavoidable uncertainty in the calibration. The situation is 
sketched in the diagrams of Fig.~\ref{fig:x1x2x3-x4x5x6x7},
drawn with the different symbols used in the text:
$X_1$ and $X_2$ are the true values; 
$X_3$ the common offset; $X_4$ and $X_5$ the independent
readings when the instrument is applied to $X_1$;
$X_6$ and $X_7$ the independent 
readings when the instrument is applied to $X_2$.
  
From this model we can easily build the transformation matrix $\bm{C}$
\begin{eqnarray}
\bm{C} &=&  \left(\! \begin{array}{ccccccc} 
            1 & 0 & 0 & 0 & 0 & 0 & 0\\
            0 & 1 & 0 & 0 & 0 & 0 & 0\\
            0 & 0 & 1 & 0 & 0 & 0 & 0\\
            1 & 0 & 1 & 1 & 0 & 0 & 0\\
            1 & 0 & 1 & 0 & 1 & 0 & 0\\
            0 & 1 & 1 & 0 & 0 & 1 & 0\\
            0 & 1 & 1 & 0 & 0 & 0 & 1 
            \end{array} \!\right)    
\end{eqnarray}
(for example it says that row 6 depends on $X_2$, $X_3$
and $e_{3}$). Applying it to the starting diagonal matrix
($v_1$, $v_2$ and $z$ are {\em initially independent};
the various errors $e_1$-$e_4$ are independent) we
get the covariance matrix of the joint multivariate normal
of interest:
\begin{equation}
\bm{V} = \left(\!\!\! \begin{array}{ccccccc} 
 \sigma_1^2 & 0           & 0 & \sigma_1^2 & \sigma_1^2 &  0   & 0\\ 
 0 & \sigma_2^2  & 0 &  0   &  0 & \sigma_2^2 &  \sigma_2^2 \\
 0 & 0           &  \sigma_3^2  & \sigma_3^2  &  \sigma_3^2
 & \sigma_3^2  &  \sigma_3^2 \\
 \sigma_1^2 & 0   &  \sigma_3^2  &
       \sigma_1^2 + \sigma_3^2 + \sigma_{e_1}^2 & 
 \sigma_1^2 + \sigma_3^2 &  \sigma_3^2 &  \sigma_3^2 \\
\sigma_1^2 & 0   &  \sigma_3^2  & \sigma_1^2 + \sigma_3^2  & 
 \sigma_1^2 + \sigma_3^2 + \sigma_{e_2}^2 & \sigma_3^2 & \sigma_3^2 \\
 0  & \sigma_2^2  &  \sigma_3^2 &  \sigma_3^2  & \sigma_3^2  & 
     \sigma_2^2 + \sigma_3^2 + \sigma_{e_3}^2 &   \sigma_2^2 + \sigma_3^2 \\
 0  & \sigma_2^2  &  \sigma_3^2 &  \sigma_3^2  & \sigma_3^2  & 
     \sigma_2^2 + \sigma_3^2 & \sigma_2^2 + \sigma_3^2 + \sigma_{e_4}^2  \\
                      \end{array} \!\right)   \nonumber 
\end{equation}
This is a very interesting covariance matrix and we leave the reader
the pleasure of exploiting all possibilities. Here we only
show a numerical example, with parameters similar to the ones
used before for a better understanding, and just discuss
a single case of conditioning.
{\small
\begin{verbatim}
> n=7; muX1=0; sigmaX1=10; muX2=0; sigmaX2=10;  # set parameters
> muZ=0; sigmaZ=1     
> mu <- c(muX1, muX2, muZ, rep(muX1+muZ,2), rep(muX2+muZ,2)) # set expected values
> ( sigma <- c(sigmaX1, sigmaX1, sigmaZ, rep(1, n-3)) )   # standard deviations  
[1] 10 10  1  1  1  1  1
> C <- matrix(rep(0, n*n), c(n,n))      # tranformation matrix
> diag(C) <- rep(1, n)
> C[4,] <- c(1, 0, 1, 1, 0, 0, 0)
> C[5,] <- c(1, 0, 1, 0, 1, 0, 0)
> C[6,] <- c(0, 1, 1, 0, 0, 1, 0)
> C[7,] <- c(0, 1, 1, 0, 0, 0, 1)
> C
     [,1] [,2] [,3] [,4] [,5] [,6] [,7]
[1,]    1    0    0    0    0    0    0
[2,]    0    1    0    0    0    0    0
[3,]    0    0    1    0    0    0    0
[4,]    1    0    1    1    0    0    0
[5,]    1    0    1    0    1    0    0
[6,]    0    1    1    0    0    1    0
[7,]    0    1    1    0    0    0    1
> V0 <- matrix(rep(0, n*n), c(n,n))     # covariance matrix   
> diag(V0) <-  sigma^2
> V <- C %*% V0 %*% t(C)
> V
     [,1] [,2] [,3] [,4] [,5] [,6] [,7]
[1,]  100    0    0  100  100    0    0
[2,]    0  100    0    0    0  100  100
[3,]    0    0    1    1    1    1    1
[4,]  100    0    1  102  101    1    1
[5,]  100    0    1  101  102    1    1
[6,]    0  100    1    1    1  102  101
[7,]    0  100    1    1    1  101  102
> su <- sqrt(diag(V))                   # standard uncertainties
> round( V/outer(su,su), 4)             # correlation matrix
       [,1]   [,2]  [,3]   [,4]   [,5]   [,6]   [,7]
[1,] 1.0000 0.0000 0.000 0.9901 0.9901 0.0000 0.0000
[2,] 0.0000 1.0000 0.000 0.0000 0.0000 0.9901 0.9901
[3,] 0.0000 0.0000 1.000 0.0990 0.0990 0.0990 0.0990
[4,] 0.9901 0.0000 0.099 1.0000 0.9902 0.0098 0.0098
[5,] 0.9901 0.0000 0.099 0.9902 1.0000 0.0098 0.0098
[6,] 0.0000 0.9901 0.099 0.0098 0.0098 1.0000 0.9902
[7,] 0.0000 0.9901 0.099 0.0098 0.0098 0.9902 1.0000
\end{verbatim}
} \noindent
Now let us assume we have applied our instrument
once on $X_1$ and once on $X_2$, obtaining the readings
$X_4=1$ and $X_6=2$, respectively. Here is how our knowledge is updated:
\mbox{}\vspace{-0.1cm}
{\small
\begin{verbatim}
> ( mu.c <- c(rep(NA, 3), 1, NA, 2, NA) )       # conditioning 
[1] NA NA NA  1 NA  2 NA
> out <- norm.mult.cond(mu, V, mu.c) 
> round( out$mu, 4)
[1] 0.9613 1.9514 0.0291 1.0000 0.9904 2.0000 1.9805
> round( out$V, 4)     
        [,1]    [,2]    [,3] [,4]    [,5] [,6]    [,7]
[1,]  1.9514  0.9613 -0.9709    0  0.9805    0 -0.0096
[2,]  0.9613  1.9514 -0.9709    0 -0.0096    0  0.9805
[3,] -0.9709 -0.9709  0.9806    0  0.0097    0  0.0097
[4,]  0.0000  0.0000  0.0000    0  0.0000    0  0.0000
[5,]  0.9805 -0.0096  0.0097    0  1.9902    0  0.0001
[6,]  0.0000  0.0000  0.0000    0  0.0000    0  0.0000
[7,] -0.0096  0.9805  0.0097    0  0.0001    0  1.9902
> round( out.s <- sqrt(diag(out$V)), 4 ) 
[1] 1.3969 1.3969 0.9902 0.0000 1.4107 0.0000 1.4107
> round( out$V / outer(out.s, out.s), 3) 
       [,1]   [,2]   [,3] [,4]   [,5] [,6]   [,7]
[1,]  1.000  0.493 -0.702  NaN  0.498  NaN -0.005
[2,]  0.493  1.000 -0.702  NaN -0.005  NaN  0.498
[3,] -0.702 -0.702  1.000  NaN  0.007  NaN  0.007
[4,]    NaN    NaN    NaN  NaN    NaN  NaN    NaN
[5,]  0.498 -0.005  0.007  NaN  1.000  NaN  0.000
[6,]    NaN    NaN    NaN  NaN    NaN  NaN    NaN
[7,] -0.005  0.498  0.007  NaN  0.000  NaN  1.000
\end{verbatim}
} \noindent
As expected, $X_4=1$ sets essentially to 1 
 the true value $X_1$ and the `future' 
-- or not yet known! -- reading $X_5$. Similarly,
$X_6$  sets essentially to 2 $X_2$ and  $X_7$. 
(The difference from the exact
value of 1 and 2, respectively, is due -- let us repeat it once again --
to the fact that we use, for didactic purposes, 
initial standard uncertainties $\sigma_1$ and 
$\sigma_2$ `relatively small', while the uncertainty  
on the common offset is `relatively large'.)
The most interest part of the result is the $3\times 3$
upper left part of the resulting correlation matrix,
which we repeat here:
\begin{equation*}
 \left(\! \begin{array}{ccc}
                  1.000 & 0.493 & -0.702 \\
                   0.493 & 1.000 &  -0.702 \\
                 -0.702 & -0.702 & 1.000
                   \end{array} 
          \!\right)
\end{equation*}
As we have learned in the previous section, the 
value of the offset gets anticorrelated to the 
true values. Moreover the two true values get
{\bf positively correlated}, as expected: a part of our uncertainty
on them is due the imprecise knowledge of the offset,
which then affects both values in the same direction.

\section{Inferences and forecasting based on mean values}
Often our inferences and forecasting are based on averages, 
instead than on individual values. 
It is rather understood 
that in Gaussian samples the inference on the Gaussian
`$\mu$' is the same if we use the mean rather than the 
detailed information, due to the so called property of
`statistical sufficiency'. It is instead less clear 
what we should expect for a next mean, based on a sample of the 
same size of the first one. For example, very often one ears
and reads\footnote{
For example, we read in Ref.~\cite{Cowan} (pp. 118-119)
\begin{quote}
{\sl 
``In reporting the measurement of $\theta$ as 
$\hat \theta_{obs}\pm \hat \sigma_{\hat \theta}$ one means that 
repeated estimates all based on $n$ observations of $x$ would be 
distributed according to a p.d.f. $g(\hat \theta)$ centered
around some true value $\theta$ and  
true standard deviation $\sigma_{\hat \theta}$, which are estimated to be
$\hat \theta_{obs}$ and  $\sigma_{\hat \theta}$.'' 
Mistakes of this kind are due to a curious ideology that refuses
to make probabilistic statements about uncertain values, in contrast
to the physicist' intuition (see extended discussions in Ref.~\cite{BR},
with hints on the way of reasoning of Gauss and Laplace).
}
\end{quote}
} 
something like
 ``if we have got a mean ($\overline{x}_p$) 
 and then imagine to repeat a large number of independent 
 samples of the same size ($n$) of the `past' one, then
 we expect about
in the interval $\overline{x}_p\pm \sigma/\sqrt{n}$\,'', 
that is
\begin{eqnarray}
P(\overline{x}_p - \frac{\sigma}{\sqrt{n}}\, \le \overline{x}_f \le \,
  \overline{x}_p + \frac{\sigma}{\sqrt{n}}) &=& 68\%\,,
\end{eqnarray}
a statement {\bf wrong} by a factor $1/\sqrt{2}$ 
in the size of the interval (or, equivalently,
about 30\% in the value of expected frequency, that should be
52\%) as we shall see in a while 
(see also Ref.~\cite{BR}). In order to do this, 
let us play with our tool, building the minimal model 
to observe the effect of interest. 

\begin{figure}
\begin{center}
\epsfig{file=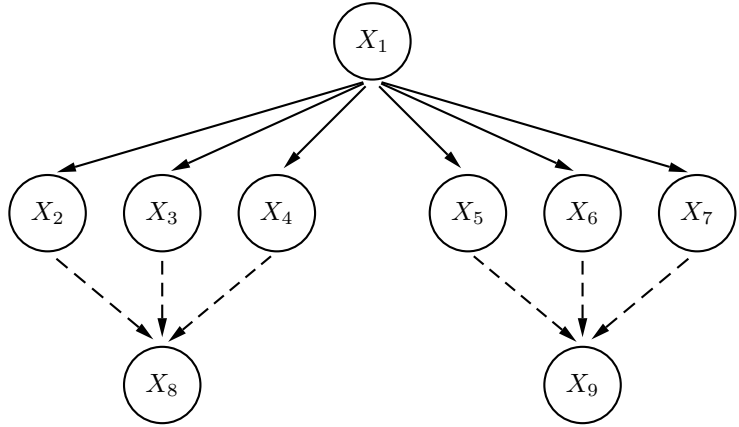,clip=} 
\end{center}
\caption{Past mean and future mean. $X_1$ is the value of the quantity,
$X_2$-$X_4$ the observations in the first samples (i.e. 
$o_1$, $o_2$ and  $o_3$), 
$X_5$-$X_7$ the observations in the second samples (i.e. 
$o_4$, $o_5$ and  $o_5$). $X_8$ and $X_9$ are the sample means.}
\label{fig:x1-2medie}
\end{figure}

Figure \ref{fig:x1-2medie} 
shows the value of a quantity ($X_1$)
and two samples, each of three observations ($X_2$-$X_4$ and 
$X_5$-$X_7$), whose mean values are $X_8$ and $X_9$ 
(the dashed arrows indicate that the links are {\em deterministic}
instead than probabilistic, since an arithmetic
mean is univocally determined by the values of the sample). 
We only need to write down the transformation rules 
to get $X_8$ and $X_9$
in order to build up the extra rows of the transformation matrix.
They are: 
\begin{eqnarray}
X_8 &=& \frac{1}{3}\,(X_2 + X_3 + X_4) 
        = X_1 +  \frac{1}{3}\,(e_1+e_2+e_3)
\label{eq:Y8-mean1} \\
X_9 &=& \frac{1}{3}\,(X_5 + X_6 + X_7) 
        = X_1 +  \frac{1}{3}\,(e_4+e_5+e_6) 
\end{eqnarray}
from which it follows
\begin{eqnarray}
\bm{C} &=&  \left(\! \begin{array}{ccccccc} 
            1 & 0 & 0 & 0 & 0 & 0 & 0\\
            1 & 1 & 0 & 0 & 0 & 0 & 0\\
            1 & 0 & 1 & 0 & 0 & 0 & 0\\
            1 & 0 & 0 & 1 & 0 & 0 & 0\\
            1 & 0 & 0 & 0 & 1 & 0 & 0\\
            1 & 0 & 0 & 0 & 0 & 1 & 0\\
            1 & 0 & 0 & 0 & 0 & 0 & 1\\
            1 & {1}/{3} & {1}/{3} & {1}/{3} & 0 & 0 & 0\\
            1 &  0 & 0 & 0  & {1}/{3} & {1}/{3} & {1}/{3}
            \end{array} \!\right)    
\end{eqnarray}
Here is our implementation in R, where we have now used 
$\sigma_1=100$, much larger than the `experimental resolution'
of 1. This choice makes, for the numerical values of the observations 
we shall use, the prior on $X_1$ practically 
irrelevant (the effect on the expected values is of the
order $10^{-5}$) so that we can better
focus on other effects:
\mbox{}\vspace{-0.1cm}
{\small
\begin{verbatim}
> n <- 7; muX1 <- 0; sigmaX1 <- 100; sigma.i <- 1     # parameters
> mu <- rep(muX1, n + 2)                              # expected values
> ( sigma <- c(sigmaX1, rep(sigma.i,n-1)) )             
> V0 <- matrix(rep(0, n*n), c(n,n))        # initial diagonal covariance matrix
> diag(V0) <- sigma^2

> C <- matrix(rep(0, n*n), c(n,n))                    transformation matrix
> diag(C) <- rep(1, n)
> C[,1] <- rep(1, n)
> C <- rbind( C, c(1, rep(1/3, 3), rep(0, 3)), c(1, rep(0, 3), rep(1/3, 3)) )
> round(C,3)
      [,1]  [,2]  [,3]  [,4]  [,5]  [,6]  [,7]
 [1,]    1 0.000 0.000 0.000 0.000 0.000 0.000
 [2,]    1 1.000 0.000 0.000 0.000 0.000 0.000
 [3,]    1 0.000 1.000 0.000 0.000 0.000 0.000
 [4,]    1 0.000 0.000 1.000 0.000 0.000 0.000
 [5,]    1 0.000 0.000 0.000 1.000 0.000 0.000
 [6,]    1 0.000 0.000 0.000 0.000 1.000 0.000
 [7,]    1 0.000 0.000 0.000 0.000 0.000 1.000
 [8,]    1 0.333 0.333 0.333 0.000 0.000 0.000
 [9,]    1 0.000 0.000 0.000 0.333 0.333 0.333

> V <- C %*% V0 %*% t(C)                       # joint covariance matrix
> ( su <- sqrt(diag(V)) )                      # initial uncertainties
[1] 100.0000 100.0050 100.0050 100.0050 100.0050 100.0050 100.0050 100.0017
[9] 100.0017
> ( m <- dim(V)[1] )             #  number of transformed variables 
[1] 9
> round( V/outer(su,su), 5)
         [,1]    [,2]    [,3]    [,4]    [,5]    [,6]    [,7]    [,8]    [,9]
 [1,] 1.00000 0.99995 0.99995 0.99995 0.99995 0.99995 0.99995 0.99998 0.99998
 [2,] 0.99995 1.00000 0.99990 0.99990 0.99990 0.99990 0.99990 0.99997 0.99993
 [3,] 0.99995 0.99990 1.00000 0.99990 0.99990 0.99990 0.99990 0.99997 0.99993
 [4,] 0.99995 0.99990 0.99990 1.00000 0.99990 0.99990 0.99990 0.99997 0.99993
 [5,] 0.99995 0.99990 0.99990 0.99990 1.00000 0.99990 0.99990 0.99993 0.99997
 [6,] 0.99995 0.99990 0.99990 0.99990 0.99990 1.00000 0.99990 0.99993 0.99997
 [7,] 0.99995 0.99990 0.99990 0.99990 0.99990 0.99990 1.00000 0.99993 0.99997
 [8,] 0.99998 0.99997 0.99997 0.99997 0.99993 0.99993 0.99993 1.00000 0.99997
 [9,] 0.99998 0.99993 0.99993 0.99993 0.99997 0.99997 0.99997 0.99997 1.00000
\end{verbatim}
} \noindent
As we see, all quantities are now highly correlated.
In particular, it is interesting to see how $X_8$ and $X_9$
are correlated with $X_1$, with any observation
of  the first sample ($X_2$-$X_4$) and with any
observation of the second sample ($X_5$-$X_7$).

\subsection{Expectations for a given value of $X_1$}
Let us now fix $X_1$ at our usual value of $2$:
\mbox{}\vspace{-0.1cm}
{\small
\begin{verbatim}
> ( mu.c <- c(2, rep(NA, m-1)) )    # X1 = 2
[1]  2 NA NA NA NA NA NA NA NA
> out <- norm.mult.cond(mu, V, mu.c)  
> out$mu
[1] 2 2 2 2 2 2 2 2 2
> round( out.s <- sqrt(diag(out$V)), 4 )
[1] 0.0000 1.0000 1.0000 1.0000 1.0000 1.0000 1.0000 0.5774 0.5774
> round(  out$V / outer(out.s, out.s), 4)
      [,1]   [,2]   [,3]   [,4]   [,5]   [,6]   [,7]   [,8]   [,9]
 [1,]  NaN    NaN    NaN    NaN    NaN    NaN    NaN    NaN    NaN
 [2,]  NaN 1.0000 0.0000 0.0000 0.0000 0.0000 0.0000 0.5774 0.0000
 [3,]  NaN 0.0000 1.0000 0.0000 0.0000 0.0000 0.0000 0.5774 0.0000
 [4,]  NaN 0.0000 0.0000 1.0000 0.0000 0.0000 0.0000 0.5774 0.0000
 [5,]  NaN 0.0000 0.0000 0.0000 1.0000 0.0000 0.0000 0.0000 0.5774
 [6,]  NaN 0.0000 0.0000 0.0000 0.0000 1.0000 0.0000 0.0000 0.5774
 [7,]  NaN 0.0000 0.0000 0.0000 0.0000 0.0000 1.0000 0.0000 0.5774
 [8,]  NaN 0.5774 0.5774 0.5774 0.0000 0.0000 0.0000 1.0000 0.0000
 [9,]  NaN 0.0000 0.0000 0.0000 0.5774 0.5774 0.5774 0.0000 1.0000
\end{verbatim}
} \noindent
As we already know, all observations become conditionally independent
with expected value 2 and standard uncertainty 1. The averages are also 
expected to be around 2, but with smaller uncertainty, namely
$1/\sqrt{3} \approx 0.5774$ And, obviously, the averages are only 
correlated with each observation of their own sample. For example,
for $X_2$ and $X_8$, we have, starting from the transformation
rules (\ref{eq:X2-O1E1}) and (\ref{eq:Y8-mean1}), 
we get, being $X_1$ certain,
\begin{eqnarray}
\mbox{Cov}[X_2, X_8] &=&   \frac{1}{3}\times \sigma_{e_1}^2 \\
                &=&  \frac{1}{3}\times 1 \approx  0.33\,,
\end{eqnarray}
and then
\begin{eqnarray}
\rho[X_2, X_8] &=& \frac{ \mbox{Cov}[X_2, X_8] }
                        { \sigma[X_2]\cdot \sigma[X_8]  } \\
 &=& \frac{1/3}{1\times 1/\sqrt{3}} = \sqrt{3} \approx 0.5774\,,
\end{eqnarray}
that is exactly what we can read in the R output.

\subsection{Reconditioning on the value of the first mean}
Let us now see what happens if we get informed about a mean value,
e.g. $X_8=2$.
\mbox{}\vspace{-0.1cm}
{\small 
\begin{verbatim}
> ( mu.c <- c(rep(NA, m-2), 2, NA) )           # first mean = 2
[1] NA NA NA NA NA NA NA  2 NA
> out <- norm.mult.cond(mu, V, mu.c)
> round(out$mu, 5)
[1] 1.99993 2.00000 2.00000 2.00000 1.99993 1.99993 1.99993 2.00000 1.99993
> round( out.s <- sqrt(diag(out$V)), 4)
[1] 0.5773 0.8165 0.8165 0.8165 1.1547 1.1547 1.1547 0.0000 0.8165
> round(out$V, 4)   
        [,1]    [,2]    [,3]    [,4]   [,5]   [,6]   [,7] [,8]   [,9]
 [1,] 0.3333  0.0000  0.0000  0.0000 0.3333 0.3333 0.3333    0 0.3333
 [2,] 0.0000  0.6667 -0.3333 -0.3333 0.0000 0.0000 0.0000    0 0.0000
 [3,] 0.0000 -0.3333  0.6667 -0.3333 0.0000 0.0000 0.0000    0 0.0000
 [4,] 0.0000 -0.3333 -0.3333  0.6667 0.0000 0.0000 0.0000    0 0.0000
 [5,] 0.3333  0.0000  0.0000  0.0000 1.3333 0.3333 0.3333    0 0.6667
 [6,] 0.3333  0.0000  0.0000  0.0000 0.3333 1.3333 0.3333    0 0.6667
 [7,] 0.3333  0.0000  0.0000  0.0000 0.3333 0.3333 1.3333    0 0.6667
 [8,] 0.0000  0.0000  0.0000  0.0000 0.0000 0.0000 0.0000    0 0.0000
 [9,] 0.3333  0.0000  0.0000  0.0000 0.6667 0.6667 0.6667    0 0.6667
>  round(  out$V / outer(out.s, out.s), 4)
        [,1] [,2] [,3] [,4]   [,5]   [,6]   [,7] [,8]   [,9]
 [1,] 1.0000  0.0  0.0  0.0 0.5000 0.5000 0.5000  NaN 0.7071
 [2,] 0.0000  1.0 -0.5 -0.5 0.0000 0.0000 0.0000  NaN 0.0000
 [3,] 0.0000 -0.5  1.0 -0.5 0.0000 0.0000 0.0000  NaN 0.0000
 [4,] 0.0000 -0.5 -0.5  1.0 0.0000 0.0000 0.0000  NaN 0.0000
 [5,] 0.5000  0.0  0.0  0.0 1.0000 0.2500 0.2500  NaN 0.7071
 [6,] 0.5000  0.0  0.0  0.0 0.2500 1.0000 0.2500  NaN 0.7071
 [7,] 0.5000  0.0  0.0  0.0 0.2500 0.2500 1.0000  NaN 0.7071
 [8,]    NaN  NaN  NaN  NaN    NaN    NaN    NaN  NaN    NaN
 [9,] 0.7071  0.0  0.0  0.0 0.7071 0.7071 0.7071  NaN 1.0000
\end{verbatim}
} \noindent
The knowledge about the first average constrains 
$X_1$ to $2.00\pm 0.58$, that is $\overline{x}\pm \sigma/\sqrt{n}$, 
while the expectations about the next average is
 $2.00\pm 0.82$, that is 
$\overline{x}\pm \sqrt{2}\,\sigma/\sqrt{n}$. The future
observations are instead expected to be  $2.00\pm 1.15$, 
where the standard uncertainty comes from 
$\sqrt{0.577^2 + 1^2}$, quadratic combination
of the uncertainty about $X_1$ and that of any of the future 
observations around $X_1$. 

And, as expected, there are correlations
among all values which are still uncertain, with the {\em exception}
of $X_1$ with $X_2$, $X_3$ and $X_4$ (the observations of the first sample). 
This on a first sight is not very intuitive. 
The reason is that $X_1$ is fully determined by the average $X_8$,
and therefore our knowledge about it cannot 
change if we are informed about the
individual values of the measurements, 
as we shall see in the next 
subsection. 

Remaining on the values of the first sample,
their expected value is exactly 2, 
instead than $1.99993$, a difference absolutely 
negligible in practice, but very
interesting indeed to understand the flow of the probabilistic updates.
Their values depend only on the average, and \underline{not} 
on the prior about $X_1$. Their uncertainty is the same 
as the uncertainty on the
future average ($\sqrt{2}\times 1/\sqrt{3}$), although not easy 
to understand at an intuitive level. Easier to understand are 
their mutual anticorrelations, since their
linear combination $X_8$ (their mean value) is fixed.

\subsection{Knowing the average and one of the values
that contribute to the first mean}
In order to better understand the role of the mean in the inference,
let us assume we also know the value of one of the three observations
contributing to it, for example $X_2=1$.
{\small 
\begin{verbatim}
> ( mu.c <- c(NA, 1, rep(NA, m-4), 2, NA) )          # first mean (X8) = 2;  X2=1 
[1] NA  1 NA NA NA NA NA  2 NA
> out <- norm.mult.cond(mu, V, mu.c, check=FALSE)
> round(out$mu, 4)
[1] 1.9999 1.0000 2.5000 2.5000 1.9999 1.9999 1.9999 2.0000 1.9999
> round(  out.s <- sqrt(diag(out$V)), 4 )
[1] 0.5773 0.0000 0.7071 0.7071 1.1547 1.1547 1.1547 0.0000 0.8165
> round(out$V, 4)               
        [,1] [,2] [,3] [,4]   [,5]   [,6]   [,7] [,8]   [,9]
 [1,] 0.3333    0  0.0  0.0 0.3333 0.3333 0.3333    0 0.3333
 [2,] 0.0000    0  0.0  0.0 0.0000 0.0000 0.0000    0 0.0000
 [3,] 0.0000    0  0.5 -0.5 0.0000 0.0000 0.0000    0 0.0000
 [4,] 0.0000    0 -0.5  0.5 0.0000 0.0000 0.0000    0 0.0000
 [5,] 0.3333    0  0.0  0.0 1.3333 0.3333 0.3333    0 0.6667
 [6,] 0.3333    0  0.0  0.0 0.3333 1.3333 0.3333    0 0.6667
 [7,] 0.3333    0  0.0  0.0 0.3333 0.3333 1.3333    0 0.6667
 [8,] 0.0000    0  0.0  0.0 0.0000 0.0000 0.0000    0 0.0000
 [9,] 0.3333    0  0.0  0.0 0.6667 0.6667 0.6667    0 0.6667
>  round(  out$V / outer(out.s, out.s), 4)
        [,1] [,2] [,3] [,4]   [,5]   [,6]   [,7] [,8]   [,9]
 [1,] 1.0000  NaN    0    0 0.5000 0.5000 0.5000  NaN 0.7071
 [2,]    NaN  NaN  NaN  NaN    NaN    NaN    NaN  NaN    NaN
 [3,] 0.0000  NaN    1   -1 0.0000 0.0000 0.0000  NaN 0.0000
 [4,] 0.0000  NaN   -1    1 0.0000 0.0000 0.0000  NaN 0.0000
 [5,] 0.5000  NaN    0    0 1.0000 0.2500 0.2500  NaN 0.7071
 [6,] 0.5000  NaN    0    0 0.2500 1.0000 0.2500  NaN 0.7071
 [7,] 0.5000  NaN    0    0 0.2500 0.2500 1.0000  NaN 0.7071
 [8,]    NaN  NaN  NaN  NaN    NaN    NaN    NaN  NaN    NaN
 [9,] 0.7071  NaN    0    0 0.7071 0.7071 0.7071  NaN 1.0000
\end{verbatim}
} \noindent
As we can see, the inference about $X_1$ does not change.
As a consequence, also the expectations about the future
observations are not affected by this extra piece of informations.
Instead, we change our knowledge about $X_3$ and $X_4$, whose
expected values become 2.5, in order to compensate $X_2=1$ [i.e
$2.5 = (3\times 2 - 1)/2$]
and they are fully anticorrelated, as more or less expected.

\section{The effect of a constrain among true values}
\begin{figure}[b]
\begin{center}
\epsfig{file=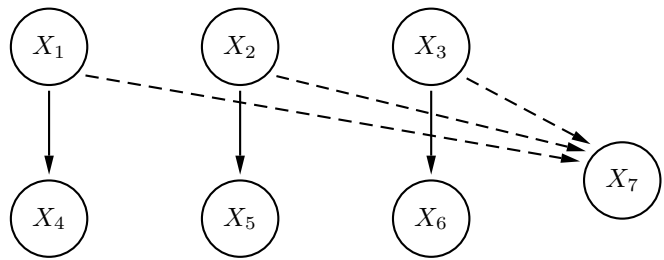,clip=}
\end{center}
\caption{Inferring the internal angles of a triangle by 
independent measurements.}
\label{fig:triangolo}
\end{figure}

Another important issue is how the knowledge that the
some quantities are intrinsically correlated 
changes the inference. Cases of this kind happen  
when several quantities are related by 
a deterministic relation, and  a well 
understood case is when measuring
the internal angles of a triangles in a flat space. 
Just to focus on a numerical example, let us imagine 
the individual angles to be determined, starting from 
very vague priors as
\begin{eqnarray}
\alpha &=& 58^\circ \pm 2^\circ \\
\beta &=& 73^\circ \pm 2^\circ \\
\gamma &=& 54^\circ \pm 2^\circ.
\end{eqnarray}
The measurements can be independent, as we have supposed (let us forget the case
of measurements with common systematics
in order to focus on the effect of the constrain), but 
nevertheless the relation $\alpha+\beta+\gamma = 180^\circ$ 
will make the results correlated.  
The graphical model is represented in 
figure \ref{fig:triangolo} 
with the extra node $X_7$ 
representing the sum of the angles and related to $X_1$, $X2$
and $X_3$ by deterministic links (dashed arrows).

\subsection{Exact solution in the case of identical resolution of the
goniometer and neglecting systematic effects}
Being the case rather simple, especially if all uncertainties
are equal, let us make the exercise of going through
the exact solution. Indicating the angles all together with the variable
$\bm{X} = \{\alpha, \beta, \gamma\}$, whose expected value
is $\mbox{E}[\bm{X}] =  \{a=58^\circ, b=73^\circ, c=54^\circ\}$.
The covariance matrix is diagonal with all terms 
equal to $\sigma^2=(2^\circ)^2$.
We make then the transformation 
to $\bm{Y} = \{\alpha, \beta, \gamma, \Sigma\}$, where 
$\Sigma = \alpha + \beta + \gamma$, and then condition on 
$\Sigma = 180^\circ$. The transformation matrix is then
\begin{eqnarray}
\bm{C} &=&  \left(\! \begin{array}{ccc} 
            1 & 0 & 0   \\
            0 & 1 & 0   \\
            0 & 0 & 1   \\
            1 & 1 & 1  
            \end{array} \!\right)    
\end{eqnarray}
from which we obtain
\begin{eqnarray}
\bm{V}_Y \!\!&=&\!\! 
\left(\! \begin{array}{ccc} 
            1 & 0 & 0   \\
            0 & 1 & 0   \\
            0 & 0 & 1   \\
            1 & 1 & 1  
            \end{array} \!\right)    
\cdot 
\left(\! \begin{array}{ccc} 
            \sigma^2 & 0 & 0   \\
            0 & \sigma^2 & 0   \\
            0 & 0 & \sigma^2  
            \end{array} \!\right)
\cdot 
\left(\! \begin{array}{cccc} 
            1 & 0 & 0 & 1  \\
            0 & 1 & 0 & 1  \\
            0 & 0 & 1 & 1 
            \end{array} \!\right)    
=   \left(\! \begin{array}{cccc} 
            \sigma^2 & 0 & 0 & \sigma^2  \\
            0 & \sigma^2 & 0 & \sigma^2   \\
            0 & 0 & \sigma^2 & \sigma^2 \\
             \sigma^2 & \sigma^2 &  \sigma^2 & 3\,\sigma^2
            \end{array} \!\right) \hspace{0.6cm}\mbox{}
\end{eqnarray}
Conditioning on $\Sigma = 180^\circ$, that is $Y_4 = 180^\circ$, 
using Eqs.~(\ref{eq:Eaton_E}) and (\ref{eq:Eaton_V}), we get
\begin{eqnarray}
\mbox{E} \left.\left(\! \begin{array}{c} 
            \alpha  \\
            \beta  \\
            \gamma 
             \end{array}
            \!\right)\right|_{\Sigma=180^\circ}
         &=& \left(\! \begin{array}{c} 
            a  \\
            b  \\
            c 
             \end{array}
            \!\right) - 
            \frac{\Delta\varphi}{3}\,
            \left(\! \begin{array}{c} 
            1  \\
            1  \\
            1
             \end{array}
            \!\right)\,,
\end{eqnarray}
with $\Delta\varphi = (a + b +c) -180^\circ$. In practice the resulting
rule is the most naive one could imagine: subtract to each value
one third of the excess of their sum above $180^\circ$. 
(If you think that this rule is to
simplistic, the reason might be that your model of uncertainty in this kind
of measurements is different than that used here, implying 
for example scale type errors. But this kind of errors are 
beyond the aim of this note, because they imply non-linear transformations.) 

This is the conditioned covariance matrix
\begin{eqnarray}
\frac{2\,\sigma^2}{3}\left(\! \begin{array}{ccc} 
            1 & -1/2 & -1/2   \\
            -1/2 & 1 & -1/2   \\
            -1/2 & -1/2 & 1  
            \end{array} \!\right)\,,    
\end{eqnarray}
written in a form that highlights the correlation matrix. 
The result is finally
\begin{eqnarray}
\alpha &=& \left(a - \frac{\Delta\varphi}{3}\right) 
           \pm \sqrt{\frac{2}{3}}\,\sigma
\end{eqnarray}
and similar expression for $\beta$ and $\gamma$, thus yielding
\begin{eqnarray}
\alpha &=& 56.33^\circ \pm 1.63^\circ \\
\beta &=& 71.33^\circ \pm 1.63^\circ \\
\gamma &=& 52.33^\circ \pm 1.63^\circ\,,
\end{eqnarray}
with $\rho(\alpha,\beta) = \rho(\alpha,\gamma) = \rho(\beta,\gamma) = -1/2$.

\subsection{Numerical implementation in R}
In analogy of what we have previously done in several cases,
we start from independent quantities $v_1$, $v_2$, $v_3$,
$e_{1}$, $e_{2}$ and $e_{3}$. 
For the true values of the angle we choose a flat prior,
modelled
with a Gaussian\footnote{Let
us remind that this does not imply we believe that the angles
could be negative or larger than $180^\circ$: it is
just a trick to have a pdf that is practically flat between 0
and  $180^\circ$. The trick allows us to use the normal multivariate
formulae of reconditioning. Obviously, one has to check that 
the final results are consistent with our assumptions and that the 
tails of the Gaussian posterior distributions are harmless, 
as it is the case in our example.} 
of central value (all values in degrees) 
$60$ and $\sigma=1000$. The expected values of the fluctuations 
of the observations around the true values are instead 0, with standard
deviations equal to the experimental resolutions, called
\code{sigma.gonio} in the code, so that it can be changed at wish. 

The transformation rules are
\begin{eqnarray}
X_1 &=& v_1 \\
X_2 &=& v_2 \\
X_3 &=& v_3 \\
X_4 &=& o_1 = v_1+e_1 \\
X_5 &=& o_2 = v_2+e_2 \\
X_6 &=& o_3 = v_3+e_3 \\
X_7 &=& v_1 + v_2 + v_3 
\end{eqnarray}
from which we get the transformation matrix
\begin{eqnarray}
\bm{C} &=&  \left(\! \begin{array}{cccccc} 
            1 & 0 & 0 & 0 & 0 & 0 \\
            0 & 1 & 0 & 0 & 0 & 0 \\
            0 & 0 & 1 & 0 & 0 & 0 \\
            1 & 0 & 0 & 1 & 0 & 0 \\
            0 & 1 & 0 & 0 & 1 & 0 \\
            0 & 0 & 1 & 0 & 0 & 1 \\
            1 & 1 & 1 & 0 & 0 & 0
            \end{array} \!\right)    
\end{eqnarray}
Here is the R code to calculate the expected 
values and covariance matrix of the three angles:
{\small 
\begin{verbatim}
mu.priors <- rep(60, 3); sigma.priors <- rep(1000, 3)     # priors
sigma.gonio <- c(2, 2, 2)               # experimental resolutions
m=6; mu0 <- c(mu.priors, rep(0, 3)) 
sigma <- c(sigma.priors, sigma.gonio)
V0 <- matrix(rep(0, m*m), c(m,m))       # diagonal matrix 
diag(V0) <- sigma^2 

C <- matrix(rep(0, m*m), c(m,m))        # tranformation matrix
diag(C) <- 1
for(i in 1:3) C[3+i, i] <- 1
C <- rbind(C, c( rep(1, 3), rep(0,3)) )

V <- C %*% V0 %*% t(C)                  # transformed matrix
mu <- as.vector(C %*% mu0)              # expected values

out <- norm.mult.cond(mu, V, c(NA, NA, NA, 58, 73, 54, 180) )
angles <- marginal.norm(out$mu, out$V, rep(1,3)) 
\end{verbatim}
} \noindent
And these are, finally, the results, shown as an R session:
{\small 
\begin{verbatim}
> angles$mu
[1] 56.33335 71.33329 52.33336
> ( sigma.angles <- sqrt(diag(angles$V)) )
[1] 1.63299 1.63299 1.63299
> ( corr <- angles$V / outer(sigma.angles, sigma.angles) )
           [,1]       [,2]       [,3]
[1,]  1.0000000 -0.4999976 -0.4999976
[2,] -0.4999976  1.0000000 -0.4999976
[3,] -0.4999976 -0.4999976  1.0000000
\end{verbatim}
} \noindent
As we see, we get the same results obtained above, with the advantage
that we can now change the experimental resolutions of the individual 
measurement. Or we can modify the model, in order 
to include the effect of common systematic, though limited
to offset type, exercise left to the reader.

\section{Fitting  linear models to data, with `known' 
standard deviations on the $y$ axes}\label{sec:fits}
\begin{figure}[!b]
\begin{center}
\epsfig{file=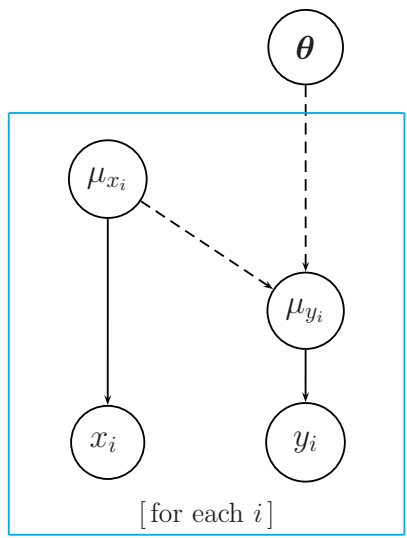,clip=,width=0.35\linewidth}
\end{center}
\caption{{\sl Graphical representation of the model in term of a
Bayesian network.~\cite{fits}}}
\label{fig:bn1}
\end{figure}

As a last example, let us see how {\em simple} fits can
be described in terms of conditioned normal multivariates.
`Simple' does not mean here linear fits, because even 
a realistic model to fit a straight line through data points
is not that `simple', if we are interested to infer also the 
standard deviations(s) describing the errors and we consider
errors on both axes (see Ref.~\cite{fits}, from which 
Fig.~\ref{fig:bn1} has been taken).
On the other hand, also fitting 
high order polynomials can be considered `simple', 
under the same assumptions. 

The meaning of Fig.~\ref{fig:bn1} is that for each data point 
we have three uncertain quantities: the true value of $x$ 
(``$\mu_{x_i}$''), the observed $x_i$ and the observed $y_i$,
while the true value $\mu_{x_i}$ is deterministically related
to $\mu_{x_i}$ and to the model parameters. So, for $n$
data points the `really' 
dimensionality of the problem (i.e. not taking 
into account the  $\mu_{y_i}$) is $3\times n+n_p$,
where $n_p$ is the number of parameters. The inference on 
the parameters $\mathbf{\theta}$ is the performed conditioning on
all $\mathbf{x}$ and $\mathbf{y}$ and marginalizing on 
$\mathbf{\mu_x}$.

A usual simplification is to ignore the errors on the $x$ values,
making then  $\mu_{y_i}$ deterministically on  $x_i$ and  $\mathbf{\theta}$.
Or, if we like, we can see each $y_i$ caused by the 
corresponding $x_i$ and the set of parameters  $\mathbf{\theta}$.
Assuming a normal error distribution with \underline{known} 
standard deviations, linear and quadratic models can be described
as
\begin{eqnarray}
Y_i &\sim& {\cal N}(c+m\,x_i,\,\sigma_i) \hspace{1.95cm}(\mbox{linear})\\
Y_i &\sim& {\cal N}(a+b\,x_i+c\,x_i^2,\,\sigma_i)\hspace{1.0cm}(\mbox{quadratic})
\end{eqnarray} 
and then expanded to all possible models of the kind
\begin{eqnarray}
Y_i &\sim& {\cal N}(\beta_1\,g_1(x_i)+\beta_2\,g_2(x_i) +\cdots,\, \sigma_i)\,,
\end{eqnarray}
where $g_1()$, $g_2()$ and so on are mathematical functions of $x_i$
not containing free parameters.\footnote{To make it even more clear,
in the case of the quadratic model we have: $\beta_1=a$,   
$\beta_2=b$ and  $\beta_3=c$; $g_1(x_i)=1$,  $g_2(x_i)=x_i$ 
and $g_3(x_i)=x_i^2$.} It is then rather clear that under
these assumptions the problem can be treated using the properties of
the multivariate normal distributions.

\begin{figure}
\begin{center}
\epsfig{file=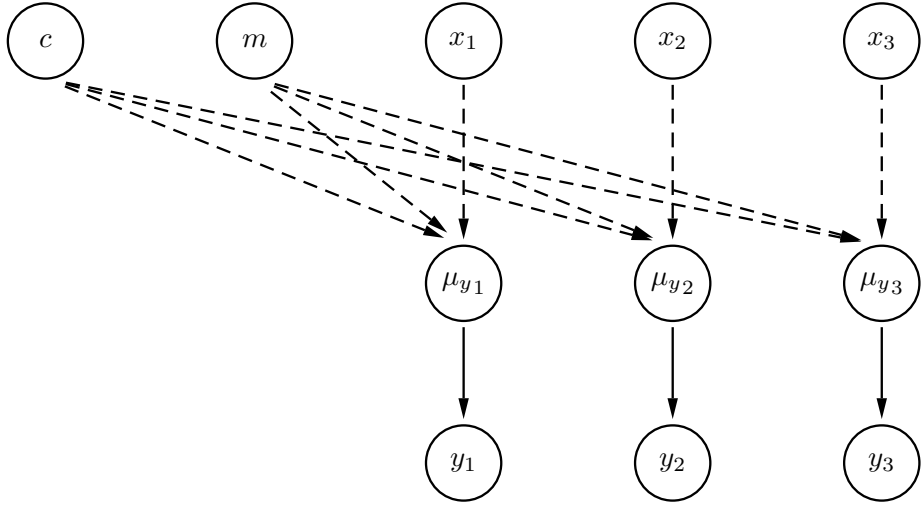,clip=,width=0.8\linewidth}
\end{center}
\caption{Linear fit to data point under usual simplifications (see text).}
\label{fig:linear_fit}
\end{figure}

The general model of Fig.~\ref{fig:bn1} becomes, for the first three
data points, that of Fig.~\ref{fig:linear_fit}.
The variables of our problem are then, indicating them with $Z_i$
\begin{eqnarray}
Z_1 &=& c \\
Z_2 &=& m \\
Z_3 &=& Y_1 = c + x_1\,m + e_1 \\
Z_4 &=& Y_2 = c + x_2\,m + e_2 \\
Z_5 &=& Y_3 = c + x_3\,m + e_3 
\end{eqnarray}
and so on.

Our usual transformation matrix transformation for the case
of three data points  is 
then\footnote{In the case of a parabolic fit
we would have instead
\begin{eqnarray*}
\bm{C} &=&  \left(\! \begin{array}{cccccc} 
            1 & 0 & 0 & 0 & 0 & 0 \\
            0 & 1 & 0 & 0 & 0 & 0 \\
            0 & 0 & 1 & 0 & 0 & 0 \\
            1 & x_1 & x_1^2 & 1 & 0 & 0 \\
            1 & x_2 & x_2^2 & 0 & 1 & 0 \\
            1 & x_3 & x_3^2 & 0 & 0 & 1 
            \end{array} \!\right)   
\end{eqnarray*}
and so on for higher order polynomials.
} 
\begin{eqnarray}
\bm{C} &=&  \left(\! \begin{array}{ccccc} 
            1 & 0 & 0 & 0 & 0  \\
            0 & 1 & 0 & 0 & 0 \\
            1 & x_1 & 1 & 0 & 0 \\
            1 & x_2 & 0 & 1 & 0 \\
            1 & x_3 & 0 & 0 & 1
            \end{array} \!\right)\,.    
\end{eqnarray}
\begin{figure}[t]
\begin{center}
\epsfig{file=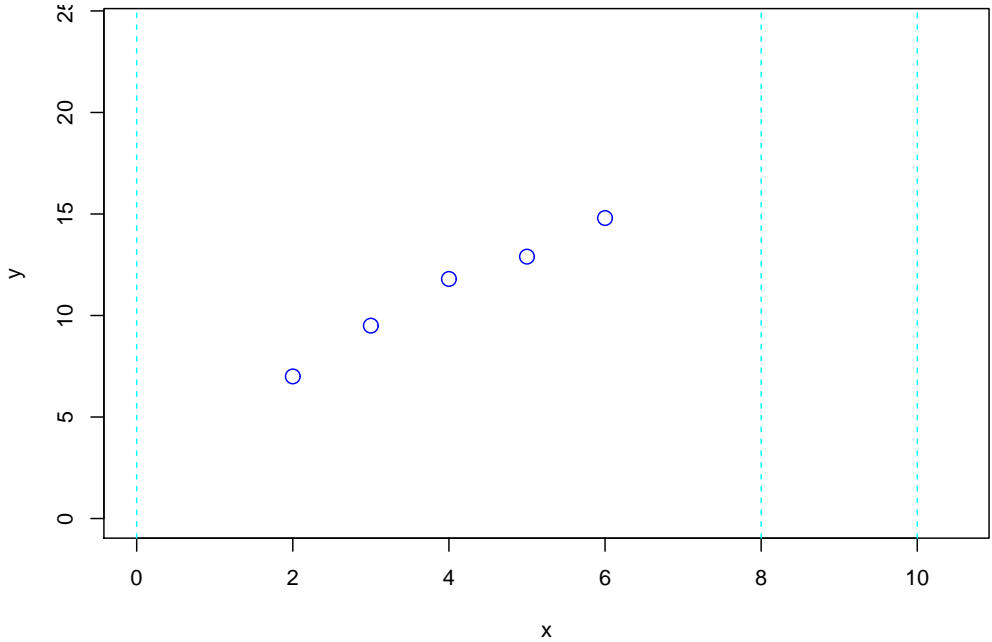,clip=,width=0.87\linewidth}  
\end{center}
\caption{Points to be fitted by a straight line. Our task is to
infer the line parameters and to make previsions about future measurements
at the $x$ points indicated by the dashed vertical lines (0, 8 and 10).
Note that no ``uncertainty bars'' have been drawn around the points,
since {\em the points are certain}(!). What are uncertain are instead
slope and intercept of the model.}
\label{fig:linear_fit_plot_0}
\end{figure}

\subsection{Numerical example with 5 `data points' and 3 previsions}
As numeric example
let us consider  (see  Fig.~\ref{fig:linear_fit_plot_0}) 
the {\bf five} $x$ values  
\code{x <- 2:6}, in correspondence of which we have `observed' 
the $y$ values 
\code{y <- c(7.0,  9.5, 11.8, 12.9, 14.8)}, in fact simulated 
by the command\\
\Rin{y <- round( 3 + 2*x + rnorm(length(x), 0, 0.5), 1)}\\
Our true values of the parameters are indeed $c=3$
and $m=2$, while the standard deviations describing the 
errors $e_i$ are all equal to 0.5. Moreover we consider 
another {\bf three} $x$ values (0, 8 and 10), about which we are interested 
in making predictions. They are indicated in
Fig.~\ref{fig:linear_fit_plot_0} by vertical dashed lines.

Having set up the problem, here is 
how we construct the initial diagonal matrix
in $R$, assigning very `uninformative priors' 
to the fit parameters.\footnote{The priors
of the numerical examples are $c=0\pm 100$ and 
$m=0\pm 100$, uncorrelated. Not that if the standard 
deviations of the priors are `quite large' then 
numerical instabilities
arise because the results depend on the sum of very 
large numbers with small ones (the most sensitive of the two is
$\sigma_0(m)$ which starts to create problems 
above 600, while $\sigma_0(c)$) is quite harmful 
up to more than 2000. \label{fn:numerical}} 
\begin{verbatim}
x <- 2:6                             # x ('predictors')
y <- c(7.0,  9.5, 11.8, 12.9, 14.8)  #  observed y
x.f <- c(8, 10, 0)                   # x of new ('future') measurements

cm.priors <- c(0,0); sigma.priors <- c(100,100)   # priors about c and m
sy <- 0.5            # standard deviation of Y values
n.points <- 8        # number of points  (5 data + 3 predictions)
mu0 <- c(cm.priors, rep(0,n.points))
sigma <- c(sigma.priors, rep(sy, n.points))
m <- n.points + 2    # dimensionality of the problem (points + parameters)
V0 <- matrix(rep(0, m*m), c(m,m))       # diagonal matrix
diag(V0) <- sigma^2  
\end{verbatim}
Then we build up the transformation matrix $C$ and calculate
the covariance matrix of the {\bf ten} quantities of the
problem (2 parameters, 5 data points and 3 points about which we want
to make predictions):
\begin{verbatim}
C <- matrix(rep(0, m*m), c(m,m))        # tranformation matrix
diag(C) <- 1
C[3:m, 1] <- 1
C[3:m, 2] <- c(x, x.f)


V <- C %*% V0 %*% t(C)                  # transformed matrix
mu <- as.vector(C %*% mu0)              # expected values
\end{verbatim}
Here are the quantities of interest
\begin{verbatim}
> sigma
 [1] 1e+03 1e+03 5e-01 5e-01 5e-01 5e-01 5e-01 5e-01 5e-01 5e-01
> C
      [,1] [,2] [,3] [,4] [,5] [,6] [,7] [,8] [,9] [,10]
 [1,]    1    0    0    0    0    0    0    0    0     0
 [2,]    0    1    0    0    0    0    0    0    0     0
 [3,]    1    2    1    0    0    0    0    0    0     0
 [4,]    1    3    0    1    0    0    0    0    0     0
 [5,]    1    4    0    0    1    0    0    0    0     0
 [6,]    1    5    0    0    0    1    0    0    0     0
 [7,]    1    6    0    0    0    0    1    0    0     0
 [8,]    1    8    0    0    0    0    0    1    0     0
 [9,]    1   10    0    0    0    0    0    0    1     0
[10,]    1    0    0    0    0    0    0    0    0     1
> mu
 [1] 0 0 0 0 0 0 0 0 0 0
> sigma.V <- sqrt(diag(V))
> round( V /outer(sigma.V,sigma.V), 4 )
        [,1]   [,2]   [,3]   [,4]   [,5]   [,6]   [,7]   [,8]   [,9]  [,10]
 [1,] 1.0000 0.0000 0.4472 0.3162 0.2425 0.1961 0.1644 0.1240 0.0995 1.0000
 [2,] 0.0000 1.0000 0.8944 0.9487 0.9701 0.9806 0.9864 0.9923 0.9950 0.0000
 [3,] 0.4472 0.8944 1.0000 0.9899 0.9762 0.9648 0.9558 0.9430 0.9345 0.4472
 [4,] 0.3162 0.9487 0.9899 1.0000 0.9971 0.9923 0.9878 0.9806 0.9754 0.3162
 [5,] 0.2425 0.9701 0.9762 0.9971 1.0000 0.9989 0.9968 0.9927 0.9895 0.2425
 [6,] 0.1961 0.9806 0.9648 0.9923 0.9989 1.0000 0.9995 0.9973 0.9952 0.1961
 [7,] 0.1644 0.9864 0.9558 0.9878 0.9968 0.9995 1.0000 0.9992 0.9979 0.1644
 [8,] 0.1240 0.9923 0.9430 0.9806 0.9927 0.9973 0.9992 1.0000 0.9997 0.1240
 [9,] 0.0995 0.9950 0.9345 0.9754 0.9895 0.9952 0.9979 0.9997 1.0000 0.0995
[10,] 1.0000 0.0000 0.4472 0.3162 0.2425 0.1961 0.1644 0.1240 0.0995 1.0000
\end{verbatim}
where the last output shows the initial correlation matrix. 
All variables are correlated, with some exceptions. In fact
intercept and slope aren't, as it should be, and the prediction
at $x=0$ (i.e. $Z_{10}$) has zero correlation with the slope (its
value is not influenced by the slope), while it is 100\% correlated
with the intercept. 

The inference on the model parameters
is finally obtained conditioning on the 
observed values of $y$ (this time we use the parameter
\code{full=FALSE} to avoid large outputs):\footnote{As 
we can see from the output, the resulting covariance matrix
is not exactly symmetrical, due to numeric effects. More
stable results  can be achieved replacing inside
\code{norm.mult.cond()} the function  \code{solve()} 
by \code{chol2inv(chol(V22))}, which makes used 
of the so called  Choleski Decomposition. 
For example \code{out\$V[2,1]} and 
\code{out\$V[1,2]}, respectively equal to $-0.09999521$ 
and $-0.09999524$, would become identical and equal to
$-0.09999498$. Nevertheless since this check has been
done only at this stage of the paper and being the result
absolutely negligible, the original 
matrix inversion function \code{solve()} has been used
also through all this section.
\label{fn:Choleski}
}
\begin{verbatim}
> ( out <- norm.mult.cond(mu, V, c(NA, NA, y, NA, NA, NA), full=FALSE ) )
$mu
[1]  3.599857  1.900031 18.800107 22.600169  3.599857

$V
            [,1]        [,2]        [,3]       [,4]        [,5]
[1,]  0.44997876 -0.09999521 -0.34998295 -0.5499734  0.44997876
[2,] -0.09999524  0.02499899  0.09999666  0.1499946 -0.09999524
[3,] -0.34998318  0.09999669  0.69999034  0.6499837 -0.34998318
[4,] -0.54997366  0.14999467  0.64998368  1.1999730 -0.54997366
[5,]  0.44997876 -0.09999521 -0.34998295 -0.5499734  0.69997876
\end{verbatim}
from which we extract standard uncertainties and correlation coefficient:
\begin{verbatim}
>  ( sigmas <- sqrt( diag(out$V) )  )
1] 0.6708046 0.1581107 0.8366543 1.0954328 0.8366473
>  ( corr <- out$V / outer(sigmas, sigmas) )
           [,1]       [,2]       [,3]       [,4]       [,5]
[1,]  1.0000000 -0.9428053 -0.6235982 -0.7484450  0.8017770
[2,] -0.9428055  1.0000000  0.7559242  0.8660217 -0.7559198
[3,] -0.6235986  0.7559244  1.0000000  0.7092032 -0.4999870
[4,] -0.7484454  0.8660219  0.7092032  1.0000000 -0.6000863
[5,]  0.8017770 -0.7559195 -0.4999867 -0.6000860  1.0000000
\end{verbatim}
Our resulting parametric inference on intercept and slope is then 
\begin{eqnarray}
c &=& 3.60 \pm 0.67 \\
m &=& 1.90 \pm 0.16 \\
\rho(c,m) &=& -0.94\,,
\end{eqnarray}
with the correlation coefficient far from being negligible,
and in fact crucial when we want to evaluate other quantities
that depend on $c$ and $m$, as we shall see in a while.

\begin{figure}[t]
\begin{center}
\epsfig{file=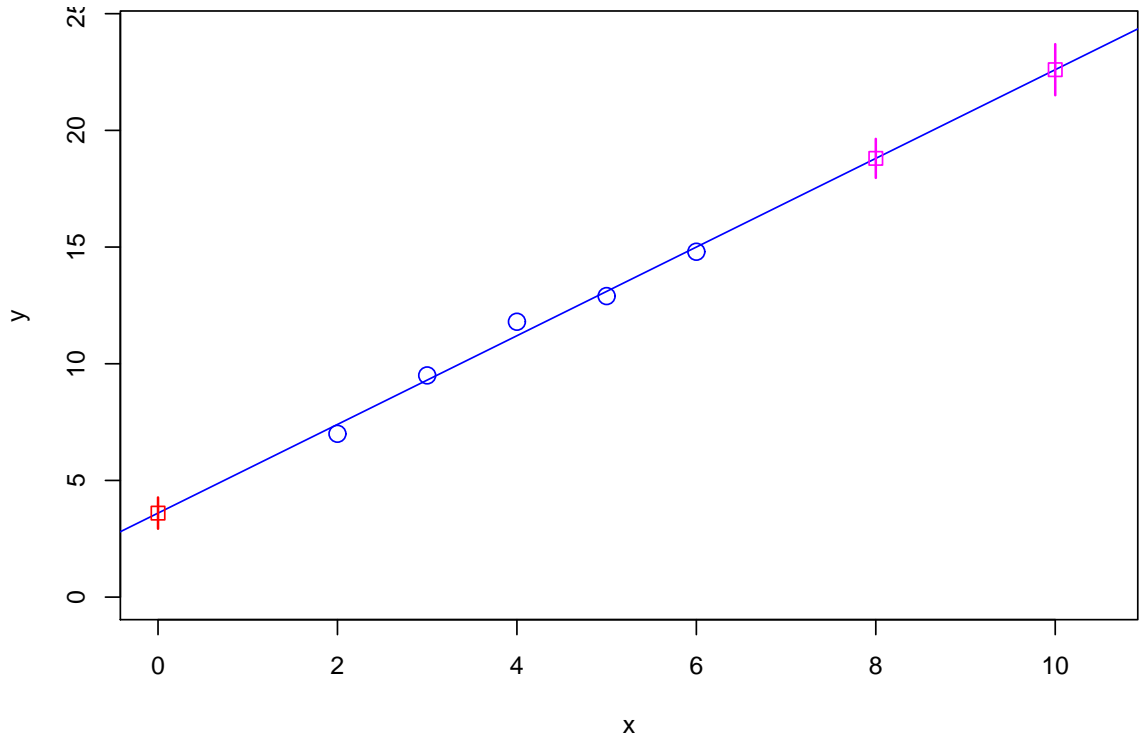,clip=,width=\linewidth}  
\end{center}
\caption{Result of the linear fit, 
including the intercept and it uncertainty. 
The prediction of possible observations at $x=8$ and $x=10$
are also reported, avoiding instead that at $x=0$ 
because it overlaps with the intercept.}
\label{fig:linear_fit_plot}
\end{figure}

We can check our result, at least as far
 expectations are concerned, against what we obtain using
the R function \code{lm()}, based on `least squares':\footnote{Under 
some conditions that usually hold in `routine' applications,
the `best estimates' of the parameters turn 
to be practically equal to those obtained using probability theory
(see e.g. \cite{BR}.)
}
\begin{verbatim}
> lm(y ~ x)

Call:
lm(formula = y ~ x)

Coefficients:
(Intercept)            x  
        3.6          1.9  
\end{verbatim}
The data points, together with the best fit line and the
intercept are reported in Fig.~\ref{fig:linear_fit_plot}. 
The expectations about the future measurements are instead
\begin{eqnarray}
Z_8 = y(x=8) &=& 18.80 \pm 0.84 \\
Z_9 = y(x=10) &=& 22.60 \pm 1.10 \\
Z_{10} = y(x=0) &=& 3.60 \pm 0.84\,,
\end{eqnarray}
with interesting correlations: 
\begin{itemize}
\item $\rho[Z_8, Z_9] = 0.71$, positive and quite high, because
      their are ``on the same side'' of the 'experimental' points
      and quite close to each other: due to the uncertainty about the slope
      they could be both smaller or larger than expected.
\item $\rho[Z_8, Z_{10}] = -0.50$, $\rho[Z_9, Z_{10}] = -0.60$ 
      negative for the opposite reason, and in absolute value 
      increasing with the distance. 
\end{itemize}
Note how the uncertainty on $Z_8$ and $Z_{10}$ are the same,
because the corresponding $x$ values (8 and 0, respectively)
are equally distant, from the 
barycenter the data along the $x$ axis. 
Instead, $\sigma(Z_{10})$ is different
from $\sigma(c)$ because {\em they are not the same thing}(!): 
the uncertainty is a parameter of the model, while 
$Z_{10}\equiv y(x=0) $ is what we would measure at $x=0$ on the base
of the information provided by the previous measurements
(and our assumptions about the model).

\subsection{Uncertainty about $\mu_y(x)$ Vs uncertainty about 
$y(x)$}
The expected $y$'s at different values of $x$ are 
simply the values $c+m\,x$ calculated at different $x$, as it is
easy to check\\
\Rin{out\$mu[1] + out\$mu[2]*x.f}\\
\Rout{18.800001 22.600001  3.599999}\\
More intriguing are the uncertainties. 
Indeed they get a contribution from the uncertainty of the 
true value  $\mu_y(x)$ and that due to the experimental error
around it. 

As far as the true values, in our simplified model they
are given by $\mu_y(x_{f_i}) = c+ x_{f_i}\,m$, which we can 
rewrite in matrix form as
\begin{eqnarray}
\left(\begin{array}{c} \mu_y(x_{f_1})\\ \mu_y(x_{f_2})\\\mu_y(x_{f_3})
      \end{array}\right) &=&  
\left(\begin{array}{cc} 1 & x_{f_1}\\  1 & x_{f_2}\\ 1 & x_{f_3}
      \end{array}\right) \cdot
\left(\begin{array}{c} c \\ m \end{array}\right) 
\end{eqnarray}
Here are then their expected values and covariance matrix
directly in R
\begin{verbatim}
> ( C.mu.f <- cbind(rep(1,3), x.f) )
       x.f
[1,] 1   8
[2,] 1  10
[3,] 1   0
\end{verbatim}
\newpage
\begin{verbatim}
>  ( mu.f <- as.vector( C.mu.f %*% out$mu[1:2] ) )
[1] 18.800107 22.600169  3.599857
> ( V.mu.f <- C.mu.f %*% out$V[1:2,1:2] %*% t(C.mu.f) )
           [,1]       [,2]       [,3]
[1,]  0.4499903  0.6499837 -0.3499832
[2,]  0.6499836  0.9499730 -0.5499737
[3,] -0.3499829 -0.5499734  0.4499788
>  ( sigma.mu.f <- sqrt(diag(V.mu.f)) )
[1] 0.6708132 0.9746656 0.6708046
\end{verbatim}
We have then 
\begin{eqnarray}
\mu_y(x=8) &=& 18.8 \pm 0.67 \\
\mu_y(x=10) &=& 22.6 \pm 0.97 \\
\mu_y(x=0) &=& 3.60 \pm 0.67,
\end{eqnarray}
with $\mu_y(x=0)$ exactly equal to the intercept. 
And again, the uncertainties on 
 $\mu_y(x=8)$ and  $\mu_y(x=10)$ are the same,
and then equal to $\sigma(c)$.\footnote{Under
the conditions we are considering here, one can prove that
\begin{eqnarray*}
\sigma^2[\mu_y(x)] &=& \frac{\sigma_y^2}{n} + 
\frac{(x-\overline{x})^2}{\overline{x^2}-\overline{x}}\cdot
\frac{\sigma_y^2}{n}\,,
\end{eqnarray*}
with the uncertainties depending on the
absolute value of $x-\overline{x}$ and on the `lever arm'
of the experimental date (the larger is their
`momentum of inertia', that is $\overline{x^2}-\overline{x}$,
the better is the determination of the slope and then
more accurate the extrapolations.
In our case this expression gives\\ 
\Rin{var.x <- sum(x\^{}2)/n\, -\, mean(x)\^{}2}\\
\Rin{sqrt(sy\^{}2/n + (x.f-mean(x))\^{}2/var.x*sy\^{}2/n )}\\
\Rout{0.6708204 0.9746794 0.6708204}\\
practically equal to the results got playing with
covariance matrices. 

Note that above formula takes into account
the correlation coefficient between $c$ and $m$. Without it
we would get\\
\Rin{sqrt(sigmas[1]\^{}2 + x.f\^{}2*sigmas[2]\^{}2)}\\
\Rout{1.4317521 1.7175207 0.6708046}\\
with $\sigma[\mu_y(x=8)]$ and 
$\sigma[\mu_y(x=10)]$ wrong by about a factor 2
(while $\sigma[\mu_y(x=0)]$ is right `by chance', 
being equal to the intercept). 
(The slight numeric difference at the 5th decimal digit
is due to the effect of the prior, not taken into account in the 
the above formula.)
} 

The reason while the uncertainties about
$y(x)$ are larger than those of  $\mu_y(x)$, for the
same $x$, is also easy to understand. To the uncertainty
about the true value we have to add that due to the 
experimental error. And in a linear model like ours
the two contributions add in quadrature, as it easy to check

while  $\mu_y(x=8)$ and  $\mu_y(x=10)$ have the standard 
uncertainty slightly smaller than those of the corresponding
 $y(x=8)$ and  $y(x=10)$. To obtain these latter standard uncertainties
it is enough to add quadratically the standard deviation of the
experimental error:
\begin{verbatim}
>  sqrt(sigma.mu.f^2 + sy^2)
[1] 0.8366542 1.0954328 0.8366473
\end{verbatim}
The effect of the experimental errors is also to dilute 
the correlations, which among the true values are
\begin{verbatim}
>  V.mu.f / outer(sigma.mu.f, sigma.mu.f)
           [,1]       [,2]       [,3]
[1,]  1.0000000  0.9941347 -0.7777671
[2,]  0.9941347  1.0000000 -0.8411826
[3,] -0.7777666 -0.8411821  1.0000000
\end{verbatim}

\subsection{Follows up}
Also in this case one do several other
instructive test, which we read to the reader.
Here is a partial list.
\begin{itemize}
\item Impose a precise value for the intercept and the slope, 
      to see how the other parameter changes. This can be
      done, for example for the intercept ($c=3$) 
      with the following conditioning:\\
      \Rin{out <- norm.mult.cond(mu, V, c(3, NA, y, NA, NA, NA), full=FALSE)}\\
      (Quick test: what would you expect for $Z_{10}$?)
\item Do the same test, but using the previous \code{out\$V}
      and \code{out\$mu}.
\item Use some informative priors for $c$, $m$ or both.
\item Make a new fit on another 5 data points generated
      from the same model, using as priors for $c$ and $m$
      the result of the previous inference (including the correlation!).
\item Make a global fit on the 10 data points of the two datasets
      (starting from uninformative priors) and compare with
      the result of the two inferences in sequence.
\end{itemize}
Then is the question of estimating the common standard deviation
of the model from the data. As told above, this cannot be done 
with the tools we are playing in this paper because the 
problem is not linear. Certainly a rough estimate can be done
by the residuals, but if the number of data points is `small'
the uncertainty on the estimated sigma do not only affect
this parameter, but also the joint pdf 
of $c$ and $m$, which is longer normal bivariate
(with consequences on the pdf's of the previsions). 
The problem has to be solved using a model without 
short cuts
and making the integrals numerically or by 
Markov Chain Monte Carlo, issues which are beyond the aim of this
paper. (And pay attention to covariance 
matrices obtained by linearization!\,\cite{CovMatrix})

\section{Propagation of evidence -- some general remarks}
\begin{figure} 
\begin{center}
\begin{tabular}{ccc}
\epsfig{file=x1-x2x3.eps,clip=} &
\epsfig{file=x1x2-x3.eps,clip=} & 
\epsfig{file=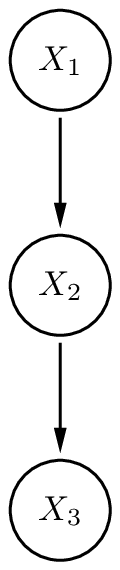,clip=} \\
& & \\
\mbox{{\large {\bf (Divergent)}}} & \mbox{{\large \bf (Convergent)}} 
& \mbox{{\large \bf (Serial)}} \\
& & \\
$ \left(\! \begin{array}{ccc} 
 \sigma_1^2 & \sigma_1^2 & \sigma_1^2 \\
 \sigma_1^2 & \sigma_1^2 + \sigma_{2|1}^2 &   \sigma_1^2 \\
 \sigma_1^2 & \sigma_1^2 &  \sigma_1^2 + \sigma_{3|1}^2 
                      \end{array} \!\right) $  &
$ \left(\! \begin{array}{ccc} 
 \sigma_1^2 & 0 & \sigma_1^2 \\ 
 0 & \sigma_2^2 &   \sigma_2^2 \\
 \sigma_1^2 & \sigma_2^2 &  \sigma_1^2 +\sigma_2^2 + \sigma_{3|1,2}^2 
                      \end{array} \!\right) $  &
$ \left(\! \begin{array}{ccc} 
 \sigma_1^2 & \sigma_1^2 & \sigma_1^2 \\ 
 \sigma_1^2 & \sigma_1^2 + \sigma_{2|1}^2 &   \sigma_1^2 + \sigma_{2|1}^2 \\
 \sigma_1^2 & \sigma_1^2 + \sigma_{2|1}^2 &  \sigma_1^2 + \sigma_{2|1}^2
+   \sigma_{3|2}^2
                      \end{array} \!\right)
$ 
\end{tabular}
\end{center}
\caption{Basics causal connections among {\em nodes}
of a {\em belief network}.}
\label{fig:propagation_evidence}
\end{figure}

Let us take again the diagrams (`{\em graphs}') 
which describe two observations
from the same true value and one observation resulting from
a true value and a systematic effect.
They are show again in Fig.~\ref{fig:propagation_evidence},
labelled with names related to the direction of the `causation'
arrows, which {\em diverge} from a single {\em node} 
or {\em converge} towards a single node. The physical interpretation
is that, as we have already seen, of a single {\em cause}
producing two {\em effects}, or two {\em causes} responsible of
a single effect, respectively.
Below each graph we have also added the covariance
matrix which characterize it, where 
$\sigma_{2|1}=\sigma[\left.X_2\right|_{X_1}]$, and so on.

For completeness
we have added in the figure also 
graph in which the effect $X_2$ is itself
 cause of another effect ({\em serial connection}). 
Sticking to the simple linear models we are dealing with, 
the transformation rules of the  graph characterize by a serial connection
are the following:
\begin{eqnarray}
X_1 &=& v_1 \\
X_2 &=& v_2 = v_1 + e_1 \\
X_3 &=& v_2 + e_2 = v_1 + e_1 + e_2 
\end{eqnarray}
from which the joint covariance matrix reported
below the diagram follows, with 
$\sigma_{2|1} = \sigma[\left.X_2\right|_{X_1} = \sigma_{e_1}$
and 
$\sigma_{3|2} = \sigma[\left.X_3\right|_{X_2} = \sigma_{e_2}$.

Analyzing the covariance matrix of the graphs
with  divergent and serial connections
we see that the variables are fully correlated:
any {\em evidence} on any of the three variables changes 
the pdf of the other two. 

Instead, in the convergent graph $X_1$ and $X_2$ are independent.
Indeed, why should the physical quantity we are going to measure 
should depend
on a calibration constant of our detector? And the other way
around.\footnote{In reality this is not impossible, but definitely unusual}
But we have already seen in the examples that 
if we observe $X_3$, then $X_1$ and $X_2$ become anticorrelated.  

The effect of the {\em propagation}  of a condition 
 (`{\em instantiation}') of one variable
to the rest of the {\em network} is very interesting also for 
its practical applications, because it allows to decompose 
a large network in subnetworks. 

\subsection{Diverging connection}
\begin{figure}
\begin{center}
\begin{tabular}{cccccc}
\epsfig{file=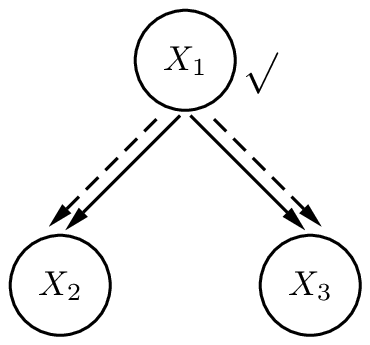,clip=} & & 
\epsfig{file=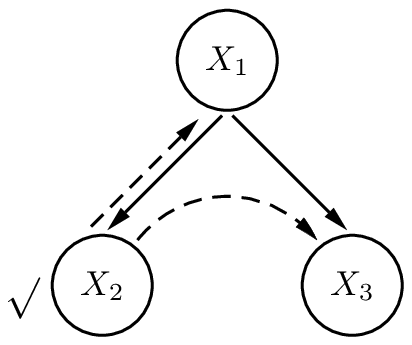,clip=} & & 
\epsfig{file=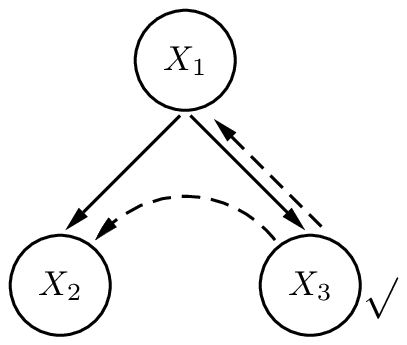,clip=} \\
&&&& \\ &&&& \\
\epsfig{file=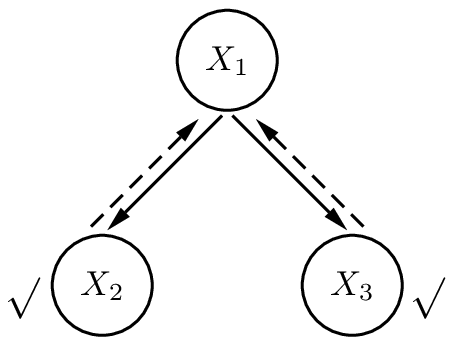,clip=} & &  
\epsfig{file=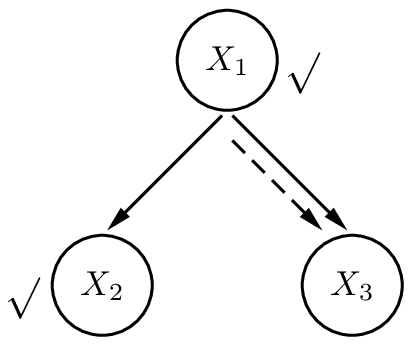,clip=} &&
\epsfig{file=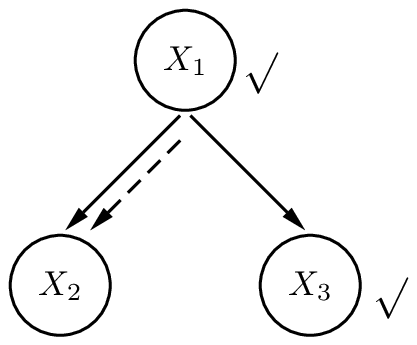,clip=}
\end{tabular}
\end{center}
\caption{Divergent connection 
with `evidence' (indicated by the symbol `$\surd$') got in some of the
variables (`instantiated nodes'). The dashed
arrows show the `flow of evidence', i.e. how
the information flows in the `network'.}
\label{fig:propagation_evidence_1}
\end{figure}
We have already seen in the numerical examples
of subsection \ref{ss:divergent_R} that if we 
condition on a value of $X_1$, then 
$X_2$ and $X_3$ become independent, and the physical reason
was very easy to be understood. This is a general property
of divergent graphs, usually stated referring to
{\em parents} and {\em children}: in a divergent graph, if
a parent is instantiated, the children become independent, i.e.
{\em evidence does not flow any longer from one child to the other}
(`an instantiated parent blocks evidence flow among children' --
we assume that there is no other connection among them!). 
The possible flows of evidence are reported in figure 
\ref{fig:propagation_evidence_1}.

Let us make the exercise to calculate the covariance matrix
of $X_2$ and $X_3$ given $X_1$. To use  Eq.~\ref{eq:Eaton_V}
we need to rewrite the three variables in a compact
form, thus defining $\bm{Y}_1 = \{X_1\}$ and  
$\bm{Y}_2 = \{X_2,\, X_3\}$. In this case
it is  convenient
to rewrite  (\ref{eq:Eaton_V}) swapping the indices,
thus obtaining: 
\begin{eqnarray}
\bm{V}\left[\left.\bm{Y}_2\right|_{\bm{Y}_1}\right] &=& 
\bm{V}_{22} - \bm{V}_{21}\,\bm{V}_{11}^{-1}\,\bm{V}_{12}\label{eq:Eaton_V2}\,,
\end{eqnarray}
with
\begin{eqnarray}
\bm{V}_{22} &=& \left(\!\! \begin{array}{cc}
                \sigma_1^2 + \sigma_{2|1}^2 &   \sigma_1^2 \\
                 \sigma_1^2 &  \sigma_1^2 + \sigma_{3|1}^2 
                 \end{array} \!\!\right) \\
\bm{V}_{21} &=&   \left(\! \begin{array}{c}
                \sigma_1^2 \\
                 \sigma_1^2 
                 \end{array}  \!\!\right) \\
\bm{V}_{11} &=&  \sigma_1^2 \\
\bm{V}_{11}^{-1} &=&  \frac{1}{\sigma_1^2} \\
\bm{V}_{12} &=&    \left(\! \! \begin{array}{cc}
                    \sigma_1^2 & \sigma_1^2
                    \end{array} \! \!\right)
\end{eqnarray}
It follows
\begin{eqnarray}
 \bm{V}_{21}\,\bm{V}_{11}^{-1}\,\bm{V}_{12}
&=&  \left(\! \begin{array}{c}
                \sigma_1^2 \\
                 \sigma_1^2 
                 \end{array}  \!\!\right) 
\cdot  \frac{1}{\sigma_1^2} \cdot 
 \left(\! \! \begin{array}{cc}
                    \sigma_1^2 & \sigma_1^2
                    \end{array} \! \!\right)
=  \left(\!\! \begin{array}{cc}
                \sigma_1^2  &   \sigma_1^2 \\
                 \sigma_1^2 &  \sigma_1^2  
                 \end{array} \!\!\right) 
\end{eqnarray}
and hence 
\begin{eqnarray}
\bm{V}\left[\left.\bm{Y}_2\right|_{\bm{Y}_1}\right] &=& 
 \left(\!\! \begin{array}{cc}
                \sigma_{2|1}^2 & 0 \\
                 0 &  \sigma_1^2 + \sigma_{3|1}^2 
                 \end{array} \!\!\right)
\end{eqnarray}
As expected, the exercise shows that $X_2$ and $X_3$
have become conditionally independent. 

\begin{figure}
\begin{center}
\begin{tabular}{cccccc}
\epsfig{file=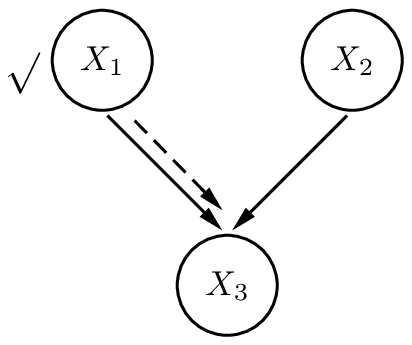,clip=} & & 
\epsfig{file=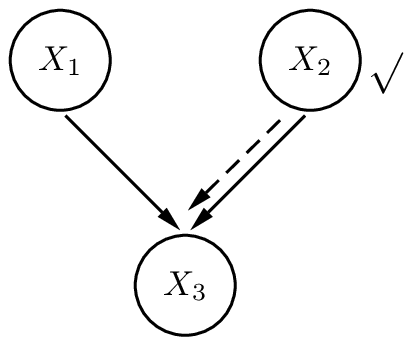,clip=} & & 
\epsfig{file=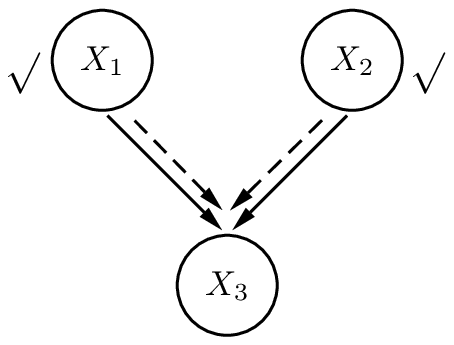,clip=} \\
&&&& \\ &&&& \\
\epsfig{file=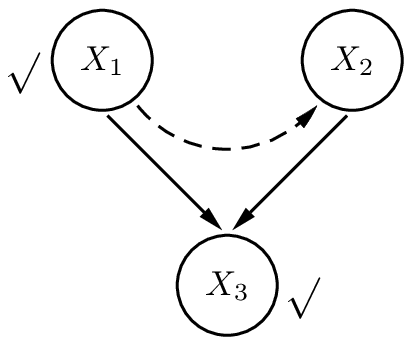,clip=} & & 
\epsfig{file=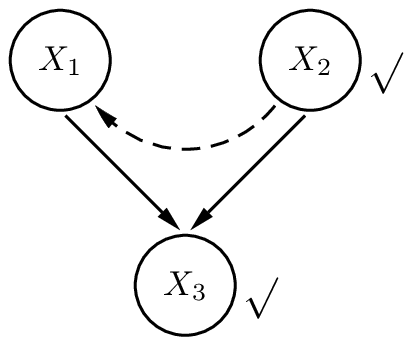,clip=} & & 
\end{tabular}
\end{center}
\caption{As Fig.~\ref{fig:propagation_evidence_1} for 
a converging connection.}
\label{fig:propagation_evidence_con}
\end{figure}

\subsection{Converging connection}
Instead, in the case of a converging connection, the 
parents are initially independent. That is
the evidence on either parent, or on both,
can influence the child(s), but cannot be transmitted
from one parent to the other, as depicted in the 
graphs of figure \ref{fig:propagation_evidence_con}.
Let us make the exercise of instantiating the child, $X_3$.

In this case the convenient partition is
$\bm{Y}_1 = \{X_1, X_2\}$ and  $\bm{Y}_2 = \{X_3\}$, 
and the conditional covariance matrix is obtained applying
directly Eq.~(\ref{eq:Eaton_V}): 
\begin{eqnarray} 
\bm{V}\left[\left.\bm{Y}_1\right|_{\bm{Y}_2}\right] &=& 
\bm{V}_{11} - \bm{V}_{12}\,\bm{V}_{22}^{-1}\,\bm{V}_{21}\label{eq:Eaton_V2}\,,
\end{eqnarray}
with
\begin{eqnarray}
\bm{V}_{11} &=& \left(\!\! \begin{array}{cc}
                \sigma_1^2  & 0 \\
                0 &  \sigma_2^2  
                 \end{array} \!\!\right) \\
\bm{V}_{12} &=&   \left(\! \begin{array}{c}
                \sigma_1^2 \\
                 \sigma_2^2 
                 \end{array}  \!\!\right) \\
\bm{V}_{22} &=&  \sigma_1^2 + \sigma_2^2 +  \sigma_{3|1,2}^2  \\
\bm{V}_{22}^{-1} &=& \left(\sigma_1^2 + \sigma_2^2 + \sigma_{3|1,2}^2\right)^{-1}\\
\bm{V}_{21} &=&    \left(\! \! \begin{array}{cc}
                    \sigma_1^2 & \sigma_2^2
                    \end{array} \! \!\right)
\end{eqnarray}
It follows
\begin{eqnarray}
 \bm{V}_{12}\,\bm{V}_{22}^{-1}\,\bm{V}_{21}
&=&  \left(\! \begin{array}{c}
                \sigma_1^2 \\ \mbox{} \\
                 \sigma_2^2 
                 \end{array}  \!\!\right) 
\cdot  \frac{1}{\sigma_1^2 + \sigma_2^2 + \sigma_{3|1,2}^2} \cdot 
 \left(\! \! \begin{array}{cc}
                    \sigma_1^2 & \sigma_2^2
                    \end{array} \! \!\right)
=  \frac{1}{\sigma_1^2 + \sigma_2^2 + \sigma_{3|1,2}^2} \cdot
    \left(\!\! \begin{array}{cc}
                \sigma_1^4  &   \sigma_1^2\sigma_2^2 \\ & \\
                 \sigma_1^2\sigma_2^2 &  \sigma_2^4  
                 \end{array} \!\!\right) \nonumber \\
&&
\end{eqnarray}
and hence 
\begin{eqnarray}
\bm{V}\left[\left.\bm{Y}_1\right|_{\bm{Y}_2}\right] &=& 
 \left(\!\! \begin{array}{cc}
 \frac{\sigma_1^2\cdot(\sigma_2^2 + \sigma_{3|1,2}^2)}
      {\sigma_1^2 + \sigma_2^2 + \sigma_{3|1,2}^2} &
   -\frac{ \sigma_1^2\sigma_2^2}{\sigma_1^2 + \sigma_2^2 + \sigma_{3|1,2}^2} \\
& \\
   -\frac{ \sigma_1^2\sigma_2^2}{\sigma_1^2 + \sigma_2^2 + \sigma_{3|1,2}^2}  &  
 \frac{\sigma_2^2\cdot (\sigma_1^2 + \sigma_{3|1,2}^2)}
      {\sigma_1^2 + \sigma_2^2 + \sigma_{3|1,2}^2} 
                 \end{array} \!\!\right)
\end{eqnarray}
As expected, the exercise shows that $X_1$ and $X_2$
become anticorrelated, although the correlation coefficient
has not a simple intuitive explanation.

\subsection{Serial connection}
\begin{figure}
\begin{center}
\begin{tabular}{cccccccccccccccc}
\epsfig{file=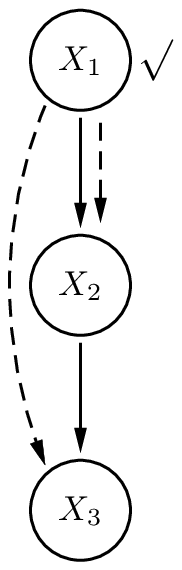,clip=}&\! & &
\epsfig{file=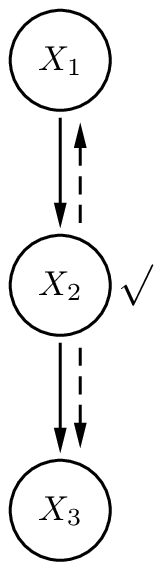,clip=}&\! & &
\epsfig{file=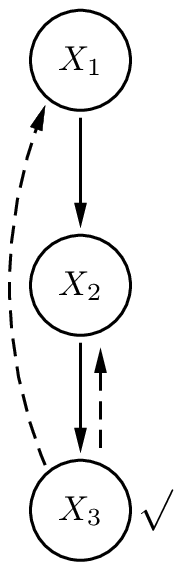,clip=}&\! & &
\epsfig{file=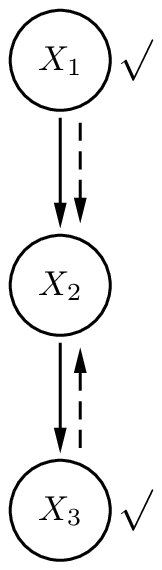,clip=}&\! & &
\epsfig{file=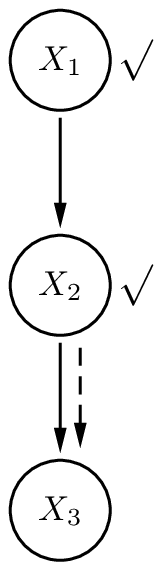,clip=}&\! & &
\epsfig{file=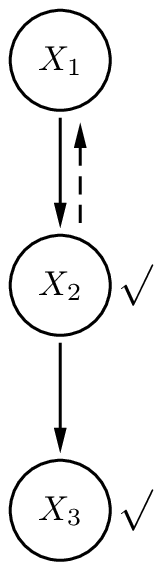,clip=}
\end{tabular}
\end{center}
\caption{As Fig.~\ref{fig:propagation_evidence_1} for 
a serial connection.}
\label{fig:propagation_evidence_1s}
\end{figure}

Let us repeat the exercise for the 
serial connection, depicted in figure \ref{fig:propagation_evidence_1s}.
The convenient partition is now
$\bm{Y}_1 = \{X_1, X_3\}$ and  $\bm{Y}_2 = \{X_2\}$. And these
are the details

\begin{eqnarray}
\bm{V}_{11} &=& \left(\!\! \begin{array}{cc}
                \sigma_1^2\ \  & \sigma_1^2 \\
               & \\
                \sigma_1^2 &  \sigma_1^2 +  \sigma_{2|1}^2 +
                               \sigma_{3|2}^2
                 \end{array} \!\!\right) \\
\bm{V}_{12} &=&   \left(\! \begin{array}{c}
                \sigma_1^2 \\
                 \sigma_1^2 +  \sigma_{2|1}^2 
                 \end{array}  \!\!\right) \\
\bm{V}_{22} &=&  \sigma_1^2 + \sigma_{2|1}^2  \\
\bm{V}_{22}^{-1} &=& \left(\sigma_1^2 + \sigma_{2|1}^2\right)^{-1}\\
\bm{V}_{21} &=&    \left(\! \! \begin{array}{cc}
                    \sigma_1^2 & \sigma_1^2 +  \sigma_{2|1}^2 
                    \end{array} \! \!\right)
\end{eqnarray}
It follows
\begin{eqnarray}
 \bm{V}_{12}\,\bm{V}_{22}^{-1}\,\bm{V}_{21}
&=&  \left(\! \begin{array}{c}
                \sigma_1^2 \\
                 \sigma_1^2 + \sigma_{2|1}^2 
                 \end{array}  \!\!\right) 
\cdot  \frac{1}{\sigma_1^2 + \sigma_{2|1}^2} \cdot 
 \left(\! \! \begin{array}{cc}
                    \sigma_1^2 & \sigma_1^2 +  \sigma_{2|1}^2
                    \end{array} \! \!\right)  \nonumber \\
&=&  \frac{1}{\sigma_1^2 + \sigma_{2|1}^2} \cdot
    \left(\!\! \begin{array}{cc}
  \sigma_1^4  & \sigma_1^2\cdot(\sigma_1^2+\sigma_{2|1}^2)  \\
& \\
  \sigma_1^2\cdot(\sigma_1^2+\sigma_{2|1}^2) & (\sigma_1^2+\sigma_{2|1}^2)^2  
                 \end{array} \!\!\right)
\end{eqnarray}
and hence 
\begin{eqnarray}
\bm{V}\left[\left.\bm{Y}_1\right|_{\bm{Y}_2}\right] &=& 
 \left(\!\! \begin{array}{cc}
 \frac{\sigma_1^2\sigma_{2|1}^2}{\sigma_1^2 + \sigma_{2|1}^2} &
 0 \\
 0  & \sigma_{3|2}^2 
  \end{array} \!\!\right)
\end{eqnarray}
As expected, the exercise shows that $X_1$ and $X_3$ 
become now independent and the uncertainty about 
$X_3$ is simply $\sigma_{3|2}$. And also in 
$\sigma[\left. X_1\right|_{X_2}]$ we recognize 
a familiar pattern (see also footnote 
\ref{fn:media_pesata}):
\begin{eqnarray}
\frac{1}{\sigma^2[\left. X_1\right|_{X_2}]} &=& 
\frac{1}{\sigma_1^2} + \frac{1}{\sigma_{2|1}^2}\,.
\end{eqnarray}

\mbox{}
\vspace{0.5cm}



\begin{thebibliography} {ref99}

\bibitem{Gauss}
C.F. Gauss, {\it ``Theoria motus corporum coelestium in sectionibus 
conicis solem ambientum''}, Hamburg 1809, n.i 172--179; reprinted
in Werke, Vol. 7 (Gota, G\"ottingen, 1871), pp 225--234.\\
(See e.g. in \url{http://www.roma1.infn.it/~dagos/history/GaussMotus/index.html})

\bibitem{R} R Core Team (2013). {\em R: A language and 
  environment for statistical
  computing}. R Foundation for Statistical Computing, Vienna, Austria.
  URL \url{http://www.R-project.org}.


\bibitem{Eaton}
M. Eaton, {\em Multivariate Statistics : A Vector Space Approach}, 
   John Wiley and Sons 1983
(available at \url{http://projecteuclid.org/}).

\bibitem{Kalman}
See e.g. \url{http://www.cs.unc.edu/~welch/kalman/}.

\bibitem{BR}
G. D'Agostini, {\it Bayesian reasoning in data analysis -- 
a critical introduction}, World Scientific 2003.

\bibitem{Cowan}
G. Cowan, {\em Statistical data analysis}, Clarendon Press, 1998. 

\bibitem{fits}
G. D'Agostini, {\it Fits, and especially linear fits, 
with errors on both axes, extra variance of the data points
 and other complications}, arXiv:physics/0511182.

\bibitem{CovMatrix}
G. D'Agostini, {\it On the use of the covariance matrix to fit correlated
data}, Nucl. Instrum. Methods. {\bf  A346} (1994) 306.

\end{thebibliography}
\end{document}